\documentclass[a4paper,12pt,fleqn]{article}
\pdfoutput=1

\makeatletter

\@addtoreset{equation}{section}
\makeatother
\usepackage{pgfplots}

\usepackage{amsfonts}
\usepackage{amsmath}
\usepackage{latexsym}
\usepackage{geometry}
\usepackage{braket}

\usepackage[affil-it]{authblk}

\usepackage{mathrsfs}

\usepackage{tikz}
\usepackage{tikz-3dplot}

\usetikzlibrary{positioning}
\usetikzlibrary{shapes.geometric}

\usepackage{float}

\usepackage{caption}
\usepackage{epigraph}
\usepackage{changepage} 

\usetikzlibrary{decorations.pathmorphing}

\usetikzlibrary{arrows.meta}

\usepackage{tikz}
\usetikzlibrary{decorations.markings}

\tikzset{
  midarrow/.style={
    postaction={decorate},
    decoration={markings, mark=at position 0.5 with {\arrow{>}}}
  },
  midarrow reversed/.style={
    postaction={decorate},
    decoration={markings, mark=at position 0.5 with {\arrow[reverse]{>}}}
  }
}

\def\be{\begin{equation}}
\def\ee{\end{equation}}
\def\bea{\begin{eqnarray}}
\def\eea{\end{eqnarray}}
\def\({\left(}
\def\){\right)}
\def\<{\left<}
\def\>{\right>}

\def\be{\begin{equation}}
\def\ee{\end{equation}}
  \def\bea{\begin{eqnarray}}
\def\eea{\end{eqnarray}}

\def\({\left(}
\def\){\right)}
\def\<{\left<}
\def\>{\right>}

\def\be{\begin{equation}}
\def\ee{\end{equation}}
\def\bea{\begin{eqnarray}}
\def\eea{\end{eqnarray}}
\def\ben{\begin{eqnarray}}
\def\een{\end{eqnarray}}
\def\({\left(}
\def\){\right)}
\def\<{\left<}
\def\>{\right>}
\def\|{\left|}

\def\x{\right|}
\def\+{\bar}
\def\mb{\mathbb}

\def\I{{\cal{I}}}

\def\t{\widetilde}

\def\O{{\cal{O}}}

\def\eps{{\cal{\varepsilon}}}

\def\x{{\vec{x}}}

\def\i{{\bar{i}}}

\def\l{{{\ell}}}

\def\*{{\times}}

\tikzset{
  midarrow/.style={
    decoration={markings, mark=at position 0.5 with {\arrow{>}}},
    postaction={decorate}
  }
}

\tikzset{
  midarrows/.style={
    decoration={markings, mark=at position 0.5 with {\arrow{<}}},
    postaction={decorate}
  }
}
 
\begin{document}

\vskip-10pt
\vskip-10pt
\hfill 
\begin{center}
\vskip 3truecm
{\Large \bf
A nonabelian Wilson surface on a lattice}
\vskip 2truecm
{\large \bf
Andreas Gustavsson}
\vspace{1cm} 
\begin{center} 
Physics Department, University of Seoul, Seoul 02504 KOREA
\end{center}
\end{center}
\vskip 2truecm
{\abstract We analyze the nonabelian surface holonomy on a bipartite hypercubic lattice following a proposal in arXiv:1002.4636 [hep-th]. The bipartite structure of the lattice enables us to introduce spike string configurations. These spikes play a crucial role for the time evolution of the string when the total number of color indices changes.}

\vfill
\vskip4pt
\eject
\pagestyle{plain}

\section{Introduction}
The 6d $(2,0)$ superconformal theory for the ADE series of possible gauge groups was first discovered in 1995 in \cite{Witten:1995zh}. Later a brane picture for the $A_{N-1}$ series of gauge groups was presented with an M2 brane ending on a stack of M5 branes along a selfdual string  in \cite{Strominger:1995ac}. For an abelian gauge group, the worldvolume theory on an M5 brane is described by an abelian tensor multiplet that includes a two-form gauge potential $B$ whose field strength $H = dB$ is selfdual. Let us now consider $N+1$ parallel M5 branes and let us separate one of these branes from a stack of $N$ coincident M5 branes. This breaks the gauge group from $U(N+1)$ down to $U(N)$. We imagine that this happens by a Higgs mechanism, but the precise details of such a Higgs mechanism are not known in six dimensions. From the brane perspective, we may have an M2 brane that stretches from the distant M5 to the stack of coincident M5 branes and it should be seen as a tensile selfdual string in the $U(N)$ gauge theory that lives on the stack of coincident M5 branes. The selfdual string appears as a central charge in the supersymmetry algebra of the M5 brane theory. We can partially understand this for a nonabelian gauge group since the supersymmetry algebra itself makes no reference to the gauge group, although if the gauge group is nonabelian, we do not have an explicit realization of these central charges. From this algebra and its central charge that corresponds to the selfdual strings, one may argue that two parallel selfdual strings shall be BPS saturated and that there shall be a zero net force between them. The BPS nature of two parallel selfdual strings was also shown explicitly for an abelian tensor multiplet minimally coupled to an effective action for the two selfdual strings in \cite{Gustavsson:2001wa}, where it was shown that for two parallel selfdual strings there is an electromagnetic repulsion force that cancels the Higgs field attraction force. If on the other hand, we have two anti-parallel selfdual strings, then this system is unstable and we expect that the two strings will come together and annihilate into electromagnetic radiation. The net electric and magnetic charges for this system are vanishing, which makes it consistent for the two strings to annihilate into charge neutral radiation particles. We expect that a closed selfdual string will behave in a similar way so that it will shrink and annihilate itself into radiation, unless there is an obstruction that stabilizes it. If there is a circle direction in the manifold, then the string will stabilize at the radius of the circle direction. Such a closed string will be perceived as a charged particle under dimensional reduction around the circle. Dimensional reduction of the selfdual string as it wraps either of the one-cycles of a two-torus was used to explain electric-magnetic duality in abelian four-dimensional Maxwell theory in \cite{Verlinde:1995mz}.

For an abelian two-form gauge potential we have an associated gauge invariant Wilson surface for any closed two-dimensional smooth surface. Its expectation value is a measurable real number, which, depending on the geometry, can be either an observable or a Weyl anomaly \cite{Gustavsson:2004gj}, \cite{Drukker:2020dcz}. On the other hand, a surface that has a boundary is not gauge invariant. However, Wilson surfaces with boundaries that we will refer to as surface holonomies, are important when we couple them to selfdual strings. In order to simplify the analysis, we will not consider selfdual strings with equal magnetic and electric charges. Instead we will consider an electrically charged tensile string with a vanishing magnetic charge. We treat this string as a first quantized string that we describe by a Schrodinger wave function. The question that we will address is how the surface holonomy acts on a tensile string as it moves in spacetime.  To this end, we need to understand how to define a nonabelian surface holonomy. 

A proposal for a nonabelian Wilson surface on a hypercubic lattice was presented in \cite{Rey:2010uz}, which was further studied in \cite{Lipstein:2014vca}. However, this construction of a nonabelian Wilson surface on a hypercubic lattice dates back already to the early \textquoteright80s  where it first appeared in \cite{Nepomechie:1982rb}. See also \cite{Orland:1982fv}, \cite{O1}, \cite{O2} for related works around that time. In this paper we will apply this proposal and show that the surface holonomy can transport a tensile electrically charged string across the lattice in a consistent manner. This will enable us to obtain the unitary compactness condition the surface holonomy shall satisfy. We will also show that the surface holonomy on a plaquette has a unit element up to a gauge transformation.

\section{Tensile and tensionless selfdual strings}
We would like to describe the tensile selfdual string that lives on a stack of $N$ coincident M5 branes. This string originates from a single M2 brane that  stretches from a distant M5 brane to the stack of $N$ coincident M5 branes. For $N>1$ this selfdual string is poorly understood. On the other hand, for $N = 1$ we have an abelian tensor multiplet that is well understood, but in this case there is no smooth solitonic selfdual string solution. However, we may introduce a dyonic string by hand as a one-dimensional submanifold in space, whose 2d worldsheet $\Sigma$ is minimally coupled to a two-form gauge potential $B$ through a coupling term $e \int_{\Sigma} B$ where $e$ is its electric charge. The dyonic string may also have a magnetic charge. To account for this charge we attach a Dirac membrane $D$ whose boundary $\partial D = \Sigma$ is the string worldsheet. In this formulation, $B$ remains globally defined single-valued gauge potential outside the Dirac membrane. The magnetic coupling is introduced by modifying the definition of the field strength as $H = dB + g\delta_D$ where $g$ is the magnetic charge and $\delta_D$ is the delta function supported on the Dirac membrane that we may define implicitly as $\int s \wedge \delta_D = \int_D s$ for an arbitrary three-form $s$ on spacetime. The selfdual string has both these two charges at equal values, $e = g$.

The selfdual string is also expected to couple to the five scalar fields $\phi^A$ where $A = 1,...,5$ label the five transverse dimensions to the M5 brane. The tension is given by $T = v$ where $v$ is related to the vacuum expectation value as $\<\phi^A\> = n^A v$ where $v = \sqrt{\<\phi^A\> \<\phi^A\>}$ and $n^A$ is a unit vector. The fluctuations of these scalar fields around the vacuum expectation value introduce a coupling $\int_{\Sigma} n^A \varphi^A$ where we defined $\phi^A = \<\phi^A\> + \varphi^A$.

The action for the combined system of gauge field, scalar fields and this dyonic string, is given by
\bea
S &=& - \frac{1}{2} \int H \wedge * H - \int d\phi^A \wedge *d\phi^A + e \int_{\Sigma} B \cr
&& - T \int_{\Sigma} d^2 \sigma \sqrt{-g} + \int_{\Sigma} d^2 \sigma \sqrt{-g} n^A \varphi^A
\eea
where $g$ denotes the induced metric on $\Sigma$. This action was presented in \cite{Arvidsson:2004xa} for $e = g$. Thus our Nambu-Goto action term is the usual Nambu-Goto action term with a nondynamical string tension. While we have no direct way of deriving the Nambu-Goto action term for this string with its tension $T$ as a fixed nondynamical factor, we do have some supporting evidence from conformal anomaly computations \cite{Drukker:2020dcz}, \cite{Gustavsson:2004gj} for the coupling term to the five scalar fields as $\sim n^A \varphi^A$. In \cite{Arvidsson:2004xa}, the following square root $\sqrt{\(v n^A + \varphi^A\) \(v n^A + \varphi^A\)} = v + n^A \varphi^A + \O(1/v)$ was used as the string tension. By expanding this square root, we see that we can recover our fixed string tension $T = v$ from the first term, and the subleading term gives the coupling between the scalar field and the string. A more rigorous derivation of this effective action would be to start from the $U(2)$ gauge theory and turning on a  Higgs vacuum expectation value. From this we would derive an effective action for the selfdual string by following the computation in the appendix of  \cite{Drukker:1999zq} for a massive W boson in $N=4$ super Yang-Mills theory. This approach is unfortunately not accessible as we do not have an action for the M5 brane theory with an unbroken $U(2)$ gauge symmetry. Our proposed effective action has a limited range of validity where $\O(1/v)$ corrections fall outside this effective theory of a tensile string entirely. We can be confident about the electric and magnetic couplings as valid effective action terms at a sufficiently far distance, much larger than the characteristic lenght scale $1/\sqrt{v}$, away from the tensile string where its precise microscopic characterization becomes irrelevant. 

We may describe the Dirac membrane, which supports the string with its magnetic charge, as an embedded surface $x^M = Y^M(\tau,\sigma,\rho)$ parametrized in some way by three worldvolume parameters $\tau,\sigma,\rho$. When we vary $Y^M$ in the action, we get a constraint equation that relates the conjugate momentum $\pi_M$ of $Y^M$ back to to $B$ and $Y^M$ and their spatial derivatives, but to no time derivatives. This means that we get a constraint rather than a dynamical equation of motion for the Dirac membrane. For tensile strings, this constraint can be quantized by a Schrodinger wave functional where we put $\pi_M = i \hbar \frac{\delta}{\delta Y^M}$. For two dyonic strings with charges $(e,g)$ and $(e',g')$ one can from this quantized constraint show that we shall have the following charge quantization condition \cite{Deser:1997se}
\bea
e g' + g e' &\in & 2 \pi \hbar \mb{Z}
\eea
by letting the two Dirac membranes wrap around the charged strings without ever intersecting or crossing each other, refered to as a `double-pass' in  \cite{Deser:1997se}. Selfdual strings have equal electric and magnetic charges. For such strings, this charge quantization condition becomes
\bea
e^2 &=& \pi \hbar \mb{Z}
\eea
The smallest unit of charge will then be $e = \sqrt{\pi \hbar}$. Other, independent considerations, such as holomorphic factorization, and commuting Wilson surfaces \cite{Gustavsson:2001wa}, \cite{Henningson:2000nu}, have shown that the single M5 brane action is characterised by a coupling constant that corresponds to $e = \sqrt{\pi \hbar}$, which shows that the selfdual string that appears in the M5 brane has its charge fixed at precisely the lowest possible value $e = \sqrt{\pi \hbar}$, consistent with the general charge quantization condition. 

Let us now study the $U(N)$ gauge theory (for $N>1$) that lives on $N$ parallel M5 branes that we will label by $M5_i$ for $i = 1,...,N$. This theory is poorly understood so it is difficult to describe it. Therefore we will slightly separate these M5 branes from each other by putting them at different locations $a_i$ so that $a_1<a_2<...<a_N$ in one transverse direction. Between $M5_i$ and $M5_j$ whose separation distance is $a_j - a_i > 0$ we can stretch an orthogonal M2 brane. The gauge group that we have on the collection of separated M5 branes is $U(1)^N$. This gauge group is generated by the abelian charges $H_i$. These charges measure the electric charge of an M2 brane boundary string as
\bea
[H_i,E_{ij}] &=& + e E_{ij}\cr
[H_j,E_{ij}] &=& - e E_{ij}\cr
[H_k,E_{ij}] &=& 0 \quad \mbox{for $k\notin \{i,j\}$}
\eea
We may think of $E_{ij}$ as the collection of two boundary string charges, or more simply, as the M2 brane charge\footnote{We use the same word `charge' here in to different contexts that should not be mixed up. First our charge may refer to the charge of a physical object (such as a string). Second, our charge may instead refer to the operator that measures that charge. Which type of charge that we a referring to in each situation should hopefully be clear from the context.}. We can also have an anti-M2 brane whose charges $E_{ji}$ are opposite to those of the M2 brane, 
\bea
[H_i,E_{ji}] &=& - e E_{ji}\cr
[H_j,E_{ji}] &=& + e E_{ji}
\eea
We may consider merging together two M2 brane charges $E_{ij}$ and $E_{jk}$ along a common boundary string in $M5_j$ if we assume that $i<j<k$. These branes carry opposite charges on M5$_j$ so that they can join and merge together there and annihilate their charges. This creates a single M2 brane $E_{ik}$ stretching between M5$_i$ and M5$_k$. This can be expressed as a commutator relation
\bea
[E_{ij},E_{jk}] &=& E_{ik}
\eea
where it should be noticed that $E_{jk} E_{ij} = 0$ reflecting the fact that for $k\neq i$ we can not cancel the abelian charges as measured by $H_k$ and $H_i$ since these live on two different M5 branes. We may also have an M2/anti-M2 brane annihilation process that we may express as the commutation relation
\bea
[E_{ij},E_{ji}] &=& H_i - H_j
\eea
Taken together, these commutation relations form the $U(N)$ Lie algebra. 
The reason why the $U(N)$ group is spontaneously broken is because the Cartan generators $H_i$ as a result of the brane separation become abelian charges that live on each of the different M5 branes, while the root generators $E_{ij}$ are different generators that we associate with M2 branes. Therefore we can not rotate the root generators and Cartan generators together while preserving this brane setup. When the M5 branes are brought together, the gauge group is enhanced to $U(N)$ which means that the Cartan generators and the root generators are rotated among each other under $U(N)$ and hence need to be completely mixed up to the extent that we no longer can make any meaningful distinction between tensor multiplet particles and selfdual strings. Instead, at this unbroken phase, we expect to see some different kind of degrees of freedom, which we may loosely refer to as tensionless strings. 

We shall now extend the above brane setup and consider $N+1$ parallel M5 branes that we place in the transverse direction along the points $a_1 = a_2 = ... = a_N << a_{N+1}$. Then we can have a very long M2 brane stretching between the stack of $N$ coincident M5 branes and the very distant M5 brane $M5_{N+1}$. The gauge group $U(N+1)$ is spontaneously broken to $U(N) \times U(1)_{N+1}$ by this brane setup. We will not discuss what happens on the distant M5 brane and instead we will only consider the $U(N)$ gauge symmetry of the coincident M5 branes. We can understand the $U(N)$ gauge charges better if we slightly separate the $N$ coincident M5 branes from one another, so that $a_1<a_2<...<a_N<<a_{N+1}$ and the gauge group is broken down to $U(1)^N$. An M2 brane that stretches from the distant $M5_{N+1}$ may end on $M5_{i}$ for some particular value $i = 1,...,N$. From the perspective of the $N$ almost coincident M5 branes, this M2 brane will be perceived as an abelian tensile string inside $M5_i$ with the $U(1)_i$ electric charge $H_i = +e$. For the other M5 branes there will be no electric charge, $H_j = 0$ for $j \neq i$. 

If we would describe an infinitely long straight string on $M5_i$ in the broken phase with $U(1)^N$ gauge symmetry by a Schrodinger wave function, then this wave function will be given by
\bea
\psi^j(t,\vec{x}) &=& \psi(t,\vec{x}) \delta^j_i
\eea
where we put a single gauge index $j$ labeling the M5 branes and $\delta^j_i$ tells us that the string is located on that specific $M5_i$ brane. The same string, when oriented in the opposite direction, will be described by the complex conjugate wave function
\bea
\psi_j(t,\vec{x}) &=& \psi^*(t,\vec{x}) \delta_j^i
\eea
We let $\vec{x}$ describe the position of the string in its transverse four-dimensional space and $t$ is time. The assumptions that we need to make about this string is that it is tensile and that its motion is nonrelativistic so that it can be described by quantum mechanics. 

These straight strings, as we consider them in a translationally symmetric state, behave like abelian particles on $M5_i$ as they move nonrelativistically in the transverse four-space. Their wave functions will transform by an abelian holonomy phase associated with their electric charges. 

In quantum mechanics, if we are given a set of possible basis quantum states, then we can also have the system in a superposition state of those basis quantum states, a so-called Schrodinger cat state. In the $U(1)^N$ gauge theory such a state would take a more general form as 
\bea
\psi^k(t,\vec{x})
\eea
The interpretation of this state should be that $\int d^4 x |\psi^k(t,x)|^2$ gives the probability that the string is confined to $M5_k$ at time $t$. We allow this state to depend on time, which means that as time evolves, we can have a changing probability for the string to end on different M5 branes. 

Let us now consider two parallel selfdual strings. These are BPS saturated strings and they can move arbitrarily close to each other. They can be on top of each other. The low energy effective field theory that lives on multiple coincident M2 branes is an ABJM theory. The boundary of the stack of coincident M2 branes is a composite selfdual string that is composed of two elementary parallel selfdual strings on top of each other. The relevant Lie algebra commutation relation for this case is 
\bea
[E_{N+1,i},E_{N+1,j}] &=& 0
\eea
which says that no decay is possible, the right-hand side being zero, of these strings, which are thus mutually stable. 

Let us next consider two anti-parallel selfdual strings. These are not BPS saturated but attract each other both electromagnetically and by the Higgs field interaction. So they will collide and annihilate into tensor multiplet particles or light selfdual strings if we study the M5 branes in the broken phase with $U(1)^N$ gauge group. That these processes may be expected can be read from the Lie algebra relations of $U(N)$,
\bea
[E_{N+1,i},E_{i,N+1}] &=& H_{N+1} - H_i
\eea
and 
\bea
[E_{N+1,i},E_{j,N+1}] &=& E_{ij} 
\eea
respectively. The former relation describes two anti-parallel strings on the same $M5_i$ that get annihilated into tensor multiplet particles on both the distant $M5_{N+1}$ and on $M5_i$. The latter relation describes two anti-parallel strings on two different $M5_i$ and $M5_j$ that get annihilated into a light selfdual string stretching between these two nearby M5 branes. 

As we bring the M5 branes together, the original tensile strings, whose tensions were $T=v$, as the two anti-parallel strings meet, they transition into tensionless strings that we may interpret as degrees of freedom  associated with the generators $H_i$ and $E_{ij}$ of the unbroken $U(N)$ gauge group. There is no discernable difference between degrees of freedom associated with $H_i$ and with $E_{ij}$ respectively, because these generators get rotated into each other by the $U(N)$ gauge symmetry.

\section{From $U(1)$ to $U(N)$ higher gauge theory} 
While we are interested in the tensile selfdual strings for an unbroken $U(N)$ gauge symmetry ($N=1,2,3,...$), we will simplify the problem. For $N = 1$ our simplified model amounts to ignoring the magnetic charge of the tensile string. We will also ignore the coupling of the tensile string to the scalar field fluctuations. For the above $U(1)$ effective description, we can see that these couplings of the selfdual string do not affect the surface holonomy that the string will pick up as it moves in spacetime, which comes entirely from its minimal electric coupling. 

Let us begin with $N=1$ where we consider Maxwell theory on $\mb{R}^{1,5}$ of a two-form gauge potential $B$ with field strength $H = dB$. In this theory we may consider an electrically charged string with tension $T$ and electric charge $e$. The action for this system is a sum
\bea
S &=& S_{Maxwell} + S_{string}
\eea
where $S_{string} = S_{NG} + S_B$ are given by
\bea
S_{Maxwell} &=& - \frac{1}{12} \int d^6 x H_{MNP} H^{MNP}\cr
S_{NG} &=& - T \int_{\Sigma} d^2 \sigma \sqrt{-g} \cr
S_B &=& e \int_{\Sigma} B\label{abelianstring}
\eea
respectively. Here $-g>0$ is minus of the determinant of the induced metric on the string worldsheet $\Sigma$ in Lorentzian signature. The two-form gauge potential is integrated over $\Sigma$ in a metric independent way. 

For a nearly straight string we may describe small fluctuations around it by choosing static gauge
\bea
X^0 &=& t\cr
X^5 &=& s\cr
X^i &=& \xi^i(t,s)
\eea
and expanding in powers of the small fluctuation $\xi^i$. To lowest order we get
\bea
S_{string} &=& \frac{T}{2} \int dt ds \(\dot{\xi}^2 - \xi'^2\) + e \int dt ds B_{ts}(t,x,\xi(t,s))
\eea
The Hamiltonian is 
\bea
H &=& \int ds \(\frac{1}{2T} \frac{\delta^2}{\delta \xi^i(s) \delta \xi^i(s)} + \frac{T}{2} \xi'(s)^2\) - e \int ds B_{ts}
\eea
The Schrodinger equation is
\bea
i \hbar \frac{\partial \psi}{\partial t} &=& H \psi
\eea
If we assume that the string moves adiabatically\footnote{One way that we may move the string adiabatically is by letting it be static and just letting it evolve in time. Another way is may take $T$ to infinity and move the string as a rigid string.} , then we solve this equation as
\bea
\psi(t_2) &=& e^{\frac{i e}{\hbar} \int_{t_1}^{t_2} \int ds B_{ts} } \psi(t_1)
\eea
We then elevate this argument to a general claim that upon quantization, the Schrodinger wave function of this string changes by a phase factor as
\bea
\psi(C) &\rightarrow & \psi(C') = U(\Sigma) \psi(C) \label{transport}
\eea
if we move the string adiabatically from the initial curve $C$ to the final curve $C'$ along the surface $\Sigma$ where the phase factor 
\bea
U(\Sigma) &=& e^{\frac{i e}{\hbar} \int_{\Sigma} B}
\eea
is the surface holonomy valued in the $U(1)$ gauge group, satisfying the compactness condition
\bea
U(\Sigma)^* U(\Sigma) &=& 1
\eea
as follows from that $B$ is real. This claim can be understood by a path integral argument where the wave function changes by the phase factor $e^{\frac{i}{\hbar} S_B}$ if we assume that adiabatic motion means that we can neglect the contribution from $S_{NG}$. From now on, we will put $\hbar = 1$ by a choice of natural units. 

It is important to notice that the phase factor that comes from the surface holonomy is topological. We can deform the spacetime metric and we can move the string as fast as we like from its initial to its final configuration. The surface holonomy is insensitive to it and will give the same result. However, for a purely electrically charged string, we do find a simplification when we move the string adiabatically. As we can ignore the dynamical phase factor coming from the kinetic energy, we get only the surface holonomy as the transformation factor if we move the string adiabatically. But even this would be impossible if we include couplings to scalar fields. Then even an adiabatic motion of the selfdual string will not isolate the topological phase factor and we will instead need to consider the metric-dependent Maldacena-Wilson surface. In this paper, we focus our attention entirely on the topological phase factor that comes from the gauge charges and we will ignore any other dynamically generated phase factors of the selfdual string. 

Under a $U(1)$ gauge transformation
\bea
B \rightarrow B + d\Lambda
\eea
the surface holonomy transforms as
\bea
U(\Sigma) \rightarrow g(\partial \Sigma) U(\Sigma)
\eea
where the $U(1)$ gauge group element $g(\partial \Sigma)$ is evaluated on the boundary $\partial \Sigma$ of the surface $\Sigma$ and is given by
\bea
g(\partial \Sigma) &=& e^{i e \int_{\partial \Sigma} \Lambda}
\eea
If $\Sigma$ has the boundary given by two loops $C'$ and $C$, then 
\bea
g(\partial \Sigma) &=& e^{i e \(\int_{C'} \Lambda - \int_C \Lambda\)}
\eea
that we may also write as
\bea
g(\partial \Sigma) &=& g(C') \(g(C)\)^*
\eea
where
\bea
g(C) &=& e^{i e \int_{C} \Lambda} 
\eea
The sign factors that appear in these two boundary integrals are important, so let us explain it in detail. On the surface $\Sigma$ we have an intrinsic orientation that we may represent by drawing small circles everywhere on the surface, each circle has its own orientation. We determine the orientation of nearby circles so that any two circles that meet tangentially at any point has opposite tangential directions. This means that once we decide the orientation of just one small circle, the orientation of all the other circles in $\Sigma$ is uniquely fixed. This collection of oriented small circles is what we may call as the orientation of $\Sigma$. Now we shall imagine the boundary circle $C'$ being connected tangentially to a small circle in $\Sigma$ with a tangential direction that is parallel to the tangent direction of $C'$. Then that small circle in turn is connected to another small circle in $\Sigma$ tangentially so that their tangents point in opposite directions and this will continue with many small circles all the way down to $C$. Then the circle that is tangential to $C$ will have a tangential direction that is anti-parallel to the tangential direction of $C$. That is the explanation for why we have a relative minus sign between the two integrals. We may summarize this long text by writing it as a compact formula,
\bea
\partial \Sigma &=& C' \cup C^\vee
\eea
where $C^\vee$ denotes the loop $C$ with its orientation reversed. We draw the surface and its two boundary loops below,
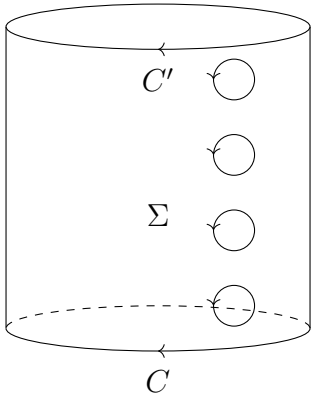
\bea
\begin{tikzpicture}  
  \draw (2,0) arc[start angle=0, end angle=180, x radius=2cm, y radius=0.3cm];
  \draw[midarrow] (2,0) arc[start angle=0, end angle=-180, x radius=2cm, y radius=0.3cm];
  \draw (-2,0) -- (-2,-4);
  \draw (2,0) -- (2,-4);
  \draw[dashed] (2,-4) arc[start angle=0, end angle=180, x radius=2cm, y radius=0.3cm];
  \draw[midarrow] (2,-4) arc[start angle=0, end angle=-180, x radius=2cm, y radius=0.3cm];
  \node at (0,-0.7) {$C'$};        
  \node at (0,-4.7) {$C$};        
  \node at (0,-2.5) {$\Sigma$};   

  \foreach \y in {-0.7,-1.7,-2.7,-3.7} {
    \draw[midarrow] (1,\y) circle (0.27);
  }
\end{tikzpicture}
\eea
The wave function needs to also transform under the gauge transformation as
\bea
\psi(C') &\rightarrow & g(C') \psi(C')\cr
\psi(C) &\rightarrow & g(C) \psi(C)
\eea
while the surface holonomy transforms as
\bea
U(\Sigma) &\rightarrow & g(C') U(\Sigma) \(g(C)\)^*
\eea
in order to ensure that the transport of the string by the surface holonomy as given by equation (\ref{transport}), is gauge covariant. Let us now consider the case when $C'$ approaches $C$ so that the area of $\Sigma$ shrinks to zero when $C'$ is on top of $C$. Let us denote this degenerate surface as $\Sigma_0$. For such a degenerate surface, the surface holonomy becomes gauge invariant,
\bea
U(\Sigma_0) \rightarrow  g(C) U(\Sigma_0) \(g(C)\)^* = U(\Sigma_0)\label{vanarea}
\eea
We state this rather trivial observation here because it turns out to be important as we turn to the nonabelian generalization. 

If we parametrize the closed string as $s\to C^M(s)$ for $s \sim s+2\pi$, then the gauge covariant differential of the wave function is given by
\bea
D \psi(C) &=& \int ds \delta C^M(s) \(\frac{\delta \psi(C)}{\delta C^M(s)} - i e B_{MN}(C(s)) \frac{dC^N(s)}{ds} \psi(C)\)
\eea
By taking the complex conjugate, we get
\bea
\(D \psi(C)\)^* &=& \int ds \delta C^M(s) \(\frac{\delta \psi(C)^*}{\delta C^M(s)} + i e B_{MN}(C(s)) \frac{d{C}^N(s)}{ds} \psi(C)^*\)
\eea
The same operator on the right-hand side is obtained if we change the orientation of the string such that $\dot{C}^M(s)$ is replaced by $-\dot{C}^M(s)$,
\bea
D \t\psi(C^\vee) &=& \int ds \delta C^M(s) \(\frac{\delta \t\psi(C^\vee)}{\delta C^M(s)} + i e B_{MN}(C(s)) \frac{d{C}^N(s)}{ds} \t\psi(C^\vee)\)
\eea
Here $\t\psi(C^\vee)$ is another wave function that is evaluated on $C^\vee$, the string whose orientation is reversed compared to $C$. We can now gauge covariantly equate these two wave functions as
\bea
\psi(C)^* &=& \t\psi(C^\vee)
\eea
Another way to argue for this relation comes from the complex conjugation relations
\bea
\psi(C')^* &=& U(\Sigma)^* \psi(C)^*\cr
U(\Sigma)^* &=& U(\Sigma^\vee)
\eea
Here $\Sigma^\vee$ is the surface $\Sigma$ with its orientation reversed. For the boundaries, we have
\bea
\partial \Sigma &=& C' \cup C^\vee\cr
\partial \Sigma^\vee &=& C'^\vee \cup C^\vee
\eea
Now since $C^\vee$ is just another closed curve, we shall also have
\bea
\t\psi(C'^\vee) &=& U(\Sigma^\vee) \t\psi(C^\vee)
\eea
In this way we again see that we may write a covariant relation as
\bea
\psi(C)^* &=& \t\psi(C^\vee) 
\eea
Below we illustrate this relation for $\Sigma$ a cylinder,
\bea
\(
\vcenter{
  \hbox{
    \begin{tikzpicture}  
  \draw (2,0) arc[start angle=0, end angle=180, x radius=2cm, y radius=0.3cm];
  \draw[midarrow] (2,0) arc[start angle=0, end angle=-180, x radius=2cm, y radius=0.3cm];
  \draw (-2,0) -- (-2,-4);
  \draw (2,0) -- (2,-4);
  \draw[dashed] (2,-4) arc[start angle=0, end angle=180, x radius=2cm, y radius=0.3cm];
  \draw[midarrow] (2,-4) arc[start angle=0, end angle=-180, x radius=2cm, y radius=0.3cm];

  \node at (0,-0.7) {$C'$};        
  \node at (0,-4.7) {$C$};        
  \node at (0,-2.5) {$\Sigma$};   

  \foreach \y in {-0.7,-1.7,-2.7,-3.7} {
    \draw[midarrow] (1,\y) circle (0.27);
  }
    \end{tikzpicture}
  }
}
\)^*
\;\;=\;\;
\vcenter{
  \hbox{
    \begin{tikzpicture}  
  \draw (2,0) arc[start angle=0, end angle=180, x radius=2cm, y radius=0.3cm];
  \draw[midarrow] (-2,0) arc[end angle=0, start angle=-180, x radius=2cm, y radius=0.3cm];
  \draw (-2,0) -- (-2,-4);
  \draw (2,0) -- (2,-4);
  \draw[dashed] (2,-4) arc[start angle=0, end angle=180, x radius=2cm, y radius=0.3cm];
  \draw[midarrow] (-2,-4) arc[end angle=0, start angle=-180, x radius=2cm, y radius=0.3cm];

  \node at (0,-0.7) {$C'^\vee$};        
  \node at (0,-4.7) {$C^\vee$};        
  \node at (0,-2.5) {$\Sigma^\vee$};   

  \foreach \y in {-0.7,-1.7,-2.7,-3.7} {
    \draw[midarrows] (1,\y) circle (0.27);
  }
    \end{tikzpicture}
  }
}
\eea

Let us now consider a spacetime where one direction is circle compactified. Then we may consider a sitution where our cylinder is extended in time and wraps around the circle direction in space. If we then perform dimensional reduction along the circle direction, the cylinder $\Sigma$ connecting the two circles $C$ and $C'$ will reduce to a line $\gamma$ connecting two endpoints $x$ and $x'$. On this dimensionally reduced line holonomy, complex conjugation acts as drawn below,
\bea
\(
\vcenter{
  \hbox{
    \begin{tikzpicture}  
[white/.style={circle, draw, fill=white, inner sep=0pt, minimum size=6pt},
 black/.style={circle, draw, fill=black, inner sep=0pt, minimum size=6pt}]
  \draw[midarrow] (0,-4) -- (0,0);
  \node at (0.5,0) {$x'$};        
  \node at (0.5,-4) {$x$};        
  \node at (0.5,-2) {$\gamma$};   
      \node[black] at (0,-4) {};
      \node[black] at (0,0) {};
    \end{tikzpicture}
  }
}
\)^*
\;\;=\;\;
\vcenter{
  \hbox{
    \begin{tikzpicture} 
[white/.style={circle, draw, fill=white, inner sep=0pt, minimum size=6pt},
 black/.style={circle, draw, fill=black, inner sep=0pt, minimum size=6pt}]
  \draw[midarrow] (0,0) -- (0,-4);
  \node at (0.5,0) {$x'$};        
  \node at (0.5,-4) {$x$};        
  \node at (0.5,-2) {$\gamma^\vee$};   
      \node[black] at (0,-4) {};
      \node[black] at (0,0) {};
     \end{tikzpicture}
  }
}
\eea
This graphical relation shows that the line holonomy satisfies 
\bea
\(U(\gamma)\)^* &=& U(\gamma^\vee) 
\eea
where $\gamma^\vee$ denotes the line with its orientation reversed. The gauge group was assumed to be abelian, but for the line holonomy this has a natural generalization to $U(N)$ gauge group where the relation becomes 
\bea
\(U^{i'}{}_i(\gamma)\)^* &=& U^i{}_{i'}(\gamma^\vee)
\eea
This relation is illustrated in the graph below,
\bea
\(
\vcenter{
  \hbox{
    \begin{tikzpicture}  
[white/.style={circle, draw, fill=white, inner sep=0pt, minimum size=6pt},
 black/.style={circle, draw, fill=black, inner sep=0pt, minimum size=6pt}]
  \draw[midarrow] (0,-4) -- (0,0);
  \node at (0.5,0) {$x',i'$};        
  \node at (0.5,-4) {$x,i$};        
  \node at (0.5,-2) {$\gamma$};   
      \node[black] at (0,-4) {};
      \node[black] at (0,0) {};
    \end{tikzpicture}
  }
}
\)^*
\;\;=\;\;
\vcenter{
  \hbox{
    \begin{tikzpicture} 
[white/.style={circle, draw, fill=white, inner sep=0pt, minimum size=6pt},
 black/.style={circle, draw, fill=black, inner sep=0pt, minimum size=6pt}]
  \draw[midarrow] (0,0) -- (0,-4);
  \node at (0.5,0) {$x',i'$};        
  \node at (0.5,-4) {$x,i$};        
  \node at (0.5,-2) {$\gamma^\vee$};   
      \node[black] at (0,-4) {};
      \node[black] at (0,0) {};
     \end{tikzpicture}
  }
}
\eea
We would now like to move back to the surface holonomy and try to understand how we shall define it for $U(N)$ gauge group. As a first attempt, we may place one color index on each closed string. Complex conjugation then appears to works fine for a cylinder, as drawn below,
\bea
\(
\vcenter{
  \hbox{
    \begin{tikzpicture}  
  \draw (2,0) arc[start angle=0, end angle=180, x radius=2cm, y radius=0.3cm];
  \draw[midarrow] (2,0) arc[start angle=0, end angle=-180, x radius=2cm, y radius=0.3cm];
  \draw (-2,0) -- (-2,-4);
  \draw (2,0) -- (2,-4);
  \draw[dashed] (2,-4) arc[start angle=0, end angle=180, x radius=2cm, y radius=0.3cm];
  \draw[midarrow] (2,-4) arc[start angle=0, end angle=-180, x radius=2cm, y radius=0.3cm];

  \node at (0,-0.7) {$C',i'$};        
  \node at (0,-4.7) {$C,i$};        
  \node at (0,-2.5) {$\Sigma$};   
    \end{tikzpicture}
  }
}
\)^*
\;\;=\;\;
\vcenter{
  \hbox{
    \begin{tikzpicture}  
  \draw (2,0) arc[start angle=0, end angle=180, x radius=2cm, y radius=0.3cm];
  \draw[midarrow] (-2,0) arc[end angle=0, start angle=-180, x radius=2cm, y radius=0.3cm];
  \draw (-2,0) -- (-2,-4);
  \draw (2,0) -- (2,-4);
  \draw[dashed] (2,-4) arc[start angle=0, end angle=180, x radius=2cm, y radius=0.3cm];
  \draw[midarrow] (-2,-4) arc[end angle=0, start angle=-180, x radius=2cm, y radius=0.3cm];

  \node at (0,-0.7) {$C'^\vee,i'$};        
  \node at (0,-4.7) {$C^\vee,i$};        
  \node at (0,-2.5) {$\Sigma^\vee$};   
    \end{tikzpicture}
  }
}
\eea
However, if we place only one color index on each string, then it is difficult to understand what shall happen to that index if the string splits into two strings,\footnote{If we have a flat line holonomy, then we may use that to parallel transport the color index $j$ from a basepoint to any other point. However, this does not apply to the M5 brane theory where we do not have a line holonomy, flat or not. Namely, suppose that we had such a line holonomy. Then we would also have a Wilson loop observable. But there is no such observable in the abelian M5 brane theory. Since the loop itself is a geometric object, it exists independently of the choice of gauge group and so it follows that we do not have a line holonomy in the nonabelian generalization either.} 
\bea
\begin{tikzpicture}  
  \draw (2-3,0) arc[start angle=0, end angle=180, x radius=2cm, y radius=0.3cm];
  \draw[midarrow] (2-3,0) arc[start angle=0, end angle=-180, x radius=2cm, y radius=0.3cm] node[midway, below] {$i'$};
  \draw (2+3,0) arc[start angle=0, end angle=180, x radius=2cm, y radius=0.3cm];
  \draw[midarrow] (2+3,0) arc[start angle=0, end angle=-180, x radius=2cm, y radius=0.3cm] node[midway, below] {$???$};
  \draw (-2-3,0) -- (-2,-3);
  \draw (-2,-3) -- (-2,-4);
  \draw (2+3,0) -- (2,-3);
  \draw (2,-3) -- (2,-4);
  \draw (-1,0) -- (0,-1);
  \draw (0,-1) -- (1,0);
  \draw[dashed] (2,-4) arc[start angle=0, end angle=180, x radius=2cm, y radius=0.3cm];
  \draw[midarrow] (2,-4) arc[start angle=0, end angle=-180, x radius=2cm, y radius=0.3cm] node[midway, below] {$i$};
\end{tikzpicture}
\eea
But in the $U(1)$ gauge theory we can have a surface holonomy that describes a string splitting process. Every quantity that is involved in such a string splitting process is well-defined. The surface holonomy on such a worldsheet is well-defined as we just need to integrate the two-form $B$ over it. For the gauge parameter, we may write it as
\bea
g(C) &=& e^{i e \int_0^{2\pi} ds \Lambda_M(C(s)) \dot{C}^M(s)}
\eea
for the string just the moment before it splits, and as
\bea
g(C) &=& e^{i e \int_0^a ds \Lambda_M(C(s)) \dot{C}^M(s)} e^{i e \int_a^{2\pi} ds \Lambda_M(C(s)) \dot{C}^M(s) \dot{C}^M(s)}
\eea
the moment after the string has split into two strings. Here the string splits after the string develops a self-intersection point at parameter value $a\in (0,2\pi)$ where 
\bea
C^M(a) &=& C^M(0)
\eea
The string splitting process can be iterated indefinitely by dividing the original parameter interval $[0,2\pi]$ into smaller and smaller sub-intervals repeatedly. That this is possible stems from the uncountable nature of the number of points in the interval $[0,2\pi]$ so that no matter how short we make a sub-interval, it still contains just as many points as the original interval. 

For the generalization to $U(N)$ gauge group, and if we put only one color index on each string,  
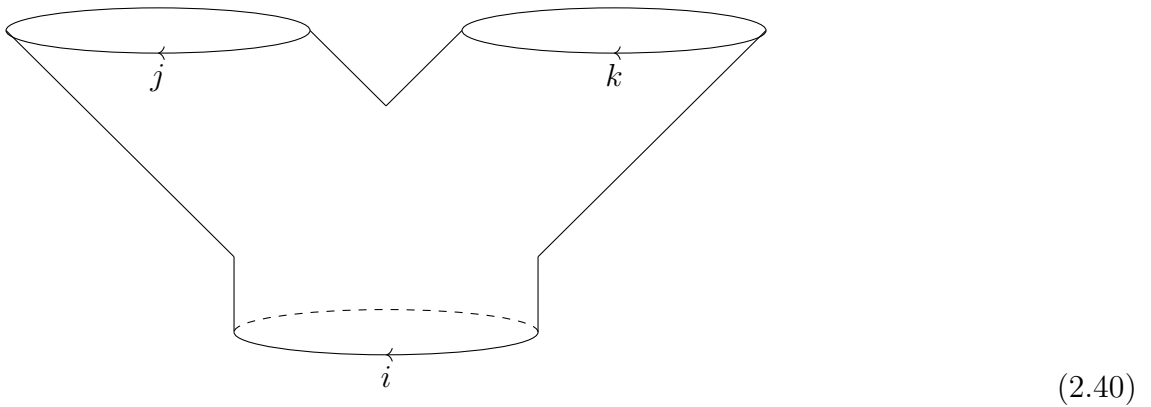
\bea
\begin{tikzpicture}  
  \draw (2-3,0) arc[start angle=0, end angle=180, x radius=2cm, y radius=0.3cm];
  \draw[midarrow] (2-3,0) arc[start angle=0, end angle=-180, x radius=2cm, y radius=0.3cm] node[midway, below] {$j$};
  \draw (2+3,0) arc[start angle=0, end angle=180, x radius=2cm, y radius=0.3cm];
  \draw[midarrow] (2+3,0) arc[start angle=0, end angle=-180, x radius=2cm, y radius=0.3cm] node[midway, below] {$k$};
  \draw (-2-3,0) -- (-2,-3);
  \draw (-2,-3) -- (-2,-4);
  \draw (2+3,0) -- (2,-3);
  \draw (2,-3) -- (2,-4);
  \draw (-1,0) -- (0,-1);
  \draw (0,-1) -- (1,0);
  \draw[dashed] (2,-4) arc[start angle=0, end angle=180, x radius=2cm, y radius=0.3cm];
  \draw[midarrow] (2,-4) arc[start angle=0, end angle=-180, x radius=2cm, y radius=0.3cm] node[midway, below] {$i$};
\end{tikzpicture}
\eea
then the surface holonomy will be a map from a wave function with one color index $i$, to a wave function describing two disjoint strings, each with one color index $j$ and $k$ respectively. It seems difficult to understand how that can be described by a unitary process.
 
If instead we put two color indices $i$ and $j$ on the incoming string, then we can preserve the total number of color indices if we put one color index on each outgoing string,
\bea
\begin{tikzpicture}  
  \draw (2-3,0) arc[start angle=0, end angle=180, x radius=2cm, y radius=0.3cm];
  \draw[midarrow] (2-3,0) arc[start angle=0, end angle=-180, x radius=2cm, y radius=0.3cm] node[midway, below] {$k$};
  \draw (2+3,0) arc[start angle=0, end angle=180, x radius=2cm, y radius=0.3cm];
  \draw[midarrow] (2+3,0) arc[start angle=0, end angle=-180, x radius=2cm, y radius=0.3cm] node[midway, below] {$\l$};
  \draw (-2-3,0) -- (-2,-3);
  \draw (-2,-3) -- (-2,-4);
  \draw (2+3,0) -- (2,-3);
  \draw (2,-3) -- (2,-4);
  \draw (-1,0) -- (0,-1);
  \draw (0,-1) -- (1,0);
  \draw[dashed] (2,-4) arc[start angle=0, end angle=180, x radius=2cm, y radius=0.3cm];
  \draw[midarrow] (2,-4) arc[start angle=0, end angle=-180, x radius=2cm, y radius=0.3cm] node[midway, below] {$i,j$};
\end{tikzpicture}
\eea
Such a process corresponds to a surface holonomy that takes two incoming color indices $i$ and $j$ and transforms them into two outgoing color indices $k$ and $\l$. Since we have equal numbers of color indices incoming as outgoing, we can describe this process with a unitary matrix. But the surface does not necessarily have to be a worldsheet surface that extends in time. It can also be a surface that is located at an instant in time and extends in spatial directions. Alternatively, we may Wick rotate the time direction so that all directions are spatial. In such a situation there is no invariant way to tell what will be an incoming and what will be an outgoing closed string. The same process can be viewed as if the string with the color index $k$ is incoming and splits into two strings, each with color indices $i,j$ and $\l$ respectively and then it is difficult to see how such a process can be unitary. 

There now seems to be only two possibilities to achieve a unitary development. Either we have an abelian gauge group and a surface holonomy carries no color indices, or else if the gauge group is nonabelian, and we will assume a $U(N)$ gauge group, then we need an infinite number of color indices that are associated to each string. In both cases, we then have the property that a string can be split into two strings in an indefinite sequence of iterated processes. Importantly, this possibility of having an indefinite iterating sequence of string splitting arises as a result of having an uncountable infinite set of color indices and of points in the parameter interval $[0,2\pi]$. Here we thus start to see a possible connection between spacetime points and color indices.

\section{From uncountable to countable}
If spacetime is a continuum, then it seems like we need an uncountable number of color indices to decorate the string. However, even in the continuum of spacetime and a smooth string, that infinity should be possible to coarse-grain. 

Let us start by looking at the loop algebra associated to the gauge group $U(N)$. The loop algebra generators will be denoted $t_a(s)$. They satisfy the Lie algebra
\bea
[t_a(s),t_b(s')] &=& i f_{ab}{}^c \delta(s-s') t_c(s)
\eea
where $f_{ab}{}^c$ are the structure constants of $U(N)$. The zero modes, 
\bea
t_a &=& \int_0^{2\pi} ds t_a(s)
\eea
are the reparametrization invariant generators of $U(N)$. The loop algebra generators have the nice property that they can easily handle a string the splits into two strings. If the string splits at $s = a$
where $a$ is a number in the range $0<a<2\pi$, then the generators $\{t_a(s)|0\leq s\leq a\}$ will act only on the first string and the generators $\{t_a(s)|a \leq s\leq 2\pi\}$ will act only on the second string and the generators from differents sets commute. Generators within each set form separate loop algebras. One might worry that a contact term could arise from the common point $s=a$. But the following argument shows that no contact term will arise there. Let us view the generators as maps from the space of continuous functions on the circle into the loop algebra, 
\bea
t_a[f] &:=& \int_0^{2\pi} ds t_a(s) f(s)
\eea
The loop algebra is given by
\bea
[t_a[f],t_b[g]] &=& i f_{ab}{}^c t_c[fg]
\eea
From this result it immediately follows that if $f$ has its support in $[0,a]$, that is, on the first string, and $g$ has its support in $[a,2\pi]$, that is, on the second string, then $fg$ vanishes in the entire interval $[0,2\pi]$ for continuous functions $f$ and $g$, so the two sets of generators commute. Within each string, we now have got an inherited loop algebra. Thus from a single parent loop algebra, we see that we can generate two children loop algebras that commute with each other and who live on each string separately. This process can be iterated indefinitely by dividing the interval $[0,2\pi]$ into smaller and smaller peaces and there is no limit on how finely we can divide this interval, the key point being that no matter how 'small' an interval is, it will still contain just as many 'points' as the original interval did. That is, an uncountable infinite number of points. 

Let us now consider a closed string and divide it into $n$ intervals $I_m = [a_{m-1},a_m]$ with $a_0 = 0$ and $0<a_1<a_2<...<a_{n-1}<2\pi$. We associate with the interval $I_m$ a wave function $\psi^i_m$. Here $i = 1,...,N$ is a fundamental index of $U(N)$. The wave function of the whole string is the product
\bea
\psi^{i_n \cdots i_1} &=& \psi^{i_n}_{n} \cdots \psi^{i_1}_1\label{stringwave}
\eea
in a simplified notation, as later we will see that half the color indices are anti-fundamentals indices that we shall put downstairs. Here we will ignore this detail though. We then define
\bea
t_a &=& \sum_{m=0}^{n-1} (t_a)^i{}_j \psi^j_m \frac{\partial}{\partial \psi^i_m} 
\eea
that act on the string wave function as
\bea
t_a \psi^{i_n \cdots i_1} &=& \sum_{m=0}^{n-1} (t_a)^{i_m}{}_{j_m}  \psi^{i_n \cdots i_{m-1} j_m i_{m+1}\cdots i_1}
\eea
These generators can be obtained from a continuum limit that is given by the loop algebra. In the continuum limit we replace the coarse-grained wave functions $\psi^i_m$ on each interval on the string with $\psi^i(s)$ at each point of the string. We then define the action of the loop algebra generators at each point of the string as
\bea
t_a(s) &=& (t_a)^i{}_j \psi^j(s) \frac{\delta}{\delta \psi^i(s)}
\eea
We then get back the above coarse-grained system from the continuum string by restricting the wave function to be stepwise constant on each interval,
\bea
\psi^i(s) &=& \psi^i_m \qquad \mbox{for} \qquad s\in I_m
\eea
It is not clear to us how to write the wave function of the entire string in the continuum limit. We do not have the proper continuum version of (\ref{stringwave}). Intuitively this would be an uncountably infinite product of wave functions
\bea
"\Pi_{s \in [0,2\pi]} \psi^{i(s)}(s)"
\eea
The difficulty in understanding the continuum limit of the wave function of the string is what motivates us to study a lattice regularization.

The wave function of the string may carry several color indices $i,j,k,...$ and we do not assume any additional structure for these color indices, such as that they form an adjoint representation of the gauge group.  Instead, we will assume that the wave function transforms in the highly reducible representation 
\bea
N \otimes \bar{N} \otimes N \otimes \bar{N} \otimes N \cdots \otimes \bar{N} 
\eea
This suggests that a continuum limit either does not exist, or if it does exist, then it will require us to develop a representation theory that goes  beyond simple loop algebras that can deal with truly infinite-dimensional representations of the $U(N)$ gauge group.

\section{Coloring the lattice}
We have encountered a problem that we seem to need an infinite number of color indices placed on a tensile string if we want to study this string in the $U(N)$ gauge theory with $N>1$.\footnote{However, another possibility that one might entertain is if finitely many indices may be sufficient but where both the number of indices and their distribution along the string will be in a quantum state determined by the quantum state of the string. Whether that would leads to an infinite regress, where the quantum state of the string determines the quantum state of the string, is not clear to us.}

In order to make progress with these many color indices, we will replace the spacetime manifold with a discrete hypercubic lattice and switch to Euclidean signature. A hypercubic lattice may be constructed as a set of vertices that are located at integer points $a \mb{Z}^6 \subset \mb{R}^6$ where $a$ is a lattice spacing parameter and where these vertices are connected by edges that go along the six Euclidean coordinate directions in $\mb{R}^6$. This is a lattice that is embedded in $\mb{R}^6$. We do not need this much structure of the lattice for our purposes, where we only will study the surface holonomy of a tensile string as it moves over the lattice since the surface holonomy does not depend on the metric of the spacetime. So we only need to know the topological structure of the lattice. The lattice can be stretched and deformed and embedded into any curved Riemannian geometry that is smoothly connected to a flat Euclidean spacetime. The topology of the lattice can be understood by looking at the 6d cube, also called the hexaract. By stacking hexaracts together we can build up the lattice. We describe the topology of the hexaract in the appendix. 

The hypercubic lattice built with hexeracts has a bipartite structure where each white vertex is connected to twelve black vertices by twelve edges and each black vertex is connected by twelve edges to twelve white vertices. We put one color index $i,j,k,\l,m,....= 1,...,N$ on each edge in this lattice. In this way, each edge gets the name of the index we put there. The most important structure in this lattice for string dynamics is the plaquette. When the string moves, a segment of the string has to cross some plaquette, regardless of how the string moves. The plaquette has one face, four edges and four vertices,
\bea
\begin{tikzpicture}[scale=3,
    white/.style={circle, draw, fill=white, inner sep=0pt, minimum size=6pt},
    black/.style={circle, draw, fill=black, inner sep=0pt, minimum size=6pt},
    label/.style={font=\small, midway, sloped, above},
    coord/.style={font=\scriptsize, anchor=south east},
    coordne/.style={font=\scriptsize, anchor=north east},
    coordnw/.style={font=\scriptsize, anchor=north west},
    coordsw/.style={font=\scriptsize, anchor=south west}
]
\coordinate (A) at (0,0,0);   
\coordinate (B) at (1,0,0);   
\coordinate (C) at (1,1,0);   
\coordinate (D) at (0,1,0);   
\node[coordne] at (A) {$$};
\node[coordnw] at (B) {$$};
\node[coordsw] at (C) {$$};
\node[coord] at (D) {$$};
\draw (A) -- (B) node[midway, below] {$\l$};
\draw (B) -- (C) node[midway, right] {$k$};
\draw (C) -- (D) node[midway, above] {$j$};
\draw (D) -- (A) node[midway, left] {$i$};
\node[white] at (A) {};
\node[black] at (B) {};
\node[white] at (C) {};
\node[black] at (D) {};
\end{tikzpicture}
\eea
A string in this lattice goes along the edges in the lattice. We define a 'segment' of the string as a portion of that string that goes along exactly one edge in the lattice. 

A string that stretches over edges in the lattice, will cross vertices of different type interchangingly each time. We will associate one color index $i,j,k,...$ of the wave function of the string to each edge that the string stretches over. If the gauge group is $U(N)$ then $i = 1,...,N$ and for $N>2$ we have two distinct $N$-dimensional representations, the fundamental $\psi^i$ and the anti-fundamental $\psi_i$ representation.\footnote{For $N=2$ we may still place the indices upstairs and downstairs, although in that case there is an invariant antisymmetric tensor $\eps_{ij}$ relating them. For $SO(N)$ gauge group the string is unoriented and we shall not distinguish between indices upstairs or downstairs. One way to see that the strings are unoriented is by recalling that for the D4 branes an $SO(N)$ gauge group arise from unoriented strings ending on these D4 branes. These strings originate correspond to unoriented M2 branes wrapping the M-theory circle. We conclude that the selfdual strings must be unoriented since they are the boundaries of these unoriented M2 branes.} The string has an orientation, for $U(N)$ gauge group, which enables us to decide whether a color index shall be fundamental or anti-fundamental from the geometry of the string on the lattice. We use the convention that if the arrow of the string points away from a white vertex of the lattice then we put the color index upstairs and otherwise downstairs. We will write these color indices reading from the right to the left as we go along the string in the direction of its orientation or its arrow direction. The color indices are indices associated to the string. But we find it convenient to put the color indices on all the edges of the lattice. This makes the indices act not only as internal indices of the gauge group, but also as geometrical markers for where the string is located. As the string moves over the lattice, the indices on the string will change so that the string picks up the indices from the edges of the lattice over which it stretches. Given a geometric view of the color indices as describing the entire string, we only need to read the color indices as we go along the string to specify its geometric shape in the lattice. Below we show a string on the lattice with three different color indices whose wave function is written by our set of conventions as $\psi^k{}_j{}^i$,
\bea
\vcenter{\hbox{\begin{tikzpicture}[scale=3,
    white/.style={circle, draw, fill=white, inner sep=0pt, minimum size=6pt},
    black/.style={circle, draw, fill=black, inner sep=0pt, minimum size=6pt},
    label/.style={font=\small, midway, sloped, above},
    coord/.style={font=\scriptsize, anchor=south east},
    coordne/.style={font=\scriptsize, anchor=north east},
    coordnw/.style={font=\scriptsize, anchor=north west},
    coordsw/.style={font=\scriptsize, anchor=south west}
]
\coordinate (A) at (0,0);   
\coordinate (B) at (1,0);   
\coordinate (C) at (2,0);   
\coordinate (D) at (3,0);   
\node[coordne] at (A) {$$};
\node[coordnw] at (B) {$$};
\node[coordsw] at (C) {$$};
\node[coord] at (D) {$$};
\draw[red,midarrow, line width=1.5pt] (A) -- (B) node[midway, below] {$i$};
\draw[red,midarrow, line width=1.5pt] (B) -- (C) node(main)[midway, below] {$j$};
\draw[red,midarrow, line width=1.5pt] (C) -- (D) node[midway, below] {$k$};
\node[white] at (A) {};
\node[black] at (B) {};
\node[white] at (C) {};
\node[black] at (D) {};
\end{tikzpicture}}}
&=& \psi^k{}_j{}^i
\eea
We will now motivate why we shall put the color indices in their conjugate representations as we go along the string following the lattice bipartite structure. 

\subsection{A tensile string that develops singularities}
Let us start with a regular closed tensile string on the hypercubic lattice. By regular we mean a string that stretches over edges in the lattice in such a way that it never meets itself on any edge or vertex. If we want to deform this string, while its segments are located only on edges, then we can decompose this deformation into its smallest units. On the lattice, these smallest units are the plaquettes moves -- different ways by which the string can move across one plaquette in the lattice. There are numerous different ways that a portion of a regular closed tensile string can move across one plaquette. If the string meets a given plaquette on only one edge, then moving it across the plaquette will take the string to the three opposite edges in this plaquette and we call that the $1\to3$ move. If the string goes over two edges in the plaquette, then moving it across means to take it to the two opposite two edges in the plaquette and we call that the $2\to 2$ move. If the string stretches over three edges in the plaquette, then we apply a $3\to 1$ move that brings the string to the opposite edge. 

Even if we start with a regular string configuration, then we will eventually encounter situations where two different segments of the string come together on one edge or in one vertex. This will eventually happen if we just iterate the three plaquette moves that we listed above to move the string over the lattice in some random fashion. That means that even if we start with a regular string, we are forced to allow the string to have singularities -- places where two segments meet on one edge or at one vertex. 

If those segments run parallel on that edge, then since the segments are locally BPS, nothing dramatic happens as the segments meet. The string may remain a tensionful nonrelativistic string even when the segments meet. This can be understood from charge conservation and specifically from the $U(N)$ commutation relation 
\bea
[E_{i,N+1},E_{j,N+1}] &=& 0
\eea
which says that no transition into new degrees of freedom can take place. 

If on the other hand the segments are anti-parallel, then the transition can occur into massless degrees of freedom -- let us say these are the tensionless strings. These are degrees of freedom that are associated to the collection of generators $H_i,E_{ij}$ that get generated by commuting $E_{i.N+1}$ with $E_{N+1,j}$. 

For the tensile string we have a wave function. As two anti-parallel segments meet on one edge, we may expect to still have a wave funcion for the entire string, including the two segments on this edge, at least for a short period of time, by the energy-time uncertainty relation, just before these two segments get annihilated and transition to tensionless strings. Since the energy of these string segments that is of the order $T a$ where $a$ is the length of each segment, the uncertainty in time, which will be the decay time, will be of the order $\triangle t \sim\frac{\hbar}{T a}$. Although the tension is very large, since it multiplies a segment of the string that we may take to be very short, this means that we can arrange so that $\triangle t$ becomes a sufficiently long duration in order for a quantum mechanical, i.e. first quantized, treatment of the singular segments of the string to be a valid approximation. There is no reason for the wave function on the singular locus of the string to take any particular form. It can still be a generic wave function. When the singular segments decay they transform into tensionless string degrees of freedom and leave our tensile string. We have charge preservation in this process only for the tensile string and the tensionless strings as a combined system, which means that the $U(N)$ charges are not conserved for the tensile string alone. We understand this by looking at the commutation relation $[E_{i,N+1},E_{j,N+1}]$, from which we see that we generate $H_i$ and $E_{ij}$ that together mix up by $U(N)$ and constitute the tensionless string degrees of freedom. Such a decay process will in general not be perceived as a unitary process from the perspective of the selfdual string wave function alone. Unitary evolution applies here to the tensile and tensionless strings together as they form a closed system.

So far we have defined segments as portions of a tensile string as it stretches over edges in the lattice. Let us now also define a junction of the tensile string as its location on a vertex in the lattice. We can then say something more specific about a particular kind of singular locus of a tensile string if we create this singular locus as a spike, where we locally deform a regular tensile string by 'pulling the string at a junction' to create a spike. This process is depicted below. We pull a junction that resides on a black vertex and move the junction to a nearby white vertex which then creates a spike on the string,
\bea
\begin{tikzpicture}[scale=3,
    white/.style={circle, draw, fill=white, inner sep=0pt, minimum size=6pt},
    black/.style={circle, draw, fill=black, inner sep=0pt, minimum size=6pt},
    label/.style={font=\small, midway, sloped, above},
    coord/.style={font=\scriptsize, anchor=south east},
    coordne/.style={font=\scriptsize, anchor=north east},
    coordnw/.style={font=\scriptsize, anchor=north west},
    coordsw/.style={font=\scriptsize, anchor=south west}
]
\coordinate (A) at (0,0);   
\coordinate (B) at (1,0);   
\coordinate (Ba) at (0.98,0);   
\coordinate (Bb) at (1.02,0);   
\coordinate (C) at (2,0);   
\coordinate (D) at (1,1);
\coordinate (Da) at (0.98,1);   
\coordinate (Db) at (1.02,1);   
\draw[midarrow, red, line width=1.5pt] (A) -- (Ba) node[midway, above] {$i$};
\draw[midarrow,red, line width=1.5pt] (Bb) -- (C) node[midway, above] {$j$};
\draw (B) -- (D) node[midway, left] {$k$};
\node[coordne] at (A) {$$};
\node[white] at (A) {};
\node[black] at (B) {};
\node[white] at (C) {};
\node[white] at (D) {};
\end{tikzpicture}
&\vcenter{\hbox{$\longrightarrow$}}&
\begin{tikzpicture}[scale=3,
    white/.style={circle, draw, fill=white, inner sep=0pt, minimum size=6pt},
    black/.style={circle, draw, fill=black, inner sep=0pt, minimum size=6pt},
    label/.style={font=\small, midway, sloped, above},
    coord/.style={font=\scriptsize, anchor=south east},
    coordne/.style={font=\scriptsize, anchor=north east},
    coordnw/.style={font=\scriptsize, anchor=north west},
    coordsw/.style={font=\scriptsize, anchor=south west}
]
\coordinate (A) at (0,0);   
\coordinate (B) at (1,0);   
\coordinate (Ba) at (0.98,0);  
\coordinate (Bb) at (1.02,0);   
\coordinate (C) at (2,0);   
\coordinate (D) at (1,1);
\coordinate (Da) at (0.98,1);   
\coordinate (Db) at (1.02,1);   
\draw[midarrow, red, line width=1.5pt] (A) -- (Ba) node[midway, above] {$i$};
\draw[midarrow,red, line width=1.5pt] (Ba) -- (Da) node[midway, left] {$k$};
\draw[midarrow,red, line width=1.5pt] (Db) -- (Bb) node[midway, right] {$k'$};
\draw[midarrow,red, line width=1.5pt] (Bb) -- (C) node[midway, above] {$j$};
\node[coordne] at (A) {$$};
\node[white] at (A) {};
\node[black] at (B) {};
\node[white] at (C) {};
\node[white] at (D) {};
\end{tikzpicture}
\eea  
As we pull the string away from the black vertex, we pull out two anti-parallel segments of the string, creating a spike on the string that stretches back and forth over the edge that we label $k$ in the figure. Under the deformation of the tensile string, its wave function gets transformed by a surface holonomy factor. Neither vertices in the lattice nor junctions on the string carry color indices. They are gauge singlets. As we pull the string at a junction, there is no place to attach a surface holonomy to the string. As we create a spike, two new color indices, $k$ and $k'$ in the figure, will be created on the singular spike segments. We might express this as a local transformation of the form
\bea
\psi\cdots{}^i{}_j\cdots &\rightarrow & \psi\cdots{}^i{}_k{}^{k'}{}_j\cdots = U^{k'}{}_k \psi\cdots{}^i{}_j\cdots
\eea
where $U^{k'}{}_k$ is the surface holonomy that we associate with the spike. In the Euclidean theory the spike creation may be viewed as a deformation of the string that covers a vanishing surface area. We expect that the surface holonomy associated to such a surface of vanishing area will be a trivial gauge singlet. One argument is that if we have a physical object (such as a junction in a tensile string) that transforms as a singlet under gauge transformations, and if we physically deform the object, then it is impossible to physically deform the object so that it starts to transform nontrivially under gauge rotations. The only $U(N)$ gauge singlet with two color indices (for $N>2$) is the Kronecker delta $\delta_k^{k'}$. This implies that the surface holonomy that we shall associate with the spike creation process depicted above shall take the form 
\ben
U^{k'}{}_k &=& \frac{1}{\sqrt{N}} \delta^{k'}_k\label{K-spike-op}
\een
We will discuss the choice of normalization factor later. Given our discussion of a time-delay $\Delta t$ as two tensile segments meet and before they annihilate, which should apply equally well to the spike, we may think of the creation of a spike as another basic 'move' that we can perform of the string as we move the string across a plaquette in the lattice.  Accordingly, we will now introduce two more moves, which are the singular moves. The first move is the 'K-spike move'\footnote{Here K stands for the Kronecker delta.}. This move creates a spike from a junction and transforms the wave function as 
\ben
\psi\cdots{}^i{}_j\cdots &\rightarrow & \psi\cdots{}^i{}_k{}^{k'}{}_j\cdots = \frac{1}{\sqrt{N}} \delta^{k'}_k \psi\cdots{}^i{}_j\cdots \label{K-move}
\een
We should also be able to reverse the process such that when we have a spike on the string, then we can remove it by another move that we will refer to as the 'R-move'\footnote{Here R stands for removal.} that transforms the wave function as
\ben
\frac{1}{\sqrt{N}} \delta^{k'}_k \psi\cdots{}^i{}_j\cdots &\rightarrow & \psi\cdots{}^i{}_j\cdots \label{R-move}
\een

Even if we start with a regular tensile string, as we keep moving the string across the lattice with basic moves, we may create singularities of various kinds. The most generic singularity will be when two random segments that are not adjacent meet. Let us focus on the case when these segments have opposite orientations when they meet. Regardless of what the wave function is on these segments, regardless of whether it is proportional to a Kronecker delta or not, these segments will be annihilated and then disappear from our tensile string as tensionless strings within a decay time period of $\Delta t \sim \frac{\hbar}{T a}$. We depict such a generic process below,
\bea
&\vcenter{
  \hbox{
    \begin{tikzpicture}[scale=3,
        white/.style={circle, draw, fill=white, inner sep=0pt, minimum size=6pt},
        black/.style={circle, draw, fill=black, inner sep=0pt, minimum size=6pt},
        coordne/.style={font=\scriptsize, anchor=north east}
    ]
\tikzset{invisible/.style={draw=none}}
      \coordinate (A1) at (0,0);
      \coordinate (B1) at (1,0);
      \coordinate (C1) at (2,0);
      \coordinate (D1) at (3,0);
      \coordinate (E1) at (4,0);
      \coordinate (A2) at (0,1);
      \coordinate (B2) at (1,1);
      \coordinate (C2) at (2,1);
      \coordinate (D2) at (3,1);
      \coordinate (E2) at (4,1);
      \draw[red, line width=1.5pt] (A1) -- (B1) node[midway, below] {$a$};
      \draw[red, line width=1.5pt] (B1) -- (C1) node[midway, below] {$b$};
      \draw[red, line width=1.5pt] (C1) -- (D1) node[midway, below] {$c$};
      \draw[red, line width=1.5pt, midarrow] (D1) -- (D2) node[midway, right] {$d$};
      \draw[red, line width=1.5pt] (D2) -- (C2) node[midway, above] {$e$};
      \draw[red, line width=1.5pt] (C2) -- (B2) node[midway, above] {$f$};
      \draw[red, line width=1.5pt] (B2) -- (A2) node[midway, above] {$g$};
      \draw[red, line width=1.5pt, midarrow] (A2) -- (A1) node[midway, left] {$h$};
      \draw (B1) -- (B2) node[midway, left] {$i$};
      \draw (C1) -- (C2) node[midway, left] {$j$};
      \node[white] at (A1) {};
      \node[black] at (B1) {};
      \node[white] at (C1) {};
      \node[black] at (D1) {};
      \node[black] at (A2) {};
      \node[white] at (B2) {};
      \node[black] at (C2) {};
      \node[white] at (D2) {};
    \end{tikzpicture}
  }
}&
\cr
&\downarrow &\cr
&\vcenter{
  \hbox{
    \begin{tikzpicture}[scale=3,
        white/.style={circle, draw, fill=white, inner sep=0pt, minimum size=6pt},
        black/.style={circle, draw, fill=black, inner sep=0pt, minimum size=6pt},
        coordne/.style={font=\scriptsize, anchor=north east}
    ]
      \coordinate (A1) at (0,0);
      \coordinate (B1) at (1,0);
      \coordinate (B1a) at (1,0.05);
      \coordinate (C1) at (2,0);
      \coordinate (C1a) at (2,0.05);
      \coordinate (D1) at (3,0);
      \coordinate (E1) at (4,0);
      \coordinate (A2) at (0,1);
      \coordinate (B2) at (1,1);
      \coordinate (C2) at (2,1);
      \coordinate (D2) at (3,1);
      \coordinate (E2) at (4,1);
      \draw[red, line width=1.5pt] (A1) -- (B1) node[midway, below] {$a$};
      \draw[red, line width=1.5pt] (B1) -- (C1) node[midway, below] {$b$};
      \draw[red, line width=1.5pt] (C1) -- (D1) node[midway, below] {$c$};
      \draw[red, line width=1.5pt, midarrow] (D1) -- (D2) node[midway, right] {$d$};
      \draw[red, line width=1.5pt] (D2) -- (C2) node[midway, below] {$e$};
      \draw[red, line width=1.5pt] (C2) -- (C1a) node[midway, left] {$j$};
      \draw[red, line width=1.5pt] (C1a) -- (B1a) node[midway, above] {$b'$};
      \draw[red, line width=1.5pt] (B1a) -- (B2) node[midway, left] {$i$};
      \draw[red, line width=1.5pt] (B2) -- (A2) node[midway, below] {$g$};
      \draw[red, line width=1.5pt, midarrow] (A2) -- (A1) node[midway, left] {$h$};
      \node[white] at (A1) {};
      \node[black] at (B1) {};
      \node[white] at (C1) {};
      \node[black] at (D1) {};
      \node[black] at (A2) {};
      \node[white] at (B2) {};
      \node[black] at (C2) {};
      \node[white] at (D2) {};
    \end{tikzpicture}
  }
}&
\cr
&\downarrow\text{}&\cr
&\vcenter{
  \hbox{\begin{tikzpicture}[scale=3,
        white/.style={circle, draw, fill=white, inner sep=0pt, minimum size=6pt},
        black/.style={circle, draw, fill=black, inner sep=0pt, minimum size=6pt},
        coordne/.style={font=\scriptsize, anchor=north east}
    ]
      \coordinate (A1) at (0,0);
      \coordinate (B1) at (1,0);
      \coordinate (B1a) at (1,0.05);
      \coordinate (C1) at (2,0);
      \coordinate (C1a) at (2,0.05);
      \coordinate (D1) at (3,0);
      \coordinate (E1) at (4,0);
      \coordinate (A2) at (0,1);
      \coordinate (B2) at (1,1);
      \coordinate (C2) at (2,1);
      \coordinate (D2) at (3,1);
      \coordinate (E2) at (4,1);
      \draw[red, line width=1.5pt] (A1) -- (B1) node[midway, below] {$a$};
      \draw[red, line width=1.5pt] (C1) -- (D1) node[midway, below] {$c$};
      \draw[red, line width=1.5pt, midarrow] (D1) -- (D2) node[midway, right] {$d$};
      \draw[red, line width=1.5pt] (D2) -- (C2) node[midway, below] {$e$};
      \draw[red, line width=1.5pt] (C2) -- (C1) node[midway, left] {$j$};
      \draw[red, line width=1.5pt] (B1) -- (B2) node[midway, left] {$i$};
      \draw[red, line width=1.5pt] (B2) -- (A2) node[midway, below] {$g$};
      \draw[red, line width=1.5pt, midarrow] (A2) -- (A1) node[midway, left] {$h$};
      \node[white] at (A1) {};
      \node[black] at (B1) {};
      \node[white] at (C1) {};
      \node[black] at (D1) {};
      \node[black] at (A2) {};
      \node[white] at (B2) {};
      \node[black] at (C2) {};
      \node[white] at (D2) {};
    \end{tikzpicture}}}&
\eea
As we can see, the closed tensile string has split into two closed tensile strings. 

Now as we further shrink the string on say the left side by the $3\to 1$ move, we may create another singular string configuration that looks like just one spike as depicted below,
\bea
\vcenter{\hbox{\begin{tikzpicture}[scale=3,
    white/.style={circle, draw, fill=white, inner sep=0pt, minimum size=6pt},
    black/.style={circle, draw, fill=black, inner sep=0pt, minimum size=6pt},
    label/.style={font=\small, midway, sloped, above},
    coord/.style={font=\scriptsize, anchor=south east},
    coordne/.style={font=\scriptsize, anchor=north east},
    coordnw/.style={font=\scriptsize, anchor=north west},
    coordsw/.style={font=\scriptsize, anchor=south west}
]
\coordinate (A) at (0,0);   
\coordinate (B) at (1,0);   
\coordinate (Ba) at (0.98,0);  
\coordinate (Bb) at (1.02,0);   
\coordinate (C) at (2,0);   
\coordinate (D) at (1,1);
\coordinate (Da) at (0.98,1);   
\coordinate (Db) at (1.02,1);   
\draw[midarrow,red, line width=1.5pt] (Da) -- (Ba) node[midway, left] {$i'$};
\draw[midarrow,red, line width=1.5pt] (Bb) -- (Db) node[midway, right] {$i$};
\node[coordne] at (B) {$$};
\node[coordne] at (D) {$$};
\node[black] at (B) {};
\node[white] at (D) {};
\end{tikzpicture}}}
\eea
and then this string will decay into tensionless strings and after that the tensile string will be completely gone. That will not be a unitary process from the perspective of the tensile string quantum mechanics obviouly. But it will be a unitary process for the closed system of both tensile and tensionless strings.

Let us now discuss the probabilistic interpretation of our tensile string wave function in a bit more detail. We assume that the string is confined to the edges in the lattice, that it is tensile with a tension $T$, and we do not consider superpositions of spatial configurations of the string. Therefore we implicitly assume that the tensile string behaves like a classical string in spacetime as it moves over the lattice. The wave function of the string is an internal characterisation of quantum probabilities that describe the color indexing of the string. Let us consider a closed string going around one plaquette, for which we have the wave function
\bea
\psi_{\l}{}^k{}_j{}^i
\eea
with four indices each ranging over $1,...,N$. We have $N^4$ complex components. The components squared and when summed over the indices, will be normalized to one,
\ben
\sum_{i,j,k,\l = 1}^N |\psi_{\l}{}^k{}_j{}^i|^2 &=& 1\label{normen}
\een
The interpretation is that the closed tensile string exists with certainty, with probability one. Each term in the sum gives the probability for each possible value of the four color indices. The unit normalization of the entire string wave function as we sum over all color indices, implies that we must normalize the K-spike surface holonomy with a factor of $1/\sqrt{N}$ in equation (\ref{K-spike-op}). This works fine for all lengths of the tensile string, where its length is counted by its number of segments, as long as this number is no less than two segments. When the string has shrunk to only two segments and these are annihilated and what is left are only tensionless strings, then we no longer have a tensile string and we can not in any meaningful way assign its wave function a probability. The string is gone, by a process that from the perspective of the tensile string is non-unitary. 

If we consider a regular string on the lattice, it can happen that the string passes through some vertex in the lattice that is connected to some plaquette in the lattice where our string stretches over none of its four edges. Then, from the perspective of this plaquette, our regular string will appear as a point residing in just one of the four vertices of the plaquette. 

Leaving out the color indices of the wave function that are not involved in the plaquette under consideration, we would describe the wave function of this regular string, at the vertex, by a wave function $\psi(x)$ with no color indices, where $x$ is the coordinate of the vertex where the regular string passes through the given plaquette. 

As we have a string residing in one vertex we are led to introduce the $0\to 4$ move. This is the move that stretches the string around all four edges of the given plaquette from the vertex. We may reverse this move, and then call that the $4\to 0$ move. This move shrinks the string that goes around one plaquette to just reside in one vertex, but this requires that the string stretches away from the same vertex into some other directions outside the plaquette under consideration, since otherwise we would not have a nonrelativistic tensile string left. 

The $0\to 4$ move creates a figure-eight string with a pointlike singularity at the vertex from which we stretch out the string. The reversed move, the $4\to 0$ move, will instead presume that we start from a figure-eight string configuration and then remove one of its closed string parts. 

Let us finally discuss the pointlike singularity at the vertex where the string self-intersects. At the self-intersection junction, the string may break apart and form two closed strings. Such a breaking would happen if the strings can reconnect each other at the junction and then move away from the vertex where they initially intersected. There are no charges that pull the segments together at the vertex because we put no color indices on the vertices. There is no portion of the string that can become annihilated at the vertex. We may therefore expect that the tensile string can remain in a figure-eight configuration for a time duration that is vastly longer than the decay time $\Delta t$ that it takes for two antiparallel string segments to annihilate each other on a common edge.

\section{The tensile string moving in the lattice}
When the string moves over the lattice, it sweeps out a two-dimensional surface, which is an embedded surface in the lattice, which is built up of plaquettes. Let us now focus on one such plaquette in the surface. We can not put an orientation on every plaquette in the hypercubic lattice in a consistent manner. But we can put an orientation on every plaquette that builds up the 2d surface. Once we fix the orientation of just one such plaquette, then the orientation of every other plaquette in the surface is uniquely fixed and this collection of oriented plaquettes then describes the orientation of the entire surface. As the string moves in the lattice over this surface, any motion can be decomposed into motions over single plaquettes. Furthermore, as far as we only are concerned with the surface holonomy that the string picks up as it moves, then motions over different plaquettes commute because different plaquettes carry different color indices. That means that we can reduce the motion of the string to its motion over a single plaquette, and reversely, we can build up any motion of the string by composing various moves of the string over various plaquettes. In short, any movement of the string across the lattice, can be reduced to just a basic move over just one plaquette in the lattice. 

Let us now put an orientation on the plaquette that the string is either going to sweep over, or has already swept over. Below we depict such a plaquette that participates specifically in such a 2d surface where we put an intrinsic orientation on it as indicated by the inner loop with an arrow pointing in one direction, which we refer to as 'the direction of the orientation' of this plaquette,
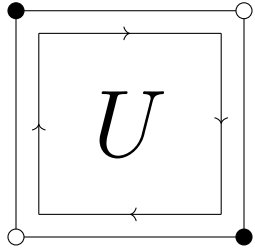
\bea
  \centering
\begin{tikzpicture}[scale=3,
    white/.style={circle, draw, fill=white, inner sep=0pt, minimum size=6pt},
    black/.style={circle, draw, fill=black, inner sep=0pt, minimum size=6pt},
    label/.style={font=\small, midway, sloped, above},
    coord/.style={font=\scriptsize, anchor=south east},
    coordne/.style={font=\scriptsize, anchor=north east},
    coordnw/.style={font=\scriptsize, anchor=north west},
    coordsw/.style={font=\scriptsize, anchor=south west}
]
\coordinate (A) at (0,0,0);   
\coordinate (B) at (1,0,0);   
\coordinate (C) at (1,1,0);   
\coordinate (D) at (0,1,0);   

\coordinate (AA) at (0.1,0.1,0);   
\coordinate (AB) at (0.9,0.1,0);   
\coordinate (AC) at (0.9,0.9,0);   
\coordinate (AD) at (0.1,0.9,0); 
\node[coordne] at (A) {$$};
\node[coordnw] at (B) {$$};
\node[coordsw] at (C) {$$};
\node[coord] at (D) {$$};
\draw (A) -- (B) node[midway, below] {$$};
\draw (B) -- (C) node[midway, right] {$$};
\draw (C) -- (D) node[midway, left] {$$};
\draw (D) -- (A) node[midway, above] {$$};

\draw[midarrow] (AB) -- (AA) node[midway, above] {$$};
\draw[midarrow] (AC) -- (AB) node[midway, right] {$$};
\draw[midarrow] (AD) -- (AC) node[midway, above] {$$};
\draw[midarrow] (AA) -- (AD) node[midway, right] {$$};
\node[white] at (A) {};
\node[black] at (B) {};
\node[white] at (C) {};
\node[black] at (D) {};

\node[canvas is xy plane at z=0, transform shape] at (0.5,0.5,0) {$U$};
\end{tikzpicture}
\eea
We have placed the letter $U$ in the middle of the plaquette. This letter represents the 'universal plaquette holonomy'. It is a unitary operator associated to this plaquette. This operator can transport a string across the plaquette in multiple different ways, which is why we refer to it as 'universal'. 

The aim of the following few sections is to show that when we move the string across a plaquette, then its wave function transforms by just one uniquely defined universal plaquette operator that takes the form $U_{\l}{}^k{}_j{}^i$. This final result is summarized in (\ref{UUU}) where we see that although we have five different ways to move the string across the plaquette, 
\begin{itemize}
\item the $0\to 4$ move
\item the $1\to 3$ move
\item the $2\to 2$ move
\item the $3\to 1$ move
\item the $4\to 0$ move
\end{itemize}
these five moves are all implemented by the same universal plaquette operator $U_{\l}{}^k{}_j{}^i$.

Below we list the ways that a string that starts at the vertex with coordinate $x$ and that are oriented along the plaquette can be transported to the other side\footnote{The other 'side' refers here to the edges around the plaquette over which the string was not extended initially.} by $U$. From the outset it is not obvious that we only need one operator for all these different moves. Therefore we will start with distinguishing between them by inserting a vertical line that separates the color indices into two groups. The indices that stand to the left of the vertical line correspond to the color indices of the transformed wave function. The indices that stand to the right of the vertical line are written in reversed order (which means reversed both in the left-right and the up-down directions) and they contract the indices of the original wave function. 
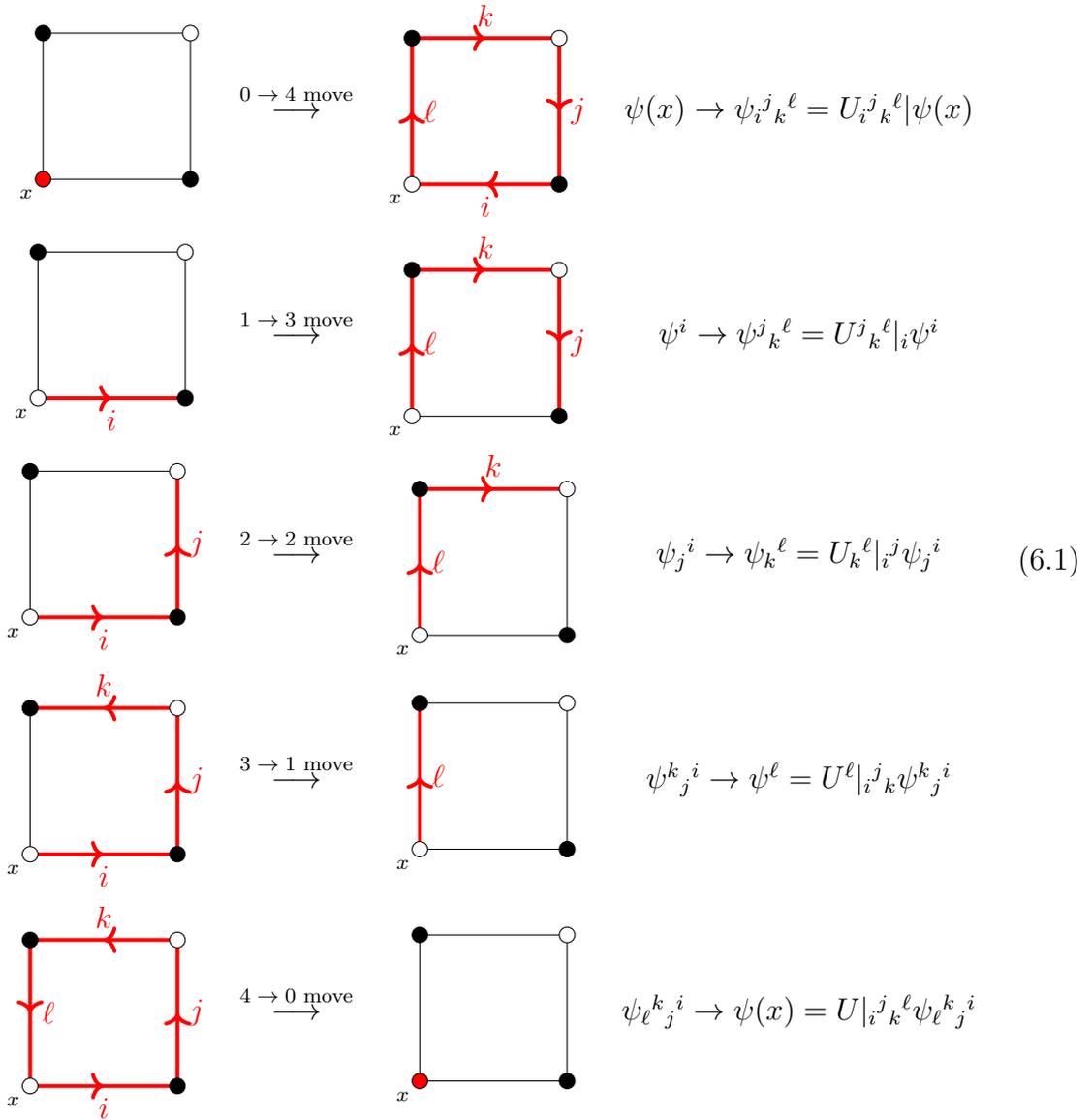
\bea
\begin{array}{cccc}
\vcenter{\hbox{
\begin{tikzpicture}[scale=2,
    white/.style={circle, draw, fill=white, inner sep=0pt, minimum size=6pt},
    black/.style={circle, draw, fill=black, inner sep=0pt, minimum size=6pt},
    red/.style={circle, draw, fill=red, inner sep=0pt, minimum size=6pt},
    label/.style={font=\small, midway, sloped, above},
    coord/.style={font=\scriptsize, anchor=south east},
    coordne/.style={font=\scriptsize, anchor=north east},
    coordnw/.style={font=\scriptsize, anchor=north west},
    coordsw/.style={font=\scriptsize, anchor=south west}
]
\coordinate (A) at (0,0);   
\coordinate (B) at (1,0);   
\coordinate (C) at (1,1);   
\coordinate (D) at (0,1);   
\node[coordne] at (A) {$x$};
\node[coordnw] at (B) {$$};
\node[coordsw] at (C) {$$};
\node[coord] at (D) {$$};
\draw (A) -- (B) node[midway, below] {$$};
\draw (C) -- (B) node[midway, right] {$$};
\draw (A) -- (D) node[midway, left] {$$};
\draw (C) -- (D) node[midway, above] {$$};
\node[red] at (A) {};
\node[black] at (B) {};
\node[white] at (C) {};
\node[black] at (D) {};
\end{tikzpicture}}}
&\overset{\text{$0\to 4$ move}}{\longrightarrow}&
\vcenter{\hbox{\begin{tikzpicture}[scale=2,
    white/.style={circle, draw, fill=white, inner sep=0pt, minimum size=6pt},
    black/.style={circle, draw, fill=black, inner sep=0pt, minimum size=6pt},
    label/.style={font=\small, midway, sloped, above},
    coord/.style={font=\scriptsize, anchor=south east},
    coordne/.style={font=\scriptsize, anchor=north east},
    coordnw/.style={font=\scriptsize, anchor=north west},
    coordsw/.style={font=\scriptsize, anchor=south west}
]
\coordinate (A) at (0,0);   
\coordinate (B) at (1,0);   
\coordinate (C) at (1,1);   
\coordinate (D) at (0,1);   
\node[coordne] at (A) {$x$};
\node[coordnw] at (B) {$$};
\node[coordsw] at (C) {$$};
\node[coord] at (D) {$$};
\draw[red, midarrow, line width=1.5pt] (B) -- (A) node[midway, below] {$i$};
\draw[red, midarrow, line width=1.5pt] (C) -- (B) node[midway, right] {$j$};
\draw[red, midarrow, line width=1.5pt] (D) -- (C) node[midway, above] {$k$};
\draw[red, midarrow, line width=1.5pt] (A) -- (D) node[midway, right] {$\l$};
\node[white] at (A) {};
\node[black] at (B) {};
\node[white] at (C) {};
\node[black] at (D) {};
\end{tikzpicture}}}
&\psi(x) \rightarrow \psi_i{}^j{}_k{}^{\l} = U_i{}^j{}_k{}^{\l}| \psi(x)\cr
\vcenter{\hbox{\begin{tikzpicture}[scale=2,
    white/.style={circle, draw, fill=white, inner sep=0pt, minimum size=6pt},
    black/.style={circle, draw, fill=black, inner sep=0pt, minimum size=6pt},
    label/.style={font=\small, midway, sloped, above},
    coord/.style={font=\scriptsize, anchor=south east},
    coordne/.style={font=\scriptsize, anchor=north east},
    coordnw/.style={font=\scriptsize, anchor=north west},
    coordsw/.style={font=\scriptsize, anchor=south west}
]
\coordinate (A) at (0,0);   
\coordinate (B) at (1,0);   
\coordinate (C) at (1,1);   
\coordinate (D) at (0,1);   
\node[coordne] at (A) {$x$};
\node[coordnw] at (B) {$$};
\node[coordsw] at (C) {$$};
\node[coord] at (D) {$$};
\draw[red, midarrow, line width=1.5pt] (A) -- (B) node[midway, below] {$i$};
\draw (C) -- (B) node[midway, right] {$$};
\draw (A) -- (D) node[midway, left] {$$};
\draw (C) -- (D) node[midway, above] {$$};
\node[white] at (A) {};
\node[black] at (B) {};
\node[white] at (C) {};
\node[black] at (D) {};
\end{tikzpicture}}}
&\overset{\text{$1\to 3$ move}}{\longrightarrow}&
\vcenter{\hbox{\begin{tikzpicture}[scale=2,
    white/.style={circle, draw, fill=white, inner sep=0pt, minimum size=6pt},
    black/.style={circle, draw, fill=black, inner sep=0pt, minimum size=6pt},
    label/.style={font=\small, midway, sloped, above},
    coord/.style={font=\scriptsize, anchor=south east},
    coordne/.style={font=\scriptsize, anchor=north east},
    coordnw/.style={font=\scriptsize, anchor=north west},
    coordsw/.style={font=\scriptsize, anchor=south west}
]
\coordinate (A) at (0,0);   
\coordinate (B) at (1,0);   
\coordinate (C) at (1,1);   
\coordinate (D) at (0,1);   
\node[coordne] at (A) {$x$};
\node[coordnw] at (B) {$$};
\node[coordsw] at (C) {$$};
\node[coord] at (D) {$$};
\draw (A) -- (B) node[midway, below] {$$};
\draw[red, midarrow, line width=1.5pt] (A) -- (D) node[midway, right] {$\l$};
\draw[red, midarrow, line width=1.5pt] (D) -- (C) node[midway, above] {$k$};
\draw[red, midarrow, line width=1.5pt] (C) -- (B) node[midway, right] {$j$};
\node[white] at (A) {};
\node[black] at (B) {};
\node[white] at (C) {};
\node[black] at (D) {};
\end{tikzpicture}}}
&\psi^i \rightarrow \psi^j{}_k{}^{\l} = U^j{}_k{}^{\l}|{}_i \psi^i\cr
\vcenter{\hbox{\begin{tikzpicture}[scale=2,
    white/.style={circle, draw, fill=white, inner sep=0pt, minimum size=6pt},
    black/.style={circle, draw, fill=black, inner sep=0pt, minimum size=6pt},
    label/.style={font=\small, midway, sloped, above},
    coord/.style={font=\scriptsize, anchor=south east},
    coordne/.style={font=\scriptsize, anchor=north east},
    coordnw/.style={font=\scriptsize, anchor=north west},
    coordsw/.style={font=\scriptsize, anchor=south west}
]
\coordinate (A) at (0,0);   
\coordinate (B) at (1,0);   
\coordinate (C) at (1,1);   
\coordinate (D) at (0,1);   
\node[coordne] at (A) {$x$};
\node[coordnw] at (B) {$$};
\node[coordsw] at (C) {$$};
\node[coord] at (D) {$$};
\draw[red, midarrow, line width=1.5pt] (A) -- (B) node[midway, below] {$i$};
\draw[red, midarrow, line width=1.5pt] (B) -- (C) node[midway, right] {$j$};
\draw (C) -- (D) node[midway, left] {$$};
\draw (D) -- (A) node[midway, above] {$$};
\node[white] at (A) {};
\node[black] at (B) {};
\node[white] at (C) {};
\node[black] at (D) {};
\end{tikzpicture}}}
&\overset{\text{$2\to 2$ move}}{\longrightarrow}&
\vcenter{\hbox{\begin{tikzpicture}[scale=2,
    white/.style={circle, draw, fill=white, inner sep=0pt, minimum size=6pt},
    black/.style={circle, draw, fill=black, inner sep=0pt, minimum size=6pt},
    label/.style={font=\small, midway, sloped, above},
    coord/.style={font=\scriptsize, anchor=south east},
    coordne/.style={font=\scriptsize, anchor=north east},
    coordnw/.style={font=\scriptsize, anchor=north west},
    coordsw/.style={font=\scriptsize, anchor=south west}
]
\coordinate (A) at (0,0);   
\coordinate (B) at (1,0);   
\coordinate (C) at (1,1);   
\coordinate (D) at (0,1);   
\node[coordne] at (A) {$x$};
\node[coordnw] at (B) {$$};
\node[coordsw] at (C) {$$};
\node[coord] at (D) {$$};
\draw (A) -- (B) node[midway, below] {$$};
\draw (B) -- (C) node[midway, right] {$$};
\draw[red, midarrow, line width=1.5pt] (A) -- (D) node[midway, right] {$\l$};
\draw[red, midarrow, line width=1.5pt] (D) -- (C) node[midway, above] {$k$};
\node[white] at (A) {};
\node[black] at (B) {};
\node[white] at (C) {};
\node[black] at (D) {};
\end{tikzpicture}}}
&\psi_j{}^i \rightarrow \psi_k{}^{\l} = U_k{}^{\l}|{}_i{}^j \psi_j{}^i\cr
\vcenter{\hbox{\begin{tikzpicture}[scale=2,
    white/.style={circle, draw, fill=white, inner sep=0pt, minimum size=6pt},
    black/.style={circle, draw, fill=black, inner sep=0pt, minimum size=6pt},
    label/.style={font=\small, midway, sloped, above},
    coord/.style={font=\scriptsize, anchor=south east},
    coordne/.style={font=\scriptsize, anchor=north east},
    coordnw/.style={font=\scriptsize, anchor=north west},
    coordsw/.style={font=\scriptsize, anchor=south west}
]
\coordinate (A) at (0,0);   
\coordinate (B) at (1,0);   
\coordinate (C) at (1,1);   
\coordinate (D) at (0,1);   
\node[coordne] at (A) {$x$};
\node[coordnw] at (B) {$$};
\node[coordsw] at (C) {$$};
\node[coord] at (D) {$$};
\draw[red, midarrow, line width=1.5pt] (A) -- (B) node[midway, below] {$i$};
\draw[red, midarrow, line width=1.5pt] (B) -- (C) node[midway, right] {$j$};
\draw[red, midarrow, line width=1.5pt] (C) -- (D) node[midway, above] {$k$};
\draw (D) -- (A) node[midway, above] {$$};
\node[white] at (A) {};
\node[black] at (B) {};
\node[white] at (C) {};
\node[black] at (D) {};
\end{tikzpicture}}}
&\overset{\text{$3\to 1$ move}}{\longrightarrow}&
\vcenter{\hbox{\begin{tikzpicture}[scale=2,
    white/.style={circle, draw, fill=white, inner sep=0pt, minimum size=6pt},
    black/.style={circle, draw, fill=black, inner sep=0pt, minimum size=6pt},
    label/.style={font=\small, midway, sloped, above},
    coord/.style={font=\scriptsize, anchor=south east},
    coordne/.style={font=\scriptsize, anchor=north east},
    coordnw/.style={font=\scriptsize, anchor=north west},
    coordsw/.style={font=\scriptsize, anchor=south west}
]
\coordinate (A) at (0,0);   
\coordinate (B) at (1,0);   
\coordinate (C) at (1,1);   
\coordinate (D) at (0,1);   
\node[coordne] at (A) {$x$};
\node[coordnw] at (B) {$$};
\node[coordsw] at (C) {$$};
\node[coord] at (D) {$$};
\draw (A) -- (B) node[midway, below] {$$};
\draw (B) -- (C) node[midway, right] {$$};
\draw (C) -- (D) node[midway, right] {$$};
\draw[red, midarrow, line width=1.5pt] (A) -- (D) node[midway, right] {$\l$};
\node[white] at (A) {};
\node[black] at (B) {};
\node[white] at (C) {};
\node[black] at (D) {};
\end{tikzpicture}}}
& \psi^k{}_j{}^i \rightarrow \psi^{\l} = U^{\l}|{}_i{}^j{}_k \psi^k{}_j{}^i\cr
\vcenter{\hbox{\begin{tikzpicture}[scale=2,
    white/.style={circle, draw, fill=white, inner sep=0pt, minimum size=6pt},
    black/.style={circle, draw, fill=black, inner sep=0pt, minimum size=6pt},
    label/.style={font=\small, midway, sloped, above},
    coord/.style={font=\scriptsize, anchor=south east},
    coordne/.style={font=\scriptsize, anchor=north east},
    coordnw/.style={font=\scriptsize, anchor=north west},
    coordsw/.style={font=\scriptsize, anchor=south west}
]
\coordinate (A) at (0,0);   
\coordinate (B) at (1,0);   
\coordinate (C) at (1,1);   
\coordinate (D) at (0,1);   
\node[coordne] at (A) {$x$};
\node[coordnw] at (B) {$$};
\node[coordsw] at (C) {$$};
\node[coord] at (D) {$$};
\draw[red, midarrow, line width=1.5pt] (A) -- (B) node[midway, below] {$i$};
\draw[red, midarrow, line width=1.5pt] (B) -- (C) node[midway, right] {$j$};
\draw[red, midarrow, line width=1.5pt] (C) -- (D) node[midway, above] {$k$};
\draw[red, midarrow, line width=1.5pt] (D) -- (A) node[midway, right] {$\l$};
\node[white] at (A) {};
\node[black] at (B) {};
\node[white] at (C) {};
\node[black] at (D) {};
\end{tikzpicture}}}
&\overset{\text{$4\to 0$ move}}{\longrightarrow}&
\vcenter{\hbox{\begin{tikzpicture}[scale=2,
    white/.style={circle, draw, fill=white, inner sep=0pt, minimum size=6pt},
    black/.style={circle, draw, fill=black, inner sep=0pt, minimum size=6pt},
    red/.style={circle, draw, fill=red, inner sep=0pt, minimum size=6pt},
    label/.style={font=\small, midway, sloped, above},
    coord/.style={font=\scriptsize, anchor=south east},
    coordne/.style={font=\scriptsize, anchor=north east},
    coordnw/.style={font=\scriptsize, anchor=north west},
    coordsw/.style={font=\scriptsize, anchor=south west}
]
\coordinate (A) at (0,0);   
\coordinate (B) at (1,0);   
\coordinate (C) at (1,1);   
\coordinate (D) at (0,1);   
\node[coordne] at (A) {$x$};
\node[coordnw] at (B) {$$};
\node[coordsw] at (C) {$$};
\node[coord] at (D) {$$};
\draw (A) -- (B) node[midway, below] {$$};
\draw (B) -- (C) node[midway, right] {$$};
\draw (C) -- (D) node[midway, right] {$$};
\draw (D) -- (A) node[midway, above] {$$};
\node[red] at (A) {};
\node[black] at (B) {};
\node[white] at (C) {};
\node[black] at (D) {};
\end{tikzpicture}}}
& \psi_{\l}{}^k{}_j{}^i \rightarrow \psi(x) = U|_i{}^j{}_k{}^{\l} \psi_{\l}{}^k{}_j{}^i
\end{array}\label{61}
\eea
All these moves, except for the $2\to 2$ move, change the length of the string. We propose that the only possible way that the length can increase by a fundamental process is through a K-spike move. Thus the first two moves, the $0\to 4$ and the $1\to 3$ moves, are composed of K-spike moves followed by a $2\to 2$ move as depicted below,
\bea
\begin{array}{ccccc}
\vcenter{\hbox{
\begin{tikzpicture}[scale=2,
    white/.style={circle, draw, fill=white, inner sep=0pt, minimum size=6pt},
    black/.style={circle, draw, fill=black, inner sep=0pt, minimum size=6pt},
    red/.style={circle, draw, fill=red, inner sep=0pt, minimum size=6pt},
    label/.style={font=\small, midway, sloped, above},
    coord/.style={font=\scriptsize, anchor=south east},
    coordne/.style={font=\scriptsize, anchor=north east},
    coordnw/.style={font=\scriptsize, anchor=north west},
    coordsw/.style={font=\scriptsize, anchor=south west}
]
\coordinate (A) at (0,0);   
\coordinate (B) at (1,0);   
\coordinate (C) at (1,1);   
\coordinate (D) at (0,1);   
\node[coordne] at (A) {$x$};
\node[coordnw] at (B) {$$};
\node[coordsw] at (C) {$$};
\node[coord] at (D) {$$};
\draw (A) -- (B) node[midway, below] {$i$};
\draw (C) -- (B) node[midway, right] {$j$};
\draw (A) -- (D) node[midway, left] {$\l$};
\draw (C) -- (D) node[midway, above] {$k$};
\node[red] at (A) {};
\node[black] at (B) {};
\node[white] at (C) {};
\node[black] at (D) {};
\end{tikzpicture}}}
&\overset{\text{}}{\longrightarrow}&
\vcenter{\hbox{\begin{tikzpicture}[scale=2,
    white/.style={circle, draw, fill=white, inner sep=0pt, minimum size=6pt},
    black/.style={circle, draw, fill=black, inner sep=0pt, minimum size=6pt},
    label/.style={font=\small, midway, sloped, above},
    coord/.style={font=\scriptsize, anchor=south east},
    coordne/.style={font=\scriptsize, anchor=north east},
    coordnw/.style={font=\scriptsize, anchor=north west},
    coordsw/.style={font=\scriptsize, anchor=south west}
]
\coordinate (A) at (0,0);   
\coordinate (Aa) at (0.04,0.04); 
\coordinate (Ba) at (1,0.04);   
\coordinate (C) at (1,1);   
\coordinate (Da) at (0.04,1);   
\node[coordne] at (A) {$x$};
\node[coordnw] at (B) {$$};
\node[coordsw] at (C) {$$};
\node[coord] at (D) {$$};
\draw[red, midarrow, line width=1.5pt] (B) -- (A) node[midway, below] {$i$};
\draw[red, midarrow, line width=1.5pt] (Aa) -- (Ba) node[midway, above] {$i'$};
\draw[red, midarrow, line width=1.5pt] (Da) -- (Aa) node[midway, right] {$\l'$};
\draw[red, midarrow, line width=1.5pt] (A) -- (D) node[midway, left] {$\l$};
\draw (C) -- (B) node[midway, right] {$j$};
\draw (C) -- (D) node[midway, above] {$k$};
\node[white] at (A) {};
\node[black] at (B) {};
\node[white] at (C) {};
\node[black] at (D) {};
\end{tikzpicture}}}
&\overset{\text{}}{\longrightarrow}&
\vcenter{\hbox{\begin{tikzpicture}[scale=2,
    white/.style={circle, draw, fill=white, inner sep=0pt, minimum size=6pt},
    black/.style={circle, draw, fill=black, inner sep=0pt, minimum size=6pt},
    label/.style={font=\small, midway, sloped, above},
    coord/.style={font=\scriptsize, anchor=south east},
    coordne/.style={font=\scriptsize, anchor=north east},
    coordnw/.style={font=\scriptsize, anchor=north west},
    coordsw/.style={font=\scriptsize, anchor=south west}
]
\coordinate (A) at (0,0);   
\coordinate (B) at (1,0);   
\coordinate (C) at (1,1);   
\coordinate (D) at (0,1);   
\node[coordne] at (A) {$x$};
\node[coordnw] at (B) {$$};
\node[coordsw] at (C) {$$};
\node[coord] at (D) {$$};
\draw[red, midarrow, line width=1.5pt] (B) -- (A) node[midway, below] {$i$};
\draw[red, midarrow, line width=1.5pt] (C) -- (B) node[midway, right] {$j$};
\draw[red, midarrow, line width=1.5pt] (D) -- (C) node[midway, above] {$k$};
\draw[red, midarrow, line width=1.5pt] (A) -- (D) node[midway, left] {$\l$};
\node[white] at (A) {};
\node[black] at (B) {};
\node[white] at (C) {};
\node[black] at (D) {};
\end{tikzpicture}}}\cr
\psi(x) &\rightarrow & \psi{}_i{}^{i'}{}_{\l'}{}^{\l} = \frac{1}{N} \delta_i^{i'} \delta^{\l}_{\l'} \psi(x) &\rightarrow & \psi_i{}^j{}_k{}^{\l} = U^j{}_k|{}^{\l'}{}_{i'} \psi{}_i{}^{i'}{}_{\l'}{}^{\l}\cr
&&&&= \frac{1}{N} U^j{}_k|{}^{\l}{}_i \psi(x)\cr
\vcenter{\hbox{\begin{tikzpicture}[scale=2,
    white/.style={circle, draw, fill=white, inner sep=0pt, minimum size=6pt},
    black/.style={circle, draw, fill=black, inner sep=0pt, minimum size=6pt},
    label/.style={font=\small, midway, sloped, above},
    coord/.style={font=\scriptsize, anchor=south east},
    coordne/.style={font=\scriptsize, anchor=north east},
    coordnw/.style={font=\scriptsize, anchor=north west},
    coordsw/.style={font=\scriptsize, anchor=south west}
]
\coordinate (A) at (0,0);   
\coordinate (B) at (1,0);   
\coordinate (C) at (1,1);   
\coordinate (D) at (0,1);   
\node[coordne] at (A) {$x$};
\node[coordnw] at (B) {$$};
\node[coordsw] at (C) {$$};
\node[coord] at (D) {$$};
\draw[red, midarrow, line width=1.5pt] (A) -- (B) node[midway, below] {$i$};
\draw (C) -- (B) node[midway, right] {$j$};
\draw (A) -- (D) node[midway, left] {$\l$};
\draw (C) -- (D) node[midway, above] {$k$};
\node[white] at (A) {};
\node[black] at (B) {};
\node[white] at (C) {};
\node[black] at (D) {};
\end{tikzpicture}}}
&\overset{\text{}}{\longrightarrow}&
\vcenter{\hbox{\begin{tikzpicture}[scale=2,
    white/.style={circle, draw, fill=white, inner sep=0pt, minimum size=6pt},
    black/.style={circle, draw, fill=black, inner sep=0pt, minimum size=6pt},
    label/.style={font=\small, midway, sloped, above},
    coord/.style={font=\scriptsize, anchor=south east},
    coordne/.style={font=\scriptsize, anchor=north east},
    coordnw/.style={font=\scriptsize, anchor=north west},
    coordsw/.style={font=\scriptsize, anchor=south west}
]
\coordinate (A) at (0,0);   
\coordinate (Aa) at (0.04,0);
\coordinate (B) at (1,0);   
\coordinate (C) at (1,1);   
\coordinate (D) at (0,1);   
\coordinate (Da) at (0.04,1);  
\node[coordne] at (A) {$x$};
\node[coordnw] at (B) {$$};
\node[coordsw] at (C) {$$};
\node[coord] at (D) {$$};
\draw[red, midarrow, line width=1.5pt] (A) -- (D) node[midway, left] {$\l$};
\draw[red, midarrow, line width=1.5pt] (Da) -- (Aa) node[midway, right] {$\l'$};
\draw[red, midarrow, line width=1.5pt] (A) -- (B) node[midway, below] {$i$};
\draw (C) -- (B) node[midway, right] {$$};
\draw (C) -- (D) node[midway, above] {$$};
\node[white] at (A) {};
\node[black] at (B) {};
\node[white] at (C) {};
\node[black] at (D) {};
\end{tikzpicture}}}
&\overset{\text{}}{\longrightarrow}&
\vcenter{\hbox{\begin{tikzpicture}[scale=2,
    white/.style={circle, draw, fill=white, inner sep=0pt, minimum size=6pt},
    black/.style={circle, draw, fill=black, inner sep=0pt, minimum size=6pt},
    label/.style={font=\small, midway, sloped, above},
    coord/.style={font=\scriptsize, anchor=south east},
    coordne/.style={font=\scriptsize, anchor=north east},
    coordnw/.style={font=\scriptsize, anchor=north west},
    coordsw/.style={font=\scriptsize, anchor=south west}
]
\coordinate (A) at (0,0);   
\coordinate (B) at (1,0);   
\coordinate (C) at (1,1);   
\coordinate (D) at (0,1);   
\node[coordne] at (A) {$x$};
\node[coordnw] at (B) {$$};
\node[coordsw] at (C) {$$};
\node[coord] at (D) {$$};
\draw (A) -- (B) node[midway, below] {$i$};
\draw[red, midarrow, line width=1.5pt] (C) -- (B) node[midway, right] {$j$};
\draw[red, midarrow, line width=1.5pt] (D) -- (C) node[midway, above] {$k$};
\draw[red, midarrow, line width=1.5pt] (A) -- (D) node[midway, left] {$\l$};
\node[white] at (A) {};
\node[black] at (B) {};
\node[white] at (C) {};
\node[black] at (D) {};
\end{tikzpicture}}}\cr
\psi^i &\rightarrow & \psi^i{}_{\l'}{}^{\l} = \frac{1}{\sqrt{N}} \delta_{\l'}^{\l} \psi^i 
&\rightarrow &\psi^j{}_k{}^{\l} = U^j{}_k|{}^{\l'}{}_{i} \psi{}^{i}{}_{\l'}{}^{\l}\cr
&&&& = \frac{1}{\sqrt{N}} U^j{}_k|{}^{\l}{}_i \psi^i
\end{array}
\eea
The remaining moves, the $3\to 1$ and $4\to 0$, require calibrated situations in order for K-spikes to arise from surface holonomies that span over a nonvanishing surface area,
\bea
\begin{array}{ccccc}
\vcenter{\hbox{\begin{tikzpicture}[scale=2,
    white/.style={circle, draw, fill=white, inner sep=0pt, minimum size=6pt},
    black/.style={circle, draw, fill=black, inner sep=0pt, minimum size=6pt},
    label/.style={font=\small, midway, sloped, above},
    coord/.style={font=\scriptsize, anchor=south east},
    coordne/.style={font=\scriptsize, anchor=north east},
    coordnw/.style={font=\scriptsize, anchor=north west},
    coordsw/.style={font=\scriptsize, anchor=south west}
]
\coordinate (A) at (0,0);   
\coordinate (B) at (1,0);   
\coordinate (C) at (1,1);   
\coordinate (D) at (0,1);   
\node[coordne] at (A) {$x$};
\node[coordnw] at (B) {$$};
\node[coordsw] at (C) {$$};
\node[coord] at (D) {$$};
\draw[red, midarrow, line width=1.5pt] (A) -- (B) node[midway, below] {$i$};
\draw[red, midarrow, line width=1.5pt] (B) -- (C) node[midway, right] {$j$};
\draw[red, midarrow, line width=1.5pt] (C) -- (D) node[midway, above] {$k$};
\draw (D) -- (A) node[midway, left] {$\l$};
\node[white] at (A) {};
\node[black] at (B) {};
\node[white] at (C) {};
\node[black] at (D) {};
\end{tikzpicture}}}
&\overset{\text{}}{\longrightarrow}&
\vcenter{\hbox{\begin{tikzpicture}[scale=2,
    white/.style={circle, draw, fill=white, inner sep=0pt, minimum size=6pt},
    black/.style={circle, draw, fill=black, inner sep=0pt, minimum size=6pt},
    label/.style={font=\small, midway, sloped, above},
    coord/.style={font=\scriptsize, anchor=south east},
    coordne/.style={font=\scriptsize, anchor=north east},
    coordnw/.style={font=\scriptsize, anchor=north west},
    coordsw/.style={font=\scriptsize, anchor=south west}
]
\coordinate (A) at (0,0);   
\coordinate (B) at (1,0);   
\coordinate (C) at (1,1);   
\coordinate (D) at (0,1);   
\coordinate (Aa) at (0,0.04); 
\coordinate (Ba) at (1,0.04); 
\node[coordne] at (A) {$x$};
\node[coordnw] at (B) {$$};
\node[coordsw] at (C) {$$};
\node[coord] at (D) {$$};
\draw[red, midarrow, line width=1.5pt] (A) -- (B) node[midway, below] {$i$};
\draw[red, midarrow, line width=1.5pt] (Ba) -- (Aa) node[midway, above] {$i'$};
\draw[red, midarrow, line width=1.5pt] (Aa) -- (D) node[midway, left] {$\l$};
\draw (D) -- (C) node[midway, above] {$k$};
\draw (C) -- (B) node[midway, right] {$j$};
\node[white] at (A) {};
\node[black] at (B) {};
\node[white] at (C) {};
\node[black] at (D) {};
\end{tikzpicture}}}
&\overset{\text{}}{\longrightarrow}&
\vcenter{\hbox{\begin{tikzpicture}[scale=2,
    white/.style={circle, draw, fill=white, inner sep=0pt, minimum size=6pt},
    black/.style={circle, draw, fill=black, inner sep=0pt, minimum size=6pt},
    label/.style={font=\small, midway, sloped, above},
    coord/.style={font=\scriptsize, anchor=south east},
    coordne/.style={font=\scriptsize, anchor=north east},
    coordnw/.style={font=\scriptsize, anchor=north west},
    coordsw/.style={font=\scriptsize, anchor=south west}
]
\coordinate (A) at (0,0);   
\coordinate (B) at (1,0);   
\coordinate (C) at (1,1);   
\coordinate (D) at (0,1);   
\node[coordne] at (A) {$x$};
\node[coordnw] at (B) {$$};
\node[coordsw] at (C) {$$};
\node[coord] at (D) {$$};
\draw (A) -- (B) node[midway, below] {$i$};
\draw (B) -- (C) node[midway, right] {$j$};
\draw (C) -- (D) node[midway, above] {$k$};
\draw[red, midarrow, line width=1.5pt] (A) -- (D) node[midway, left] {$\l$};
\node[white] at (A) {};
\node[black] at (B) {};
\node[white] at (C) {};
\node[black] at (D) {};
\end{tikzpicture}}}\cr
\psi^k{}_j{}^i &\rightarrow & \psi^{\l}{}_{i'}{}^i = \frac{1}{\sqrt{N}} \psi^{\l} \delta^i_{i'} &\rightarrow & \psi^{\l}\cr 
&& = U^{\l}{}_{i'}|{}^j{}_k\psi^k{}_j{}^i &&\sqrt{N} \psi^{\l} = U^{\l}{}_i|{}^j{}_k \psi^k{}_j{}^i\cr
\vcenter{\hbox{\begin{tikzpicture}[scale=2,
    white/.style={circle, draw, fill=white, inner sep=0pt, minimum size=6pt},
    black/.style={circle, draw, fill=black, inner sep=0pt, minimum size=6pt},
    label/.style={font=\small, midway, sloped, above},
    coord/.style={font=\scriptsize, anchor=south east},
    coordne/.style={font=\scriptsize, anchor=north east},
    coordnw/.style={font=\scriptsize, anchor=north west},
    coordsw/.style={font=\scriptsize, anchor=south west}
]
\coordinate (A) at (0,0);   
\coordinate (B) at (1,0);   
\coordinate (C) at (1,1);   
\coordinate (D) at (0,1);   
\node[coordne] at (A) {$x$};
\node[coordnw] at (B) {$$};
\node[coordsw] at (C) {$$};
\node[coord] at (D) {$$};
\draw[red, midarrow, line width=1.5pt] (A) -- (B) node[midway, below] {$i$};
\draw[red, midarrow, line width=1.5pt] (B) -- (C) node[midway, right] {$j$};
\draw[red, midarrow, line width=1.5pt] (C) -- (D) node[midway, above] {$k$};
\draw[red, midarrow, line width=1.5pt] (D) -- (A) node[midway, left] {$\l$};
\node[white] at (A) {};
\node[black] at (B) {};
\node[white] at (C) {};
\node[black] at (D) {};
\end{tikzpicture}}}
&\overset{\text{}}{\longrightarrow}&
\vcenter{\hbox{\begin{tikzpicture}[scale=2,
    white/.style={circle, draw, fill=white, inner sep=0pt, minimum size=6pt},
    black/.style={circle, draw, fill=black, inner sep=0pt, minimum size=6pt},
    label/.style={font=\small, midway, sloped, above},
    coord/.style={font=\scriptsize, anchor=south east},
    coordne/.style={font=\scriptsize, anchor=north east},
    coordnw/.style={font=\scriptsize, anchor=north west},
    coordsw/.style={font=\scriptsize, anchor=south west}
]
\coordinate (A) at (0,0);   
\coordinate (B) at (1,0);   
\coordinate (C) at (1,1);   
\coordinate (D) at (0,1);   
\coordinate (Aa) at (0.04,0.04);
\coordinate (Ba) at (1,0.04);
\coordinate (Da) at (0.04,1);
\node[coordne] at (A) {$x$};
\node[coordnw] at (B) {$$};
\node[coordsw] at (C) {$$};
\node[coord] at (D) {$$};
\draw[red, midarrow, line width=1.5pt] (A) -- (B) node[midway, below] {$i$};
\draw[red, midarrow, line width=1.5pt] (Ba) -- (Aa) node[midway, above] {$i'$};
\draw[red, midarrow, line width=1.5pt] (Aa) -- (Da) node[midway, right] {$\l'$};
\draw[red, midarrow, line width=1.5pt] (D) -- (A) node[midway, left] {$\l$};
\draw (B) -- (C) node[midway, right] {$j$};
\draw (C) -- (D) node[midway, above] {$k$};
\node[white] at (A) {};
\node[black] at (B) {};
\node[white] at (C) {};
\node[black] at (D) {};
\end{tikzpicture}}}
&\overset{\text{}}{\longrightarrow}&
\vcenter{\hbox{\begin{tikzpicture}[scale=2,
    white/.style={circle, draw, fill=white, inner sep=0pt, minimum size=6pt},
    black/.style={circle, draw, fill=black, inner sep=0pt, minimum size=6pt},
    red/.style={circle, draw, fill=red, inner sep=0pt, minimum size=6pt},
    label/.style={font=\small, midway, sloped, above},
    coord/.style={font=\scriptsize, anchor=south east},
    coordne/.style={font=\scriptsize, anchor=north east},
    coordnw/.style={font=\scriptsize, anchor=north west},
    coordsw/.style={font=\scriptsize, anchor=south west}
]
\coordinate (A) at (0,0);   
\coordinate (B) at (1,0);   
\coordinate (C) at (1,1);   
\coordinate (D) at (0,1);   
\node[coordne] at (A) {$x$};
\node[coordnw] at (B) {$$};
\node[coordsw] at (C) {$$};
\node[coord] at (D) {$$};
\draw (A) -- (B) node[midway, below] {$i$};
\draw (B) -- (C) node[midway, right] {$j$};
\draw (C) -- (D) node[midway, above] {$k$};
\draw (D) -- (A) node[midway, left] {$\l$};
\node[red] at (A) {};
\node[black] at (B) {};
\node[white] at (C) {};
\node[black] at (D) {};
\end{tikzpicture}}}\cr
\psi_{\l}{}^k{}_j{}^i &\rightarrow &\psi_{\l}{}^{\l'}{}_{i'}{}^i = \frac{1}{N} \delta^i_{i'} \delta^{\l}_{\l'} \psi(x)&\rightarrow & \psi(x)\cr
&&= U^{\l'}{}_{i'}|{}^j{}_k \psi_{\l}{}^k{}_j{}^i&& N \psi(x) = U^{\l}{}_{i}|{}^j{}_k \psi_{\l}{}^k{}_j{}^i
\end{array}
\eea
We need to calibrate here in order to get a K-spike that we can remove by an R-move. 

By comparing with (\ref{61}) we get
\ben
U_i{}^j{}_k{}^{\l}| &=& \frac{1}{N} U^j{}_k|{}^{\l}{}_i\cr
U^j{}_k{}^{\l}|{}_i &=& \frac{1}{\sqrt{N}} U^j{}_k|{}^{\l}{}_i\cr
U^{\l}|{}_i{}^j{}_k &=& \sqrt{N} U^{\l}{}_i|{}^j{}_k\cr
U|{}_i{}^j{}_k{}^{\l} &=& N U^{\l}{}_i|{}^j{}_k\label{n61}
\een
These relations show that the vertical line separating the indices can be essentially removed because each of these move-operators are related to one another. We have to be a little bit careful when we remove the vertical line since there are the factors of $\sqrt{N}$ that we need to take into account. The above relations are an important first step towards defining what we call as the 'universal plaquette operator' in the next section.

\section{The universal plaquette operator}
Let us consider a closed string that has a portion that stretches over two edges as drawn below, 
\bea
\begin{tikzpicture}[scale=3,
    white/.style={circle, draw, fill=white, inner sep=0pt, minimum size=6pt},
    black/.style={circle, draw, fill=black, inner sep=0pt, minimum size=6pt},
    label/.style={font=\small, midway, sloped, above},
    coord/.style={font=\scriptsize, anchor=south east},
    coordne/.style={font=\scriptsize, anchor=north east},
    coordnw/.style={font=\scriptsize, anchor=north west},
    coordsw/.style={font=\scriptsize, anchor=south west}
]
\coordinate (A) at (0,0);   
\coordinate (B) at (1,0);   
\coordinate (Ba) at (0.98,0);   
\coordinate (Bb) at (1.02,0);   
\coordinate (C) at (2,0);   
\coordinate (D) at (1,1);
\coordinate (Da) at (0.98,1);   
\coordinate (Db) at (1.02,1);   
\coordinate (x) at (0,1);  
\coordinate (y) at (2,1);  
\draw[midarrow, red, line width=1.5pt] (A) -- (Ba) node[midway, above] {$i$};
\draw[midarrow,red, line width=1.5pt] (Bb) -- (C) node[midway, above] {$j$};
\draw (B) -- (D) node[midway, left] {$k$};
\draw (A) -- (x) node[midway, left] {$m$};
\draw (C) -- (y) node[midway, left] {$n$};
\draw (x) -- (D) node[midway, above] {$p$};
\draw (D) -- (y) node[midway, above] {$q$};
\node[coordne] at (A) {$$};
\node[white] at (A) {};
\node[black] at (B) {};
\node[white] at (C) {};
\node[white] at (D) {};
\end{tikzpicture}
\eea
Let us now move up the string by a $1\to 3$ move on the left side. Then the string configuration becomes as below,
\bea
\begin{tikzpicture}[scale=3,
    white/.style={circle, draw, fill=white, inner sep=0pt, minimum size=6pt},
    black/.style={circle, draw, fill=black, inner sep=0pt, minimum size=6pt},
    label/.style={font=\small, midway, sloped, above},
    coord/.style={font=\scriptsize, anchor=south east},
    coordne/.style={font=\scriptsize, anchor=north east},
    coordnw/.style={font=\scriptsize, anchor=north west},
    coordsw/.style={font=\scriptsize, anchor=south west}
]
\coordinate (A) at (0,0);   
\coordinate (B) at (1,0);   
\coordinate (Ba) at (0.98,0);   
\coordinate (Bb) at (1.02,0);   
\coordinate (C) at (2,0);   
\coordinate (D) at (1,1);
\coordinate (Da) at (0.98,1);   
\coordinate (Db) at (1.02,1);   
\coordinate (x) at (0,1);  
\coordinate (y) at (2,1);  
\draw (A) -- (B) node[midway, above] {$i$};
\draw (B) -- (D) node[midway, left] {$$};
\draw (A) -- (x) node[midway, left] {$$};
\draw (C) -- (y) node[midway, left] {$n$};
\draw (x) -- (D) node[midway, above] {$$};
\draw (D) -- (y) node[midway, above] {$q$};
\draw[midarrow, red, line width=1.5pt] (A) -- (x) node[midway, left] {$m$};
\draw[midarrow, red, line width=1.5pt] (x) -- (D) node[midway, above] {$p$};
\draw[midarrow, red, line width=1.5pt] (D) -- (B) node[midway, left] {$k$};
\draw[midarrow,red, line width=1.5pt] (Bb) -- (C) node[midway, above] {$j$};
\node[coordne] at (A) {$$};
\node[white] at (A) {};
\node[black] at (B) {};
\node[white] at (C) {};
\node[white] at (D) {};
\end{tikzpicture}
\eea
Under this move, the wave function transforms as
\bea
\psi_j{}^i \rightarrow U^k{}_p{}^m|{}_i \psi_j{}^i
\eea
By using a relation that we obtained in (\ref{n61}), we can rewrite this as
\bea
\psi_j{}^i \rightarrow \frac{1}{\sqrt{N}} U^k{}_p|{}^m{}_i \psi_j{}^i
\eea
Let us now instead first create a spike in the middle as below,
\bea
\begin{tikzpicture}[scale=3,
    white/.style={circle, draw, fill=white, inner sep=0pt, minimum size=6pt},
    black/.style={circle, draw, fill=black, inner sep=0pt, minimum size=6pt},
    label/.style={font=\small, midway, sloped, above},
    coord/.style={font=\scriptsize, anchor=south east},
    coordne/.style={font=\scriptsize, anchor=north east},
    coordnw/.style={font=\scriptsize, anchor=north west},
    coordsw/.style={font=\scriptsize, anchor=south west}
]
\coordinate (A) at (0,0);   
\coordinate (B) at (1,0);   
\coordinate (Ba) at (0.98,0);  
\coordinate (Bb) at (1.02,0);   
\coordinate (C) at (2,0);   
\coordinate (D) at (1,1);
\coordinate (Da) at (0.98,1);   
\coordinate (Db) at (1.02,1);   
\draw (A) -- (B) node[midway, above] {$$};
\draw (B) -- (D) node[midway, left] {$$};
\draw (A) -- (x) node[midway, left] {$m$};
\draw (C) -- (y) node[midway, left] {$n$};
\draw (x) -- (D) node[midway, above] {$p$};
\draw (D) -- (y) node[midway, above] {$q$};
\draw[midarrow, red, line width=1.5pt] (A) -- (Ba) node[midway, above] {$i$};
\draw[midarrow,red, line width=1.5pt] (Ba) -- (Da) node[midway, left] {$k$};
\draw[midarrow,red, line width=1.5pt] (Db) -- (Bb) node[midway, right] {$k'$};
\draw[midarrow,red, line width=1.5pt] (Bb) -- (C) node[midway, above] {$j$};
\node[coordne] at (A) {$$};
\node[white] at (A) {};
\node[black] at (B) {};
\node[white] at (C) {};
\node[white] at (D) {};
\end{tikzpicture}
\eea  
and then apply a $2\to 2$ move to get the same final string configuration as before. In this case, the wave function transforms as
\bea
\psi_j{}^i \rightarrow \psi_j{}^{k}{}_{k'}{}^i = \frac{1}{\sqrt{N}} \delta_{k'}^{k} \psi_j{}^i
\eea
for the K-spike move, and then as 
\bea
\psi_j^{k}{}_{k'}{}^i \rightarrow U_p{}^m|_i{}^{k'} \psi_j{}^{k}{}_{k'}{}^i = \frac{1}{\sqrt{N}} U_p{}^m|_i{}^{k} \psi_j{}^i
\eea
for the $2\to 2$ move. For this to be consistent, the two results shall agree and since $\psi_j{}^i$ can be an arbitrary wave function, this forces us to impose the constraint relation
\ben
U_p{}^m|{}_i{}^k &=& U^k{}_p|{}^m{}_i\label{it1}
\een
on the $2\to 2$ move operator. By shifting the entire string configuration one step to the right, we flip black and white vertices and there we discover a corresponding relation
\bea
U^k{}_p|{}^m{}_i &=& U_i{}^k|{}_p{}^m\label{it2}
\eea
and now we can iterate (\ref{it1}) one more step to get the relation
\bea
U_i{}^k|{}_p{}^m &=& U^m{}_i|{}^k{}_p
\eea
and finally we can iterate (\ref{it2}) to connect back to where we started,
\bea
U^m{}_i|{}^k{}_p &=& U_p{}^m|{}_i{}^k
\eea
In this way, we have shown a cyclic property of the $2\to 2$ move operator, which could seem to be quite unexpected, since apriori, why would it be cyclically symmetric? In contrast, for example the line holonomy operator in a lattice one-form Yang-Mills gauge theory is not cyclically symmetric. If we write this line holonomy operator as $U^i{}_j$ then we do not have any relation such as $U^i{}_j = U_j{}^i$. So that we see a cyclic symmetry for the $2\to 2$ plaquette operator is non-trivial.

Now finally we are ready to introduce the 'universal plaquette operator'. We will define this operator as
\bea
U_p{}^m{}_i{}^k &:=& U_p{}^m|{}_i{}^k
\eea
where we simply removed the vertical bar. This operator is cyclically symmetric and it is related to any of the $p \to (4-p)$ move operator for $p = 0,1,2,3,4$ by factors of $\sqrt{N}$ to various powers as depicted below, 
\bea
\begin{array}{cccc}
\vcenter{\hbox{
\begin{tikzpicture}[scale=2,
    white/.style={circle, draw, fill=white, inner sep=0pt, minimum size=6pt},
    black/.style={circle, draw, fill=black, inner sep=0pt, minimum size=6pt},
    red/.style={circle, draw, fill=red, inner sep=0pt, minimum size=6pt},
    label/.style={font=\small, midway, sloped, above},
    coord/.style={font=\scriptsize, anchor=south east},
    coordne/.style={font=\scriptsize, anchor=north east},
    coordnw/.style={font=\scriptsize, anchor=north west},
    coordsw/.style={font=\scriptsize, anchor=south west}
]
\coordinate (A) at (0,0);   
\coordinate (B) at (1,0);   
\coordinate (C) at (1,1);   
\coordinate (D) at (0,1);   
\node[coordne] at (A) {$x$};
\node[coordnw] at (B) {$$};
\node[coordsw] at (C) {$$};
\node[coord] at (D) {$$};
\draw (A) -- (B) node[midway, below] {$$};
\draw (C) -- (B) node[midway, right] {$$};
\draw (A) -- (D) node[midway, left] {$$};
\draw (C) -- (D) node[midway, above] {$$};
\node[red] at (A) {};
\node[black] at (B) {};
\node[white] at (C) {};
\node[black] at (D) {};
\end{tikzpicture}}}
&\longrightarrow &
\vcenter{\hbox{\begin{tikzpicture}[scale=2,
    white/.style={circle, draw, fill=white, inner sep=0pt, minimum size=6pt},
    black/.style={circle, draw, fill=black, inner sep=0pt, minimum size=6pt},
    label/.style={font=\small, midway, sloped, above},
    coord/.style={font=\scriptsize, anchor=south east},
    coordne/.style={font=\scriptsize, anchor=north east},
    coordnw/.style={font=\scriptsize, anchor=north west},
    coordsw/.style={font=\scriptsize, anchor=south west}
]
\coordinate (A) at (0,0);   
\coordinate (B) at (1,0);   
\coordinate (C) at (1,1);   
\coordinate (D) at (0,1);   
\node[coordne] at (A) {$x$};
\node[coordnw] at (B) {$$};
\node[coordsw] at (C) {$$};
\node[coord] at (D) {$$};
\draw[red, midarrow, line width=1.5pt] (B) -- (A) node[midway, below] {$i$};
\draw[red, midarrow, line width=1.5pt] (C) -- (B) node[midway, right] {$j$};
\draw[red, midarrow, line width=1.5pt] (D) -- (C) node[midway, above] {$k$};
\draw[red, midarrow, line width=1.5pt] (A) -- (D) node[midway, right] {$\l$};
\node[white] at (A) {};
\node[black] at (B) {};
\node[white] at (C) {};
\node[black] at (D) {};
\end{tikzpicture}}}
&\psi(x) \rightarrow \psi_i{}^j{}_k{}^{\l} = N U_i{}^j{}_k{}^{\l} \psi(x)\cr
\vcenter{\hbox{\begin{tikzpicture}[scale=2,
    white/.style={circle, draw, fill=white, inner sep=0pt, minimum size=6pt},
    black/.style={circle, draw, fill=black, inner sep=0pt, minimum size=6pt},
    label/.style={font=\small, midway, sloped, above},
    coord/.style={font=\scriptsize, anchor=south east},
    coordne/.style={font=\scriptsize, anchor=north east},
    coordnw/.style={font=\scriptsize, anchor=north west},
    coordsw/.style={font=\scriptsize, anchor=south west}
]
\coordinate (A) at (0,0);   
\coordinate (B) at (1,0);   
\coordinate (C) at (1,1);   
\coordinate (D) at (0,1);   
\node[coordne] at (A) {$x$};
\node[coordnw] at (B) {$$};
\node[coordsw] at (C) {$$};
\node[coord] at (D) {$$};
\draw[red, midarrow, line width=1.5pt] (A) -- (B) node[midway, below] {$i$};
\draw (C) -- (B) node[midway, right] {$$};
\draw (A) -- (D) node[midway, left] {$$};
\draw (C) -- (D) node[midway, above] {$$};
\node[white] at (A) {};
\node[black] at (B) {};
\node[white] at (C) {};
\node[black] at (D) {};
\end{tikzpicture}}}
& \longrightarrow &
\vcenter{\hbox{\begin{tikzpicture}[scale=2,
    white/.style={circle, draw, fill=white, inner sep=0pt, minimum size=6pt},
    black/.style={circle, draw, fill=black, inner sep=0pt, minimum size=6pt},
    label/.style={font=\small, midway, sloped, above},
    coord/.style={font=\scriptsize, anchor=south east},
    coordne/.style={font=\scriptsize, anchor=north east},
    coordnw/.style={font=\scriptsize, anchor=north west},
    coordsw/.style={font=\scriptsize, anchor=south west}
]
\coordinate (A) at (0,0);   
\coordinate (B) at (1,0);   
\coordinate (C) at (1,1);   
\coordinate (D) at (0,1);   
\node[coordne] at (A) {$x$};
\node[coordnw] at (B) {$$};
\node[coordsw] at (C) {$$};
\node[coord] at (D) {$$};
\draw (A) -- (B) node[midway, below] {$$};
\draw[red, midarrow, line width=1.5pt] (A) -- (D) node[midway, right] {$\l$};
\draw[red, midarrow, line width=1.5pt] (D) -- (C) node[midway, above] {$k$};
\draw[red, midarrow, line width=1.5pt] (C) -- (B) node[midway, right] {$j$};
\node[white] at (A) {};
\node[black] at (B) {};
\node[white] at (C) {};
\node[black] at (D) {};
\end{tikzpicture}}}
&\psi^i \rightarrow \psi^j{}_k{}^{\l} = \sqrt{N} U_i{}^j{}_k{}^{\l} \psi^i\cr
\vcenter{\hbox{\begin{tikzpicture}[scale=2,
    white/.style={circle, draw, fill=white, inner sep=0pt, minimum size=6pt},
    black/.style={circle, draw, fill=black, inner sep=0pt, minimum size=6pt},
    label/.style={font=\small, midway, sloped, above},
    coord/.style={font=\scriptsize, anchor=south east},
    coordne/.style={font=\scriptsize, anchor=north east},
    coordnw/.style={font=\scriptsize, anchor=north west},
    coordsw/.style={font=\scriptsize, anchor=south west}
]
\coordinate (A) at (0,0);   
\coordinate (B) at (1,0);   
\coordinate (C) at (1,1);   
\coordinate (D) at (0,1);   
\node[coordne] at (A) {$x$};
\node[coordnw] at (B) {$$};
\node[coordsw] at (C) {$$};
\node[coord] at (D) {$$};
\draw[red, midarrow, line width=1.5pt] (A) -- (B) node[midway, below] {$i$};
\draw[red, midarrow, line width=1.5pt] (B) -- (C) node[midway, right] {$j$};
\draw (C) -- (D) node[midway, left] {$$};
\draw (D) -- (A) node[midway, above] {$$};
\node[white] at (A) {};
\node[black] at (B) {};
\node[white] at (C) {};
\node[black] at (D) {};
\end{tikzpicture}}}
& \longrightarrow &
\vcenter{\hbox{\begin{tikzpicture}[scale=2,
    white/.style={circle, draw, fill=white, inner sep=0pt, minimum size=6pt},
    black/.style={circle, draw, fill=black, inner sep=0pt, minimum size=6pt},
    label/.style={font=\small, midway, sloped, above},
    coord/.style={font=\scriptsize, anchor=south east},
    coordne/.style={font=\scriptsize, anchor=north east},
    coordnw/.style={font=\scriptsize, anchor=north west},
    coordsw/.style={font=\scriptsize, anchor=south west}
]
\coordinate (A) at (0,0);   
\coordinate (B) at (1,0);   
\coordinate (C) at (1,1);   
\coordinate (D) at (0,1);   
\node[coordne] at (A) {$x$};
\node[coordnw] at (B) {$$};
\node[coordsw] at (C) {$$};
\node[coord] at (D) {$$};
\draw (A) -- (B) node[midway, below] {$$};
\draw (B) -- (C) node[midway, right] {$$};
\draw[red, midarrow, line width=1.5pt] (A) -- (D) node[midway, right] {$\l$};
\draw[red, midarrow, line width=1.5pt] (D) -- (C) node[midway, above] {$k$};
\node[white] at (A) {};
\node[black] at (B) {};
\node[white] at (C) {};
\node[black] at (D) {};
\end{tikzpicture}}}
&\psi_j{}^i \rightarrow \psi_k{}^{\l} = U{}_i{}^j{}_k{}^{\l} \psi_j{}^i\cr
\vcenter{\hbox{\begin{tikzpicture}[scale=2,
    white/.style={circle, draw, fill=white, inner sep=0pt, minimum size=6pt},
    black/.style={circle, draw, fill=black, inner sep=0pt, minimum size=6pt},
    label/.style={font=\small, midway, sloped, above},
    coord/.style={font=\scriptsize, anchor=south east},
    coordne/.style={font=\scriptsize, anchor=north east},
    coordnw/.style={font=\scriptsize, anchor=north west},
    coordsw/.style={font=\scriptsize, anchor=south west}
]
\coordinate (A) at (0,0);   
\coordinate (B) at (1,0);   
\coordinate (C) at (1,1);   
\coordinate (D) at (0,1);   
\node[coordne] at (A) {$x$};
\node[coordnw] at (B) {$$};
\node[coordsw] at (C) {$$};
\node[coord] at (D) {$$};
\draw[red, midarrow, line width=1.5pt] (A) -- (B) node[midway, below] {$i$};
\draw[red, midarrow, line width=1.5pt] (B) -- (C) node[midway, right] {$j$};
\draw[red, midarrow, line width=1.5pt] (C) -- (D) node[midway, above] {$k$};
\draw (D) -- (A) node[midway, above] {$$};
\node[white] at (A) {};
\node[black] at (B) {};
\node[white] at (C) {};
\node[black] at (D) {};
\end{tikzpicture}}}
& \longrightarrow &
\vcenter{\hbox{\begin{tikzpicture}[scale=2,
    white/.style={circle, draw, fill=white, inner sep=0pt, minimum size=6pt},
    black/.style={circle, draw, fill=black, inner sep=0pt, minimum size=6pt},
    label/.style={font=\small, midway, sloped, above},
    coord/.style={font=\scriptsize, anchor=south east},
    coordne/.style={font=\scriptsize, anchor=north east},
    coordnw/.style={font=\scriptsize, anchor=north west},
    coordsw/.style={font=\scriptsize, anchor=south west}
]
\coordinate (A) at (0,0);   
\coordinate (B) at (1,0);   
\coordinate (C) at (1,1);   
\coordinate (D) at (0,1);   
\node[coordne] at (A) {$x$};
\node[coordnw] at (B) {$$};
\node[coordsw] at (C) {$$};
\node[coord] at (D) {$$};
\draw (A) -- (B) node[midway, below] {$$};
\draw (B) -- (C) node[midway, right] {$$};
\draw (C) -- (D) node[midway, right] {$$};
\draw[red, midarrow, line width=1.5pt] (A) -- (D) node[midway, right] {$\l$};
\node[white] at (A) {};
\node[black] at (B) {};
\node[white] at (C) {};
\node[black] at (D) {};
\end{tikzpicture}}}
& \psi^k{}_j{}^i \rightarrow \psi^{\l} = \frac{1}{\sqrt{N}} U{}_i{}^j{}_k{}^{\l} \psi^k{}_j{}^i\cr
\vcenter{\hbox{\begin{tikzpicture}[scale=2,
    white/.style={circle, draw, fill=white, inner sep=0pt, minimum size=6pt},
    black/.style={circle, draw, fill=black, inner sep=0pt, minimum size=6pt},
    label/.style={font=\small, midway, sloped, above},
    coord/.style={font=\scriptsize, anchor=south east},
    coordne/.style={font=\scriptsize, anchor=north east},
    coordnw/.style={font=\scriptsize, anchor=north west},
    coordsw/.style={font=\scriptsize, anchor=south west}
]
\coordinate (A) at (0,0);   
\coordinate (B) at (1,0);   
\coordinate (C) at (1,1);   
\coordinate (D) at (0,1);   
\node[coordne] at (A) {$x$};
\node[coordnw] at (B) {$$};
\node[coordsw] at (C) {$$};
\node[coord] at (D) {$$};
\draw[red, midarrow, line width=1.5pt] (A) -- (B) node[midway, below] {$i$};
\draw[red, midarrow, line width=1.5pt] (B) -- (C) node[midway, right] {$j$};
\draw[red, midarrow, line width=1.5pt] (C) -- (D) node[midway, above] {$k$};
\draw[red, midarrow, line width=1.5pt] (D) -- (A) node[midway, right] {$\l$};
\node[white] at (A) {};
\node[black] at (B) {};
\node[white] at (C) {};
\node[black] at (D) {};
\end{tikzpicture}}}
& \longrightarrow &
\vcenter{\hbox{\begin{tikzpicture}[scale=2,
    white/.style={circle, draw, fill=white, inner sep=0pt, minimum size=6pt},
    black/.style={circle, draw, fill=black, inner sep=0pt, minimum size=6pt},
    red/.style={circle, draw, fill=red, inner sep=0pt, minimum size=6pt},
    label/.style={font=\small, midway, sloped, above},
    coord/.style={font=\scriptsize, anchor=south east},
    coordne/.style={font=\scriptsize, anchor=north east},
    coordnw/.style={font=\scriptsize, anchor=north west},
    coordsw/.style={font=\scriptsize, anchor=south west}
]
\coordinate (A) at (0,0);   
\coordinate (B) at (1,0);   
\coordinate (C) at (1,1);   
\coordinate (D) at (0,1);   
\node[coordne] at (A) {$x$};
\node[coordnw] at (B) {$$};
\node[coordsw] at (C) {$$};
\node[coord] at (D) {$$};
\draw (A) -- (B) node[midway, below] {$$};
\draw (B) -- (C) node[midway, right] {$$};
\draw (C) -- (D) node[midway, right] {$$};
\draw (D) -- (A) node[midway, above] {$$};
\node[red] at (A) {};
\node[black] at (B) {};
\node[white] at (C) {};
\node[black] at (D) {};
\end{tikzpicture}}}
& \psi_{\l}{}^k{}_j{}^i \rightarrow \psi = \frac{1}{N} U_i{}^j{}_k{}^{\l} \psi_{\l}{}^k{}_j{}^i
\end{array}\label{UUU}
\eea

\section{Unitary compactness conditions}
As a string moves in the lattice, it sweeps out a two-dimensional surface. To this surface we associate a surface holonomy. The string has an orientation. But in order to define the surface holonomy, we need to specify an orientation of the surface itself, independently of the orientation of the string that moves over the surface. Below we picture one plaquette that we assign an orientation as indicated by an inner black square with an arrow on it. We refer to this as the orientation of the surface, as opposed to the orientation of the string. Since there are two choices of this orientation, there are correspondingly, two kinds of surface holonomies that we denote by $U$ and $\t{U}$ respectively. These are related as in the picture below,
\bea
\begin{array}{ccc}
\vcenter{\hbox{\begin{tikzpicture}[scale=3,
    white/.style={circle, draw, fill=white, inner sep=0pt, minimum size=6pt},
    black/.style={circle, draw, fill=black, inner sep=0pt, minimum size=6pt},
    label/.style={font=\small, midway, sloped, above},
    coord/.style={font=\scriptsize, anchor=south east},
    coordne/.style={font=\scriptsize, anchor=north east},
    coordnw/.style={font=\scriptsize, anchor=north west},
    coordsw/.style={font=\scriptsize, anchor=south west}
]
\coordinate (A) at (0,0);   
\coordinate (B) at (1,0);   
\coordinate (C) at (1,1);   
\coordinate (D) at (0,1);   
\node[coordne] at (A) {$x$};
\node[coordnw] at (B) {$$};
\node[coordsw] at (C) {$$};
\node[coord] at (D) {$$};
\draw (A) -- (B) node[midway, below] {$$};
\draw (B) -- (C) node[midway, right] {$$};
\draw (C) -- (D) node[midway, left] {$$};
\draw (D) -- (A) node[midway, above] {$$};
\node[white] at (A) {};
\node[black] at (B) {};
\node[white] at (C) {};
\node[black] at (D) {};

\coordinate (AA) at (0.1,0.1,0);   
\coordinate (AB) at (0.9,0.1,0);   
\coordinate (AC) at (0.9,0.9,0);   
\coordinate (AD) at (0.1,0.9,0); 

\draw[midarrow] (AB) -- (AA) node[midway, above] {$$};
\draw[midarrow] (AC) -- (AB) node[midway, right] {$$};
\draw[midarrow] (AD) -- (AC) node[midway, above] {$$};
\draw[midarrow] (AA) -- (AD) node[midway, right] {$$};

\node[canvas is xy plane at z=0, transform shape] at (0.5,0.5,0) {$U$};
\end{tikzpicture}}}
&=& 
\vcenter{\hbox{\begin{tikzpicture}[scale=3,
    white/.style={circle, draw, fill=white, inner sep=0pt, minimum size=6pt},
    black/.style={circle, draw, fill=black, inner sep=0pt, minimum size=6pt},
    label/.style={font=\small, midway, sloped, above},
    coord/.style={font=\scriptsize, anchor=south east},
    coordne/.style={font=\scriptsize, anchor=north east},
    coordnw/.style={font=\scriptsize, anchor=north west},
    coordsw/.style={font=\scriptsize, anchor=south west}
]
\coordinate (A) at (0,0);   
\coordinate (B) at (1,0);   
\coordinate (C) at (1,1);   
\coordinate (D) at (0,1);   
\node[coordne] at (A) {$x$};
\node[coordnw] at (B) {$$};
\node[coordsw] at (C) {$$};
\node[coord] at (D) {$$};
\draw (A) -- (B) node[midway, below] {$$};
\draw (B) -- (C) node[midway, right] {$$};
\draw (C) -- (D) node[midway, left] {$$};
\draw (D) -- (A) node[midway, above] {$$};
\node[white] at (A) {};
\node[black] at (B) {};
\node[white] at (C) {};
\node[black] at (D) {};

\coordinate (AA) at (0.1,0.1,0);   
\coordinate (AB) at (0.9,0.1,0);   
\coordinate (AC) at (0.9,0.9,0);   
\coordinate (AD) at (0.1,0.9,0); 

\coordinate (AA) at (0.1,0.1,0);   
\coordinate (AB) at (0.9,0.1,0);   
\coordinate (AC) at (0.9,0.9,0);   
\coordinate (AD) at (0.1,0.9,0); 

\draw[midarrow] (AA) -- (AB) node[midway, above] {$$};
\draw[midarrow] (AB) -- (AC) node[midway, right] {$$};
\draw[midarrow] (AC) -- (AD) node[midway, above] {$$};
\draw[midarrow] (AD) -- (AA) node[midway, right] {$$};

\node[canvas is xy plane at z=0, transform shape] at (0.5,0.5,0) {$\t{U}$};
\end{tikzpicture}}}
\end{array}
\eea
The operators $U$ and $\t{U}$ may now pull out a string that is residing at the vertex $x$ (and extending outside the plaquette in some other directions not drawn) and give the string the orientation that is drawn on the plaquette. We illustrate this below,
\bea
\vcenter{\hbox{\begin{tikzpicture}[scale=3,
    white/.style={circle, draw, fill=white, inner sep=0pt, minimum size=6pt},
   black/.style={circle, draw, fill=black, inner sep=0pt, minimum size=6pt},
   red/.style={circle, draw, fill=red, inner sep=0pt, minimum size=6pt},
   label/.style={font=\small, midway, sloped, above},
    coord/.style={font=\scriptsize, anchor=south east},
    coordne/.style={font=\scriptsize, anchor=north east},
    coordnw/.style={font=\scriptsize, anchor=north west},
    coordsw/.style={font=\scriptsize, anchor=south west}
]
\coordinate (A) at (0,0,0);   
\coordinate (B) at (1,0,0);   
\coordinate (C) at (1,1,0);   
\coordinate (D) at (0,1,0);   

\coordinate (AA) at (0.1,0.1,0);   
\coordinate (AB) at (0.9,0.1,0);   
\coordinate (AC) at (0.9,0.9,0);   
\coordinate (AD) at (0.1,0.9,0); 
\node[coordnw] at (B) {$$};
\node[coordsw] at (C) {$$};
\node[coord] at (D) {$$};
\draw (D) -- (A) node[midway, left] {$\l$};
\draw (C) -- (D) node[midway, above] {$k$};
\draw (B) -- (C) node[midway, right] {$j$};
\draw (A) -- (B) node[midway, below] {$i$};
\node[red] at (A) {};
\node[black] at (B) {};
\node[white] at (C) {};
\node[black] at (D) {};
\end{tikzpicture}}}
&\rightarrow & 
\vcenter{\hbox{\begin{tikzpicture}[scale=3,
    white/.style={circle, draw, fill=white, inner sep=0pt, minimum size=6pt},
    black/.style={circle, draw, fill=black, inner sep=0pt, minimum size=6pt},
    label/.style={font=\small, midway, sloped, above},
    coord/.style={font=\scriptsize, anchor=south east},
    coordne/.style={font=\scriptsize, anchor=north east},
    coordnw/.style={font=\scriptsize, anchor=north west},
    coordsw/.style={font=\scriptsize, anchor=south west}
]
\coordinate (A) at (0,0,0);   
\coordinate (B) at (1,0,0);   
\coordinate (C) at (1,1,0);   
\coordinate (D) at (0,1,0);   

\coordinate (AA) at (0.1,0.1,0);   
\coordinate (AB) at (0.9,0.1,0);   
\coordinate (AC) at (0.9,0.9,0);   
\coordinate (AD) at (0.1,0.9,0); 
\node[coordnw] at (B) {$$};
\node[coordsw] at (C) {$$};
\node[coord] at (D) {$$};
\draw[red, midarrow, line width=1.5pt] (B) -- (A) node[midway, below] {$i$};
\draw[red, midarrow, line width=1.5pt] (C) -- (B) node[midway, right] {$j$};
\draw[red, midarrow, line width=1.5pt] (D) -- (C) node[midway, above] {$k$};
\draw[red, midarrow, line width=1.5pt] (A) -- (D) node[midway, left] {$\l$};
\node[white] at (A) {};
\node[black] at (B) {};
\node[white] at (C) {};
\node[black] at (D) {};

\coordinate (AA) at (0.1,0.1,0);   
\coordinate (AB) at (0.9,0.1,0);   
\coordinate (AC) at (0.9,0.9,0);   
\coordinate (AD) at (0.1,0.9,0); 

\draw[midarrow] (AB) -- (AA) node[midway, above] {$$};
\draw[midarrow] (AC) -- (AB) node[midway, right] {$$};
\draw[midarrow] (AD) -- (AC) node[midway, above] {$$};
\draw[midarrow] (AA) -- (AD) node[midway, right] {$$};

\node[canvas is xy plane at z=0, transform shape] at (0.5,0.5,0) {$U$};
\end{tikzpicture}}}\cr
\vcenter{\hbox{\begin{tikzpicture}[scale=3,
    white/.style={circle, draw, fill=white, inner sep=0pt, minimum size=6pt},
    black/.style={circle, draw, fill=black, inner sep=0pt, minimum size=6pt},
    red/.style={circle, draw, fill=red, inner sep=0pt, minimum size=6pt},
    label/.style={font=\small, midway, sloped, above},
    coord/.style={font=\scriptsize, anchor=south east},
    coordne/.style={font=\scriptsize, anchor=north east},
    coordnw/.style={font=\scriptsize, anchor=north west},
    coordsw/.style={font=\scriptsize, anchor=south west}
]
\coordinate (A) at (0,0,0);   
\coordinate (B) at (1,0,0);   
\coordinate (C) at (1,1,0);   
\coordinate (D) at (0,1,0);   

\coordinate (AA) at (0.1,0.1,0);   
\coordinate (AB) at (0.9,0.1,0);   
\coordinate (AC) at (0.9,0.9,0);   
\coordinate (AD) at (0.1,0.9,0); 
\node[coordnw] at (B) {$$};
\node[coordsw] at (C) {$$};
\node[coord] at (D) {$$};
\draw (A) -- (D) node[midway, left] {$\l$};
\draw (D) -- (C) node[midway, above] {$k$};
\draw (C) -- (B) node[midway, right] {$j$};
\draw (B) -- (A) node[midway, below] {$i$};
\node[red] at (A) {};
\node[black] at (B) {};
\node[white] at (C) {};
\node[black] at (D) {};
\end{tikzpicture}}}
&\rightarrow & 
\vcenter{\hbox{\begin{tikzpicture}[scale=3,
    white/.style={circle, draw, fill=white, inner sep=0pt, minimum size=6pt},
    black/.style={circle, draw, fill=black, inner sep=0pt, minimum size=6pt},
    label/.style={font=\small, midway, sloped, above},
    coord/.style={font=\scriptsize, anchor=south east},
    coordne/.style={font=\scriptsize, anchor=north east},
    coordnw/.style={font=\scriptsize, anchor=north west},
    coordsw/.style={font=\scriptsize, anchor=south west}
]
\coordinate (A) at (0,0,0);   
\coordinate (B) at (1,0,0);   
\coordinate (C) at (1,1,0);   
\coordinate (D) at (0,1,0);   
\coordinate (AA) at (0.1,0.1,0);   
\coordinate (AB) at (0.9,0.1,0);   
\coordinate (AC) at (0.9,0.9,0);   
\coordinate (AD) at (0.1,0.9,0); 
\node[coordnw] at (B) {$$};
\node[coordsw] at (C) {$$};
\node[coord] at (D) {$$};
\draw[red, midarrow, line width=1.5pt] (A) -- (B) node[midway, below] {$i$};
\draw[red, midarrow, line width=1.5pt] (B) -- (C) node[midway, right] {$j$};
\draw[red, midarrow, line width=1.5pt] (C) -- (D) node[midway, above] {$k$};
\draw[red, midarrow, line width=1.5pt] (D) -- (A) node[midway, left] {$\l$};
\node[white] at (A) {};
\node[black] at (B) {};
\node[white] at (C) {};
\node[black] at (D) {};

\coordinate (AA) at (0.1,0.1,0);   
\coordinate (AB) at (0.9,0.1,0);   
\coordinate (AC) at (0.9,0.9,0);   
\coordinate (AD) at (0.1,0.9,0); 

\draw[midarrow] (AA) -- (AB) node[midway, above] {$$};
\draw[midarrow] (AB) -- (AC) node[midway, right] {$$};
\draw[midarrow] (AC) -- (AD) node[midway, above] {$$};
\draw[midarrow] (AD) -- (AA) node[midway, right] {$$};
\node[canvas is xy plane at z=0, transform shape] at (0.5,0.5,0) {$\t{U}$};
\end{tikzpicture}}}
\eea
 
The wave function of our string is complex-valued. We will assume that complex conjugation reverses the orientation of the string,
\bea
\(\vcenter{\hbox{\begin{tikzpicture}[scale=3,
    white/.style={circle, draw, fill=white, inner sep=0pt, minimum size=6pt},
    black/.style={circle, draw, fill=black, inner sep=0pt, minimum size=6pt},
    label/.style={font=\small, midway, sloped, above},
    coord/.style={font=\scriptsize, anchor=south east},
    coordne/.style={font=\scriptsize, anchor=north east},
    coordnw/.style={font=\scriptsize, anchor=north west},
    coordsw/.style={font=\scriptsize, anchor=south west}
]
\coordinate (A) at (0,0,0);   
\coordinate (B) at (1,0,0);   
\coordinate (C) at (1,1,0);   
\coordinate (D) at (0,1,0);   

\coordinate (AA) at (0.1,0.1,0);   
\coordinate (AB) at (0.9,0.1,0);   
\coordinate (AC) at (0.9,0.9,0);   
\coordinate (AD) at (0.1,0.9,0); 
\node[coordnw] at (B) {$$};
\node[coordsw] at (C) {$$};
\node[coord] at (D) {$$};
\draw[red, midarrow, line width=1.5pt] (A) -- (D) node[midway, left] {$\l$};
\draw[red, midarrow, line width=1.5pt] (D) -- (C) node[midway, above] {$k$};
\draw[red, midarrow, line width=1.5pt] (C) -- (B) node[midway, right] {$j$};
\draw[red, midarrow, line width=1.5pt] (B) -- (A) node[midway, below] {$i$};
\node[white] at (A) {};
\node[black] at (B) {};
\node[white] at (C) {};
\node[black] at (D) {};
\end{tikzpicture}}}\)^*
&=& 
\vcenter{\hbox{\begin{tikzpicture}[scale=3,
    white/.style={circle, draw, fill=white, inner sep=0pt, minimum size=6pt},
    black/.style={circle, draw, fill=black, inner sep=0pt, minimum size=6pt},
    label/.style={font=\small, midway, sloped, above},
    coord/.style={font=\scriptsize, anchor=south east},
    coordne/.style={font=\scriptsize, anchor=north east},
    coordnw/.style={font=\scriptsize, anchor=north west},
    coordsw/.style={font=\scriptsize, anchor=south west}
]
\coordinate (A) at (0,0,0);   
\coordinate (B) at (1,0,0);   
\coordinate (C) at (1,1,0);   
\coordinate (D) at (0,1,0);   

\coordinate (AA) at (0.1,0.1,0);   
\coordinate (AB) at (0.9,0.1,0);   
\coordinate (AC) at (0.9,0.9,0);   
\coordinate (AD) at (0.1,0.9,0); 
\node[coordnw] at (B) {$$};
\node[coordsw] at (C) {$$};
\node[coord] at (D) {$$};
\draw[red, midarrow, line width=1.5pt] (A) -- (B) node[midway, below] {$i$};
\draw[red, midarrow, line width=1.5pt] (B) -- (C) node[midway, right] {$j$};
\draw[red, midarrow, line width=1.5pt] (C) -- (D) node[midway, above] {$k$};
\draw[red, midarrow, line width=1.5pt] (D) -- (A) node[midway, left] {$\l$};
\node[white] at (A) {};
\node[black] at (B) {};
\node[white] at (C) {};
\node[black] at (D) {};
\end{tikzpicture}}}
\eea

For the wave function, this means that complex conjugation shall act as
\bea
\(\psi_i{}^j{}_k{}^{\l}\)^* &=& \t\psi_{\l}{}^k{}_j{}^i
\eea
where we use tilde to emphasise that this can be some other wave function, which can be different from the original wave function. For a  string that passes through the plaquette in one of its four vertices, say at the vertex with the coordinate $x$, we will define
\bea
\psi(x)^* &=& \t\psi(x)
\eea
Let us next assume that we have the $0\to 4$ move under which these wave functions transform as
\bea
\psi(x) &\to & \psi_i{}^j{}_k{}^{\l} = U_i{}^j{}_k{}^{\l} \psi(x)\cr
\t\psi(x) &\to & \t\psi_{\l}{}^k{}_j{}^i = \t{U}_{\l}{}^k{}_j{}^i \t\psi(x)
\eea
In the second line we use $\t{U}$ since the final string has the opposite orientation. By taking complex conjugation of the first relation, we discover a relation between $U$ and $\t{U}$. Namely, they are related by complex conjugation as
\bea
\t{U}_{\l}{}^k{}_j{}^i &=& \(U_i{}^j{}_k{}^{\l}\)^*
\eea
We will now study the $2\to 2$ move. We start with a string that goes along two edges of the plaquette, which are labeled by the color indices $i$ and $j$. What the string does as it continues outside the plaquette to form a closed string in the end, is of no concern to us here. We will just assume that the string remains static outside the plaquette under consideration. We then move the string to the other side of the plaquette by a $2\to 2$ move. It then stretches over the edges that carry the color indices $\l$ and $k$. Then we move the string back to its original position, but we will put primes on the color indices at this stage, so they will read $i'$ and $j'$. The sequence of $2\to 2$ movements is pictured below,
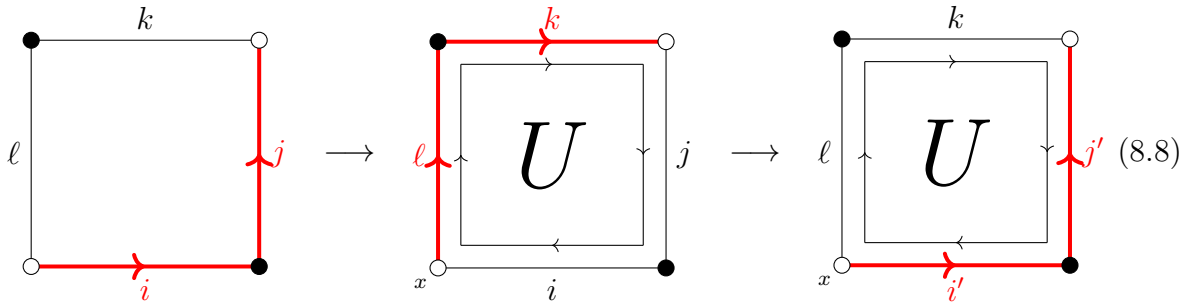
\bea
\begin{array}{ccccc}
\vcenter{\hbox{\begin{tikzpicture}[scale=3,
    white/.style={circle, draw, fill=white, inner sep=0pt, minimum size=6pt},
    black/.style={circle, draw, fill=black, inner sep=0pt, minimum size=6pt},
    label/.style={font=\small, midway, sloped, above},
    coord/.style={font=\scriptsize, anchor=south east},
    coordne/.style={font=\scriptsize, anchor=north east},
    coordnw/.style={font=\scriptsize, anchor=north west},
    coordsw/.style={font=\scriptsize, anchor=south west}
]
\coordinate (A) at (0,0);   
\coordinate (B) at (1,0);   
\coordinate (C) at (1,1);   
\coordinate (D) at (0,1);   
\node[coordne] at (A) {$$};
\node[coordnw] at (B) {$$};
\node[coordsw] at (C) {$$};
\node[coord] at (D) {$$};
\draw[red, midarrow, line width=1.5pt] (A) -- (B) node[midway, below] {$i$};
\draw[red, midarrow, line width=1.5pt] (B) -- (C) node[midway, right] {$j$};
\draw (C) -- (D) node[midway, above] {$k$};
\draw (D) -- (A) node[midway, left] {$\l$};
\node[white] at (A) {};
\node[black] at (B) {};
\node[white] at (C) {};
\node[black] at (D) {};
\end{tikzpicture}}}
& \longrightarrow &
\vcenter{\hbox{\begin{tikzpicture}[scale=3,
    white/.style={circle, draw, fill=white, inner sep=0pt, minimum size=6pt},
    black/.style={circle, draw, fill=black, inner sep=0pt, minimum size=6pt},
    label/.style={font=\small, midway, sloped, above},
    coord/.style={font=\scriptsize, anchor=south east},
    coordne/.style={font=\scriptsize, anchor=north east},
    coordnw/.style={font=\scriptsize, anchor=north west},
    coordsw/.style={font=\scriptsize, anchor=south west}
]
\coordinate (A) at (0,0);   
\coordinate (B) at (1,0);   
\coordinate (C) at (1,1);   
\coordinate (D) at (0,1);   
\node[coordne] at (A) {$x$};
\node[coordnw] at (B) {$$};
\node[coordsw] at (C) {$$};
\node[coord] at (D) {$$};
\draw (A) -- (B) node[midway, below] {$i$};
\draw (B) -- (C) node[midway, right] {$j$};
\draw[red, midarrow, line width=1.5pt] (A) -- (D) node[midway, left] {$\l$};
\draw[red, midarrow, line width=1.5pt] (D) -- (C) node[midway, above] {$k$};
\node[white] at (A) {};
\node[black] at (B) {};
\node[white] at (C) {};
\node[black] at (D) {};

\coordinate (AA) at (0.1,0.1,0);   
\coordinate (AB) at (0.9,0.1,0);   
\coordinate (AC) at (0.9,0.9,0);   
\coordinate (AD) at (0.1,0.9,0); 

\draw[midarrow] (AB) -- (AA) node[midway, above] {$$};
\draw[midarrow] (AC) -- (AB) node[midway, right] {$$};
\draw[midarrow] (AD) -- (AC) node[midway, above] {$$};
\draw[midarrow] (AA) -- (AD) node[midway, right] {$$};
\node[canvas is xy plane at z=0, transform shape] at (0.5,0.5,0) {$U$};
\end{tikzpicture}}}
& \longrightarrow &
\vcenter{\hbox{\begin{tikzpicture}[scale=3,
    white/.style={circle, draw, fill=white, inner sep=0pt, minimum size=6pt},
    black/.style={circle, draw, fill=black, inner sep=0pt, minimum size=6pt},
    label/.style={font=\small, midway, sloped, above},
    coord/.style={font=\scriptsize, anchor=south east},
    coordne/.style={font=\scriptsize, anchor=north east},
    coordnw/.style={font=\scriptsize, anchor=north west},
    coordsw/.style={font=\scriptsize, anchor=south west}
]
\coordinate (A) at (0,0);   
\coordinate (B) at (1,0);   
\coordinate (C) at (1,1);   
\coordinate (D) at (0,1);   
\node[coordne] at (A) {$x$};
\node[coordnw] at (B) {$$};
\node[coordsw] at (C) {$$};
\node[coord] at (D) {$$};
\draw[red, midarrow, line width=1.5pt] (A) -- (B) node[midway, below] {$i'$};
\draw[red, midarrow, line width=1.5pt] (B) -- (C) node[midway, right] {$j'$};
\draw (C) -- (D) node[midway, above] {$k$};
\draw (D) -- (A) node[midway, left] {$\l$};
\node[white] at (A) {};
\node[black] at (B) {};
\node[white] at (C) {};
\node[black] at (D) {};

\coordinate (AA) at (0.1,0.1,0);   
\coordinate (AB) at (0.9,0.1,0);   
\coordinate (AC) at (0.9,0.9,0);   
\coordinate (AD) at (0.1,0.9,0); 

\draw[midarrow] (AB) -- (AA) node[midway, above] {$$};
\draw[midarrow] (AC) -- (AB) node[midway, right] {$$};
\draw[midarrow] (AD) -- (AC) node[midway, above] {$$};
\draw[midarrow] (AA) -- (AD) node[midway, right] {$$};

\node[canvas is xy plane at z=0, transform shape] at (0.5,0.5,0) {$U$};
\end{tikzpicture}}}
\end{array}
\eea
We have drawn the orientation of the surface along with the letter $U$ to emphasise that $U$ is defined with respect to this particular orientation. Since the string goes parallel with the orientation of the surface under the first move but antiparallel with the orientation of the surface under the second move, the wave function will transform differently under the two moves according to 
\bea
\psi{}_j{}^i &\rightarrow & \psi{}_k{}^{\l} = U{}_k{}^{\l}{}_i{}^j \psi{}_j{}^i\cr
&\rightarrow & \psi{}_{j'}{}^{i'} = \t{U}{}_{j'}{}^{i'}{}_{\l}{}^k \psi{}_k{}^{\l} = \t{U}{}_{j'}{}^{i'}{}_{\l}{}^k U{}_k{}^{\l}{}_i{}^j \psi{}_j{}^i
\eea
As the string moves over the plaquette and then back, its surface holonomy shall cancel as it comes back. This is the two-dimensional version of the zig-zag symmetry of a backtracking Wilson line in a Yang-Mills gauge theory. As this shall be true for an arbitrary initial wave function, we are led to the requirement that
\bea
\t{U}{}_{j'}{}^{i'}{}_{\l}{}^k U{}_k{}^{\l}{}_i{}^j &=& \delta^{i'}_i \delta_{j'}^j
\eea
If we now recall the complex conjugation rule for the plaquette holonomy, then we can also write this condition as a unitary compactness condition,
\bea
\(U{}_k{}^{\l}{}_{i'}{}^{j'}\)^* U{}_k{}^{\l}{}_i{}^j &=& \delta^{i'}_i \delta_{j'}^j
\eea
We have introduced color indices with primes. This could seem to contradict our view that color indices serve as locations of edges in the lattice. However, we may still use indices as geometric locations of edges while attaching primes to distinguish their different values at that edge. So if we need to specify the edge $i$ using two different strings, then we will denote these indices as $i$ and $i'$ where both refer to the same edge in the lattice but where we allow for them to take two different index values, within the range $1,...,N$.

\section{The $N^2$ degrees of freedom in $U_{\l}{}^k{}_j{}^i$}
The universal plaquette operator is represented as a tensor $U_{\l}{}^k{}_j{}^i$. If we just consider this as an unconstrained tensor, then it will have $N^4$ complex components. There is another version $U^k{}_j{}^i{}_{\l}$ with another set of $N^4$ complex components. The conditions
\bea
U_{\l}{}^k{}_j{}^i &=& U^k{}_j{}^i{}_{\l}
\eea
reduce this totality set of $2N^4$ complex components to $N^4$ complex components $\{U_{\l}{}^k{}_j{}^i\}$. The reality conditions
\bea
\(U_{\l}{}^k{}_j{}^i\)^* &=& \t{U}_i{}^j{}_k{}^{\l}
\eea
due to the tilde operator being independent, thus define another set of components and are thus still leaving us with $N^4$ complex components. The symmetric conditions
\bea
U_{\l}{}^k{}_j{}^i - U_j{}^i{}_{\l}{}^k &=& 0
\eea
comprise $N^2 (N^2 - 1)/2$ conditions which leave us with $N^2(N^2+1)/2$ complex components. The unitarity conditions
\bea
\(U_{\l}{}^k{}_j{}^i\)^* U_{\l}{}^k{}_{j'}{}^{i'} &=& \delta_i^{i'} \delta_{j'}^j
\eea
comprise $N^4$ real conditions leaving us with
\bea
2 \frac{N^2\(N^2+1\)}{2} - N^4 &=& N^2
\eea
real components in the universal plaquette operator at the end of the day.

\section{The unit plaquette operator}
In this section we will discuss the unit plaquette operator that we will denote as $\I_i{}^j{}_k{}^{\l}$. Obviously it shall be cyclically symmetric and unitary,
\bea
\I_i{}^j{}_k{}^{\l} &=& \I^j{}_k{}^{\l}{}_i\cr
\(\I_i{}^j{}_k{}^{\l}\)^* \I_i{}^j{}_a{}^b &=& \delta_a^k \delta^b_{\l}
\eea
just as any universal plaquette operator. One may then try some Kronecker combinations and check whether they can satisfy both these conditions. The obvious guess would be to take the $U(N)$ invariant combination $\delta_i^j \delta_k^{\l} + \delta_i^{\l} \delta^j_k$ which is cyclically symmetric. But this does not satisfy the unitary constraint. Instead, we find that a solution that satisfies both the cyclicity and unitarity constraints is of the form
\bea
\I_i{}^j{}_k{}^{\l} &=& S_{ik} S^{j\l}
\eea
where $S_{ij}$ is any symmetric and unitary matrix,
\bea
S_{ij} &=& S_{ji}\cr
\(S_{ij}\)^* S_{jk} &=& \delta^i_k
\eea
We define 
\bea
S^{ij} &:= & \(S_{ij}\)^*
\eea 
We claim that this is the best candidate for a unit plaquette operator that we can possibly find. It is not uniquely defined, it can not be since there is no invariant tensor $S_{ij}$ in the $U(N)$ gauge group. Thus our claim must be that any solution $S_{ij}$ works equally well for defining the unit. 

We will now show in what sense this plaquette operator can act like a unit. To this end, we will lift a closed string that goes around a plaquette in the basement floor up to the first floor, and then up to the second floor. Why we need to lift the string all the way up to the second floor is related to the bipartite structure of the lattice. The process of lifting the string two floors up is shown below,

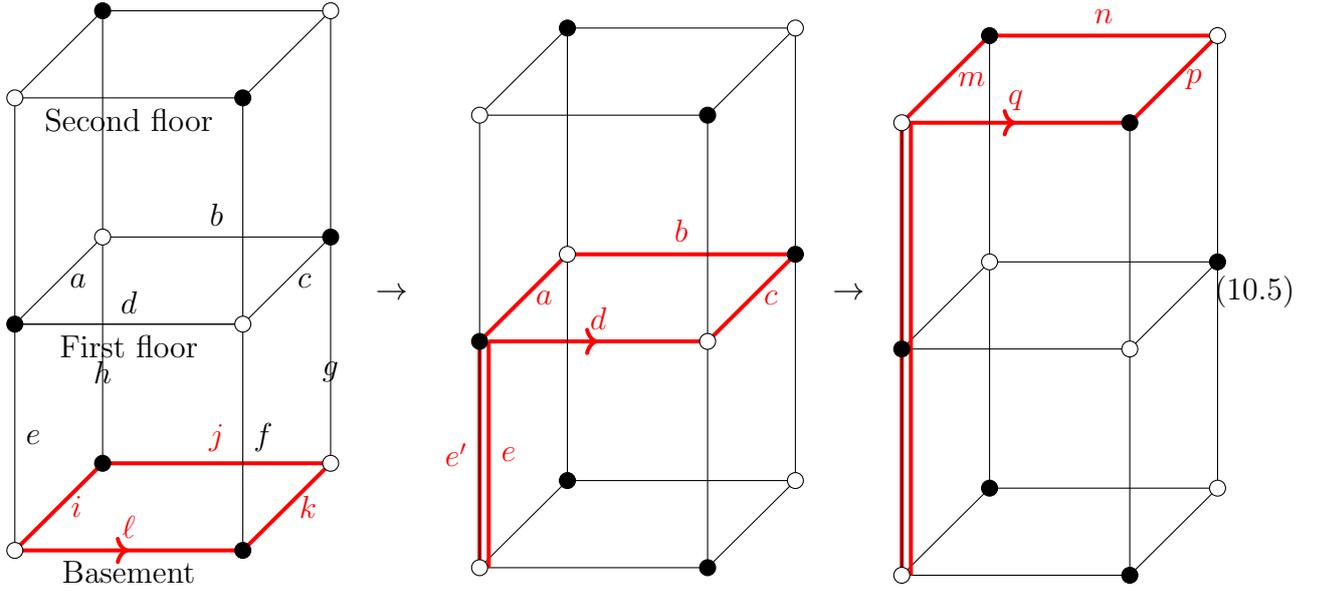
\bea
\begin{array}{ccccc}
\vcenter{\hbox{\begin{tikzpicture}[scale=3,
    white/.style={circle, draw, fill=white, inner sep=0pt, minimum size=6pt},
    black/.style={circle, draw, fill=black, inner sep=0pt, minimum size=6pt},
    label/.style={font=\small, midway, sloped, above},
    coord/.style={font=\scriptsize, anchor=south east}
]
\coordinate (A) at (0,0,0);   
\coordinate (B) at (1,0,0);   
\coordinate (C) at (1,0,1);   
\coordinate (D) at (0,0,1);   

\coordinate (E) at (0,1,0);   
\coordinate (F) at (1,1,0);   
\coordinate (G) at (1,1,1);   
\coordinate (H) at (0,1,1);   

\coordinate (I) at (0,2,0);   
\coordinate (J) at (1,2,0);   
\coordinate (K) at (1,2,1);   
\coordinate (L) at (0,2,1);   
\draw (D) -- (C) node[midway, below] {Basement};
\draw[midarrow, red, line width=1.5pt] (D) -- (C) node[midway, above] {$\l$}; 
\draw[red, line width=1.5pt] (C) -- (B) node[midway, right] {$k$};
\draw[red, line width=1.5pt] (B) -- (A) node[midway, above] {$j$};
\draw[red, line width=1.5pt] (A) -- (D) node[midway, right] {$i$};

\draw (H) -- (G) node[midway, below] {First floor};
\draw (H) -- (G) node[midway, above] {$d$};
\draw (G) -- (F) node[midway, right] {$c$};
\draw (F) -- (E) node[midway, above] {$b$};
\draw (E) -- (H) node[midway, right] {$a$};

\draw (L) -- (K) node[midway, below] {Second floor};
\draw (K) -- (J) node[midway, above] {$$};
\draw (J) -- (I) node[midway, above] {$$};
\draw (I) -- (L) node[midway, above] {$$};

\draw (A) -- (E) node[midway, right, below] {$h$};
\draw (F) -- (B) node[midway, below] {$g$};
\draw (C) -- (G) node[midway, right] {$f$};
\draw (H) -- (D) node[midway, right] {$e$};
\draw (E) -- (I) node[midway, above] {$$};
\draw (F) -- (J) node[midway, above] {$$};
\draw (G) -- (K) node[midway, above] {$$};
\draw (H) -- (L) node[midway, above] {$$};

\node[black] at (A) {};
\node[white] at (B) {};
\node[black] at (C) {};
\node[white] at (D) {};
\node[white] at (E) {};
\node[black] at (F) {};
\node[white] at (G) {};
\node[black] at (H) {};
\node[black] at (I) {};
\node[white] at (J) {};
\node[black] at (K) {};
\node[white] at (L) {};
\end{tikzpicture}}}
&\rightarrow &
\vcenter{\hbox{\begin{tikzpicture}[scale=3,
    white/.style={circle, draw, fill=white, inner sep=0pt, minimum size=6pt},
    black/.style={circle, draw, fill=black, inner sep=0pt, minimum size=6pt},
    label/.style={font=\small, midway, sloped, above},
    coord/.style={font=\scriptsize, anchor=south east}
]
\coordinate (A) at (0,0,0);   
\coordinate (B) at (1,0,0);   
\coordinate (C) at (1,0,1);   
\coordinate (D) at (0,0,1);   

\coordinate (E) at (0,1,0);   
\coordinate (F) at (1,1,0);   
\coordinate (G) at (1,1,1);   
\coordinate (H) at (0,1,1);   

\coordinate (Ha) at (0.04,1,1);
\coordinate (Da) at (0.04,0,1);  

\coordinate (I) at (0,2,0);   
\coordinate (J) at (1,2,0);   
\coordinate (K) at (1,2,1);   
\coordinate (L) at (0,2,1);   
\draw (D) -- (C) node[midway, above] {$$};
\draw (C) -- (B) node[midway, right] {$$};
\draw (B) -- (A) node[midway, above] {$$};
\draw (A) -- (D) node[midway, right] {$$};

\draw[red, line width=1.5pt] (Da) -- (Ha) node[midway, right] {$e$};
\draw[midarrow, red, line width=1.5pt] (Ha) -- (G) node[midway, above] {$d$};
\draw[red, line width=1.5pt] (G) -- (F) node[midway, right] {$c$};
\draw[red, line width=1.5pt] (F) -- (E) node[midway, above] {$b$};
\draw[red, line width=1.5pt] (E) -- (H) node[midway, right] {$a$};
\draw[red, line width=1.5pt] (H) -- (D) node[midway, left] {$e'$};

\draw (L) -- (K) node[midway, above] {$$};
\draw (K) -- (J) node[midway, above] {$$};
\draw (J) -- (I) node[midway, above] {$$};
\draw (I) -- (L) node[midway, above] {$$};

\draw (A) -- (E) node[midway, above] {$$};
\draw (F) -- (B) node[midway, above] {$$};
\draw (C) -- (G) node[midway, above] {$$};
\draw (H) -- (D) node[midway, above] {$$};
\draw (E) -- (I) node[midway, above] {$$};
\draw (F) -- (J) node[midway, above] {$$};
\draw (G) -- (K) node[midway, above] {$$};
\draw (H) -- (L) node[midway, above] {$$};
\node[black] at (A) {};
\node[white] at (B) {};
\node[black] at (C) {};
\node[white] at (D) {};
\node[white] at (E) {};
\node[black] at (F) {};
\node[white] at (G) {};
\node[black] at (H) {};
\node[black] at (I) {};
\node[white] at (J) {};
\node[black] at (K) {};
\node[white] at (L) {};
\end{tikzpicture}}}
&\rightarrow &
\vcenter{\hbox{\begin{tikzpicture}[scale=3,
    white/.style={circle, draw, fill=white, inner sep=0pt, minimum size=6pt},
    black/.style={circle, draw, fill=black, inner sep=0pt, minimum size=6pt},
    label/.style={font=\small, midway, sloped, above},
    coord/.style={font=\scriptsize, anchor=south east}
]
\coordinate (A) at (0,0,0);   
\coordinate (B) at (1,0,0);   
\coordinate (C) at (1,0,1);   
\coordinate (D) at (0,0,1);   

\coordinate (Da) at (0.04,0,1);   
\coordinate (La) at (0.04,2,1);   

\coordinate (E) at (0,1,0);   
\coordinate (F) at (1,1,0);   
\coordinate (G) at (1,1,1);   
\coordinate (H) at (0,1,1);   

\coordinate (I) at (0,2,0);   
\coordinate (J) at (1,2,0);   
\coordinate (K) at (1,2,1);   
\coordinate (L) at (0,2,1);   
\draw (D) -- (C) node[midway, above] {$$};
\draw (C) -- (B) node[midway, right] {$$};
\draw (B) -- (A) node[midway, above] {$$};
\draw (A) -- (D) node[midway, right] {$$};

\draw (H) -- (G) node[midway, above] {$$};
\draw (G) -- (F) node[midway, right] {$$};
\draw (F) -- (E) node[midway, above] {$$};
\draw (E) -- (H) node[midway, right] {$$};

\draw[red, line width=1.5pt] (Da) -- (La) node[midway, above] {$$};
\draw[midarrow, red, line width=1.5pt] (L) -- (K) node[midway, above] {$q$};
\draw[red, line width=1.5pt] (K) -- (J) node[midway, right] {$p$};
\draw[red, line width=1.5pt] (J) -- (I) node[midway, above] {$n$};
\draw[red, line width=1.5pt]  (I) -- (L) node[midway, right] {$m$};
\draw[red, line width=1.5pt]  (L) -- (D) node[midway, right] {$$};

\draw (A) -- (E) node[midway, above] {$$};
\draw (F) -- (B) node[midway, above] {$$};
\draw (C) -- (G) node[midway, above] {$$};
\draw (H) -- (D) node[midway, above] {$$};
\draw (E) -- (I) node[midway, above] {$$};
\draw (F) -- (J) node[midway, above] {$$};
\draw (G) -- (K) node[midway, above] {$$};
\draw (H) -- (L) node[midway, above] {$$};
\node[black] at (A) {};
\node[white] at (B) {};
\node[black] at (C) {};
\node[white] at (D) {};
\node[white] at (E) {};
\node[black] at (F) {};
\node[white] at (G) {};
\node[black] at (H) {};
\node[black] at (I) {};
\node[white] at (J) {};
\node[black] at (K) {};
\node[white] at (L) {};
\end{tikzpicture}}}
\end{array}
\eea

We may think on the process as follows. We start by the first and in some sense the most heavy lift\footnote{Stated more precisely, as one segment become three segments, since the string has everywhere the same tension set by a Higgs scalar vacuum expectation value, the total mass of the string increases. The energy comes from tensionless strings that create a string-antistring pair forming a K-spike. The K-spike costs the energy to create (the mass of two new string segments), while the subsequent $2\to 2$ move can be performed with almost no energy input if we move the string adiabatically. Specifically, the $2\to 2$ can not change the mass of the string because it does not change the length nor the tension of the string.} where we lift the string segment $\l$ by a $1\to 3$ move. This is implemented by acting with 
\bea
\frac{1}{\sqrt{N}} \I_{\l}{}^f{}_d{}^e
\eea
After that has been done, the subsequent lifts are in some sense not so heavy as they only require us to lift two segments using the $2\to 2$ moves that we use to lift up the string segments $k$, $j$ and $i$ in turn by acting with 
\bea
 \I^i{}_{e'}{}^a{}_h \I_j{}^h{}_b{}^g \I^k{}_g{}^c{}_f
\eea
Now when we make the last lift job, that is by acting with $ \I^i{}_{e}{}^a{}_h$, which is lifting up the segment $i$, we discover that the edge $e$ is already occupied by the string. Namely, it is occupied by the string that was lifted up with the initial and heaviest $1\to 3$ move. That means we shall put a prime for the color index $e$ for this final $2\to 2$ move as  $\I^i{}_{e'}{}^a{}_h$. The wave function of the string under this lift will transform as
\bea
(\psi_1)_{e'}{}^a{}_b{}^c{}_d{}^e &=& \frac{1}{\sqrt{N}} \I^i{}_{e'}{}^a{}_h \I_j{}^h{}_b{}^g \I^k{}_g{}^c{}_f \I_{\l}{}^f{}_d{}^e (\psi_0)_i{}^j{}_k{}^{\l} 
\eea
If we plug in our expression for the unit, we get
\bea
(\psi_1)_{e'}{}^a{}_b{}^c{}_d{}^e &=& \frac{1}{\sqrt{N}} \delta^e_{e'} S^{ai} S_{bj} S^{ck} S_{d\l} (\psi_0)_i{}^j{}_k{}^{\l}
\eea
The wave function on the left-hand side has two segments on the same edge. In order to be able to proceed, let us assume that a K-spike is obtained there,
\bea
(\psi_1)_{e'}{}^a{}_b{}^c{}_d{}^e &=& \frac{1}{\sqrt{N}} \delta_{e'}^e (\psi_1)^a{}_b{}^c{}_d
\eea
Then we find that, after an R-move,
\bea
(\psi_1)^a{}_b{}^c{}_d &=& S^{ai} S_{bj} S^{ck} S_{d\l} (\psi_0)_i{}^j{}_k{}^{\l}
\eea
We can now iterate the process one more time and lift the string up to the second floor. Repeating the full procedure once more, that is, first the heavy $1\to 3$ move, followed by three $2\to 2$ moves and finally an R-move of an assumed K-spike, the wave function transforms as
\bea
(\psi_2)_m{}^n{}_p{}^q &=& S_{ma} S^{nb} S_{pc} S^{qd} (\psi_1)^a{}_b{}^c{}_d\cr
&=& \(S_{ma} S^{nb} S_{pc} S^{qd}\) \(S^{ai} S_{bj} S^{ck} S_{d\l}\) (\psi_0)_i{}^j{}_k{}^{\l}\cr
&=& (\psi_0)_m{}^n{}_p{}^q
\eea
We now see that we indeed moved to the second floor without any change, thus justifying the name of 'unit' for our plaquette operator $\I_i{}^j{}_k{}^{\l}$. 

Admittedly, we had to calibrate at one step in order to generate a K-spike. That means that this operator acts like a unit in situations where we can carry out the computation. That is of course as much as we should ask for. It would be rather pointless to ask if this operator still would act like a unit in a regime where we can not compute anything. In that regime we need a completely different computational framework and we expect everything to change to such a degree that nothing of what we do here can be carried over to that regime where we may have some sort of tensionless strings. 

Let us briefly comment on other possible uses of the unit. The most basic move one may think of is that of stretching out a right-angled two-spike string so that it goes around one plaquette. 
\bea
\begin{array}{cccc}
\vcenter{\hbox{\begin{tikzpicture}[scale=2,
    white/.style={circle, draw, fill=white, inner sep=0pt, minimum size=6pt},
    black/.style={circle, draw, fill=black, inner sep=0pt, minimum size=6pt},
    label/.style={font=\small, midway, sloped, above},
    coord/.style={font=\scriptsize, anchor=south east},
    coordne/.style={font=\scriptsize, anchor=north east},
    coordnw/.style={font=\scriptsize, anchor=north west},
    coordsw/.style={font=\scriptsize, anchor=south west}
]
\coordinate (A) at (0,0);   
\coordinate (Aa) at (0.04,0.04); 
\coordinate (Ba) at (1,0.04);   
\coordinate (B) at (1,0);   
\coordinate (C) at (1,1);   
\coordinate (Da) at (0.04,1);   
\coordinate (D) at (0,1); 
\node[coordne] at (A) {$x$};
\node[coordnw] at (B) {$$};
\node[coordsw] at (C) {$$};
\node[coord] at (D) {$$};
\draw[red, midarrow, line width=1.5pt] (Da) -- (Aa) node[midway, right] {$k'$};
\draw[red, midarrow, line width=1.5pt] (Aa) -- (Ba) node[midway, above] {$j'$};
\draw[red, midarrow, line width=1.5pt] (B) -- (A) node[midway, below] {$i$};
\draw[red, midarrow, line width=1.5pt] (A) -- (D) node[midway, left] {$\l$};
\draw (C) -- (B) node[midway, right] {$$};
\draw (C) -- (D) node[midway, above] {$$};
\node[white] at (A) {};
\node[black] at (B) {};
\node[white] at (C) {};
\node[black] at (D) {};
\end{tikzpicture}}}
&\longrightarrow &
\vcenter{\hbox{\begin{tikzpicture}[scale=2,
    white/.style={circle, draw, fill=white, inner sep=0pt, minimum size=6pt},
    black/.style={circle, draw, fill=black, inner sep=0pt, minimum size=6pt},
    label/.style={font=\small, midway, sloped, above},
    coord/.style={font=\scriptsize, anchor=south east},
    coordne/.style={font=\scriptsize, anchor=north east},
    coordnw/.style={font=\scriptsize, anchor=north west},
    coordsw/.style={font=\scriptsize, anchor=south west}
]
\coordinate (A) at (0,0);   
\coordinate (B) at (1,0);   
\coordinate (C) at (1,1);   
\coordinate (D) at (0,1);   
\node[coordne] at (A) {$x$};
\node[coordnw] at (B) {$$};
\node[coordsw] at (C) {$$};
\node[coord] at (D) {$$};
\draw[red, midarrow, line width=1.5pt] (B) -- (A) node[midway, below] {$i$};
\draw[red, midarrow, line width=1.5pt] (C) -- (B) node[midway, right] {$j$};
\draw[red, midarrow, line width=1.5pt] (D) -- (C) node[midway, above] {$k$};
\draw[red, midarrow, line width=1.5pt] (A) -- (D) node[midway, left] {$\l$};
\node[white] at (A) {};
\node[black] at (B) {};
\node[white] at (C) {};
\node[black] at (D) {};
\end{tikzpicture}}}
& \begin{array}{cc}
&\psi{}_i{}^{j'}{}_{k'}{}^{\l} = \frac{1}{N} \delta_i^{j'} \delta^{\l}_{k'} \psi(x) \cr
&\rightarrow \psi_i{}^j{}_k{}^{\l} = \I^j{}_k{}^{k'}{}_{j'} \psi{}_i{}^{j'}{}_{k'}{}^{\l} \cr
&= \frac{1}{N} \I^j{}_k{}^{\l}{}_i \psi(x)
\end{array}
\end{array}
\eea
The spike-string wave function is given by
\bea
\psi{}_i{}^{j'}{}_{k'}{}^{\l} &=& \frac{1}{N} \delta_i^{j'} \delta^{\l}_{k'} \psi(x) 
\eea
We then stretch out the segments $j'$ and $k'$ by the unit, which transforms the wave function as
\bea
\psi{}_i{}^{j'}{}_{k'}{}^{\l} \to \psi_i{}^j{}_k{}^{\l} = \frac{1}{N}\I^j{}_k{}^{\l}{}_i \psi(x) = \frac{1}{N} S_{\l j} S^{ki} \psi(x)
\eea
The transformed wave function $\frac{1}{N} S_{\l j} S^{ki} \psi(x)$ is not the same as the spike-string wave function $\frac{1}{N} \delta_i^{j'} \delta^{\l}_{k'} \psi(x)$ that we started with. But that could not be expected because the two string configurations are completely different from one another. The unit can be made to act like a unit only when it acts on things that we can equate. Specifically, it can be made to act like a unit when the initial and the final string configurations are exactly the same when embedded in the lattice and indistinguishable from each other up to translation and rotation.

Now having this unit, we may expand around it as
\bea
U_i{}^j{}_k{}^{\l} &=& \I_i{}^j{}_k{}^{\l} + i e T_i{}^j{}_k{}^{\l} + \O(e^2)
\eea
and in this way study the set-up perturbatively in the dimensionless parameter $e$. This is not possible for a selfdual string for which $e$ is fixed at its selfdual value. But if our string is not magnetically charged, then $e$ can be choosen arbitrarily small and it can then serve as an expansion parameter. Unitarity implies that 
\bea
\(T_i{}^j{}_k{}^{\l}\)^* &=& T_{\l}{}^k{}_j{}^i
\eea
An interesting explicit realization that also satisfies the cyclic condition is given by  
\bea
U_i{}^j{}_k{}^{\l} &=& \I_i{}^j{}_k{}^{\l} + i e \(S_{ij} P^{k\l} + P_{ij} S^{k\l}\) + \O(e^2)
\eea
where $(P_{ij})^* = P^{ij}$ and $P_{ij} = P_{ji}$ but which is otherwise an  unconstrained matrix that captures $\O(N^2)$ degrees of freedom of the plaquette holonomy.

\section{The disallowed spike moves}
Let us consider a string that wraps around four edges of one plaquette, and then let us try to pull out a spike from one of its vertices $x$, such that the spike runs parallel with the edge $i$ in the plaquette. After that, we apply a $2\to 2$ move. These moves are pictured below,
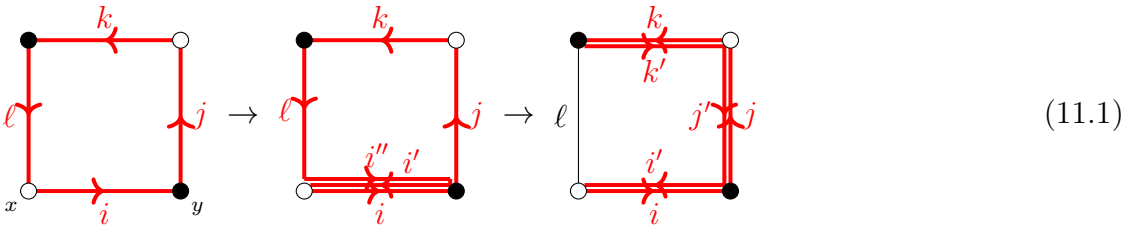
\bea
\vcenter{\hbox{\begin{tikzpicture}[scale=2,
    white/.style={circle, draw, fill=white, inner sep=0pt, minimum size=6pt},
    black/.style={circle, draw, fill=black, inner sep=0pt, minimum size=6pt},
    label/.style={font=\small, midway, sloped, above},
    coord/.style={font=\scriptsize, anchor=south east},
    coordne/.style={font=\scriptsize, anchor=north east},
    coordnw/.style={font=\scriptsize, anchor=north west},
    coordsw/.style={font=\scriptsize, anchor=south west}
]
\coordinate (A) at (0,0);  
\coordinate (B) at (1,0); 
\coordinate (C) at (1,1);
\coordinate (D) at (0,1); 
\coordinate (Aa) at (0.04,0.04); 
\coordinate (Ba) at (0.96,0.04);  
\coordinate (Ca) at (0.96,0.96);  
\coordinate (Da) at (0.04,0.96);  
\node[coordne] at (A) {$x$};
\node[coordnw] at (B) {$y$};
\node[coordsw] at (C) {$$};
\node[coord] at (D) {$$};
\draw[red, midarrow, line width=1.5pt] (A) -- (B) node[midway, below] {$i$};
\draw[red, midarrow, line width=1.5pt] (B) -- (C) node[midway, right] {$j$};
\draw[red, midarrow, line width=1.5pt] (C) -- (D) node[midway, above] {$k$};
\draw[red, midarrow, line width=1.5pt] (D) -- (A) node[midway, left] {$\l$};
\node[white] at (A) {};
\node[black] at (B) {};
\node[white] at (C) {};
\node[black] at (D) {};
\end{tikzpicture}}}
\rightarrow 
\vcenter{\hbox{\begin{tikzpicture}[scale=2,
    white/.style={circle, draw, fill=white, inner sep=0pt, minimum size=6pt},
    black/.style={circle, draw, fill=black, inner sep=0pt, minimum size=6pt},
    label/.style={font=\small, midway, sloped, above},
    coord/.style={font=\scriptsize, anchor=south east},
    coordne/.style={font=\scriptsize, anchor=north east},
    coordnw/.style={font=\scriptsize, anchor=north west},
    coordsw/.style={font=\scriptsize, anchor=south west}
]
\coordinate (A) at (0,0);  
\coordinate (B) at (1,0); 
\coordinate (C) at (1,1);
\coordinate (D) at (0,1); 
\coordinate (Aa) at (0.04,0.04); 
\coordinate (Ba) at (0.96,0.04);  
\coordinate (Ab) at (0.00,0.08); 
\coordinate (Bb) at (0.96,0.08); 
\coordinate (Ca) at (0.96,0.96);  
\coordinate (Da) at (0.04,0.96);  
\node[coordne] at (A) {$$};
\node[coordnw] at (B) {$$};
\node[coordsw] at (C) {$$};
\node[coord] at (D) {$$};
\draw[red, midarrow, line width=1.5pt] (A) -- (B) node[midway, below] {$i$};
\draw[red, midarrow, line width=1.5pt] (B) -- (C) node[midway, right] {$j$};
\draw[red, midarrow, line width=1.5pt] (C) -- (D) node[midway, above] {$k$};
\draw[red, midarrow, line width=1.5pt] (D) -- (Ab) node[midway, left] {$\l$};
\draw[red, midarrow, line width=1.5pt] (Ab) -- (Bb) node[midway, above] {$i''$};
\draw[red, line width=1.5pt] (Bb) -- (Ba) ;
\draw[red, midarrow, line width=1.5pt] (Ba) -- (Aa) node[midway, above] {$\qquad i'$};

\node[white] at (A) {};
\node[black] at (B) {};
\node[white] at (C) {};
\node[black] at (D) {};
\end{tikzpicture}}}
\rightarrow 
\vcenter{\hbox{\begin{tikzpicture}[scale=2,
    white/.style={circle, draw, fill=white, inner sep=0pt, minimum size=6pt},
    black/.style={circle, draw, fill=black, inner sep=0pt, minimum size=6pt},
    label/.style={font=\small, midway, sloped, above},
    coord/.style={font=\scriptsize, anchor=south east},
    coordne/.style={font=\scriptsize, anchor=north east},
    coordnw/.style={font=\scriptsize, anchor=north west},
    coordsw/.style={font=\scriptsize, anchor=south west}
]
\coordinate (A) at (0,0);  
\coordinate (B) at (1,0); 
\coordinate (C) at (1,1);
\coordinate (D) at (0,1); 
\coordinate (Aa) at (0.04,0.04); 
\coordinate (Ba) at (0.96,0.04);  
\coordinate (Ca) at (0.96,0.96);  
\coordinate (Da) at (0.04,0.96);  
\node[coordne] at (A) {$$};
\node[coordnw] at (B) {$$};
\node[coordsw] at (C) {$$};
\node[coord] at (D) {$$};
\draw[red, midarrow, line width=1.5pt] (A) -- (B) node[midway, below] {$i$};
\draw[red, midarrow, line width=1.5pt] (Ba) -- (Aa) node[midway, above] {$i'$};
\draw[red, midarrow, line width=1.5pt] (B) -- (C) node[midway, right] {$j$};
\draw[red, midarrow, line width=1.5pt] (Ca) -- (Ba) node[midway, left] {$j'$};
\draw[red, midarrow, line width=1.5pt] (C) -- (D) node[midway, above] {$k$};
\draw[red, midarrow, line width=1.5pt] (Da) -- (Ca) node[midway, below] {$k'$};
\draw (D) -- (A) node[midway, left] {$\l$};
\node[white] at (A) {};
\node[black] at (B) {};
\node[white] at (C) {};
\node[black] at (D) {};
\end{tikzpicture}}}\label{dis}
\eea
Under the first spike-move from $x$ the wave function transforms as
\bea
\psi{}_{\l}{}^k{}_j{}^i \rightarrow \frac{1}{\sqrt{N}} \delta^{i''}_{i'} \psi{}_{\l}{}^k{}_j{}^i
\eea
A K-spike is generated here since the surface holonomy that creates this spike has vanishing area by pulling out a spike from the vertex $x$. Hence this surface holonomy needs to be an $U(N)$ invariant and therefore the K-spike is the only possibility. However, we could get the same string configuration if instead we pull out a spike from $y$, in which case the wave function transforms as
\bea
\psi{}_{\l}{}^k{}_j{}^i \rightarrow \frac{1}{\sqrt{N}} \delta^{i}_{i'} \psi{}_{\l}{}^k{}_j{}^{i''}
\eea
By the uniqueness of the wave function, we can not have two different wave functions assigned to this same graph. We conclude that creating a K-spike on an edge that is already occupied by a string is disallowed. From this it follows that it is disallowed to have three spikes in one plaquette, as it was created from a disallowed string configuration in the second move in (\ref{dis}). However, there is no limit on how many spikes that a string can have as long as no more than two spikes are located on one plaquette. 

Below we draw a sequence of allowed moves,
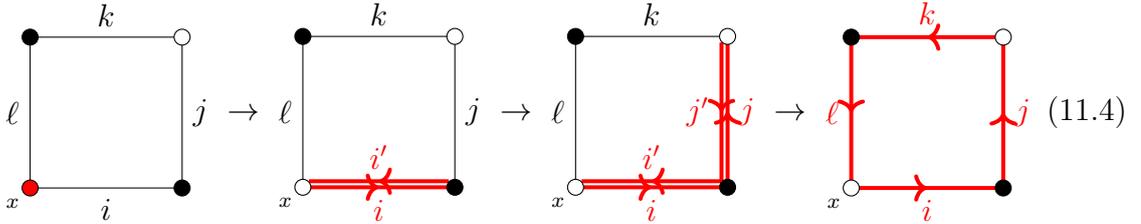
\bea
\vcenter{\hbox{\begin{tikzpicture}[scale=2,
    white/.style={circle, draw, fill=white, inner sep=0pt, minimum size=6pt},
    black/.style={circle, draw, fill=black, inner sep=0pt, minimum size=6pt},
    red/.style={circle, draw, fill=red, inner sep=0pt, minimum size=6pt},
    label/.style={font=\small, midway, sloped, above},
    coord/.style={font=\scriptsize, anchor=south east},
    coordne/.style={font=\scriptsize, anchor=north east},
    coordnw/.style={font=\scriptsize, anchor=north west},
    coordsw/.style={font=\scriptsize, anchor=south west}
]
\coordinate (A) at (0,0);  
\coordinate (B) at (1,0); 
\coordinate (C) at (1,1);
\coordinate (D) at (0,1); 
\coordinate (Aa) at (0.04,0.04); 
\coordinate (Ba) at (0.96,0.04);  
\coordinate (Ca) at (0.96,0.96);  
\coordinate (Da) at (0.04,0.96);  
\node[coordne] at (A) {$x$};
\node[coordnw] at (B) {$$};
\node[coordsw] at (C) {$$};
\node[coord] at (D) {$$};
\draw (A) -- (B) node[midway, below] {$i$};
\draw (B) -- (C) node[midway, right] {$j$};
\draw (C) -- (D) node[midway, above] {$k$};
\draw (D) -- (A) node[midway, left] {$\l$};
\node[red] at (A) {};
\node[black] at (B) {};
\node[white] at (C) {};
\node[black] at (D) {};
\end{tikzpicture}}}
\rightarrow 
\vcenter{\hbox{\begin{tikzpicture}[scale=2,
    white/.style={circle, draw, fill=white, inner sep=0pt, minimum size=6pt},
    black/.style={circle, draw, fill=black, inner sep=0pt, minimum size=6pt},
    label/.style={font=\small, midway, sloped, above},
    coord/.style={font=\scriptsize, anchor=south east},
    coordne/.style={font=\scriptsize, anchor=north east},
    coordnw/.style={font=\scriptsize, anchor=north west},
    coordsw/.style={font=\scriptsize, anchor=south west}
]
\coordinate (A) at (0,0);  
\coordinate (B) at (1,0); 
\coordinate (C) at (1,1);
\coordinate (D) at (0,1); 
\coordinate (Aa) at (0.04,0.04); 
\coordinate (Ba) at (0.96,0.04);  
\coordinate (Ca) at (0.96,0.96);  
\coordinate (Da) at (0.04,0.96);  
\node[coordne] at (A) {$x$};
\node[coordnw] at (B) {$$};
\node[coordsw] at (C) {$$};
\node[coord] at (D) {$$};
\draw[red, midarrow, line width=1.5pt] (A) -- (B) node[midway, below] {$i$};
\draw[red, midarrow, line width=1.5pt] (Ba) -- (Aa) node[midway, above] {$i'$};
\draw (B) -- (C) node[midway, right] {$j$};
\draw (C) -- (D) node[midway, above] {$k$};
\draw (D) -- (A) node[midway, left] {$\l$};
\node[white] at (A) {};
\node[black] at (B) {};
\node[white] at (C) {};
\node[black] at (D) {};
\end{tikzpicture}}}
\rightarrow 
\vcenter{\hbox{\begin{tikzpicture}[scale=2,
    white/.style={circle, draw, fill=white, inner sep=0pt, minimum size=6pt},
    black/.style={circle, draw, fill=black, inner sep=0pt, minimum size=6pt},
    label/.style={font=\small, midway, sloped, above},
    coord/.style={font=\scriptsize, anchor=south east},
    coordne/.style={font=\scriptsize, anchor=north east},
    coordnw/.style={font=\scriptsize, anchor=north west},
    coordsw/.style={font=\scriptsize, anchor=south west}
]
\coordinate (A) at (0,0);  
\coordinate (B) at (1,0); 
\coordinate (C) at (1,1);
\coordinate (D) at (0,1); 
\coordinate (Aa) at (0.04,0.04); 
\coordinate (Ba) at (0.96,0.04);  
\coordinate (Ca) at (0.96,0.96);  
\coordinate (Da) at (0.04,0.96);  
\node[coordne] at (A) {$x$};
\node[coordnw] at (B) {$$};
\node[coordsw] at (C) {$$};
\node[coord] at (D) {$$};
\draw[red, midarrow, line width=1.5pt] (A) -- (B) node[midway, below] {$i$};
\draw[red, midarrow, line width=1.5pt] (Ba) -- (Aa) node[midway, above] {$i'$};
\draw[red, midarrow, line width=1.5pt] (B) -- (C) node[midway, right] {$j$};
\draw[red, midarrow, line width=1.5pt] (Ca) -- (Ba) node[midway, left] {$j'$};
\draw (C) -- (D) node[midway, above] {$k$};
\draw (D) -- (A) node[midway, left] {$\l$};
\node[white] at (A) {};
\node[black] at (B) {};
\node[white] at (C) {};
\node[black] at (D) {};
\end{tikzpicture}}}
\rightarrow 
\vcenter{\hbox{\begin{tikzpicture}[scale=2,
    white/.style={circle, draw, fill=white, inner sep=0pt, minimum size=6pt},
    black/.style={circle, draw, fill=black, inner sep=0pt, minimum size=6pt},
    label/.style={font=\small, midway, sloped, above},
    coord/.style={font=\scriptsize, anchor=south east},
    coordne/.style={font=\scriptsize, anchor=north east},
    coordnw/.style={font=\scriptsize, anchor=north west},
    coordsw/.style={font=\scriptsize, anchor=south west}
]
\coordinate (A) at (0,0);  
\coordinate (B) at (1,0); 
\coordinate (C) at (1,1);
\coordinate (D) at (0,1); 
\coordinate (Aa) at (0.04,0.04); 
\coordinate (Ba) at (0.96,0.04);  
\coordinate (Ca) at (0.96,0.96);  
\coordinate (Da) at (0.04,0.96);  
\node[coordne] at (A) {$x$};
\node[coordnw] at (B) {$$};
\node[coordsw] at (C) {$$};
\node[coord] at (D) {$$};
\draw[red, midarrow, line width=1.5pt] (A) -- (B) node[midway, below] {$i$};
\draw[red, midarrow, line width=1.5pt] (B) -- (C) node[midway, right] {$j$};
\draw[red, midarrow, line width=1.5pt] (C) -- (D) node[midway, above] {$k$};
\draw[red, midarrow, line width=1.5pt] (D) -- (A) node[midway, left] {$\l$};
\node[white] at (A) {};
\node[black] at (B) {};
\node[white] at (C) {};
\node[black] at (D) {};
\end{tikzpicture}}}
\eea
Below is a corresponding sequence where the last move is disallowed,
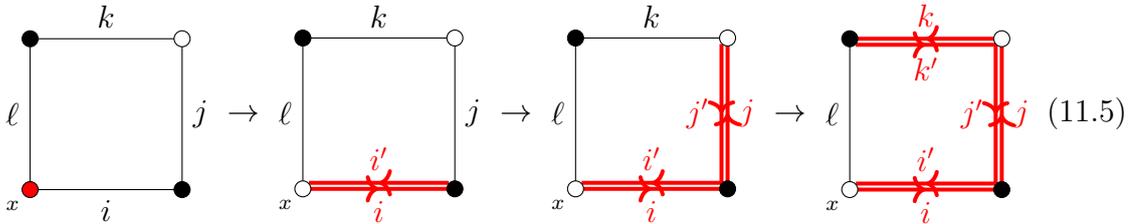
\bea
\vcenter{\hbox{\begin{tikzpicture}[scale=2,
    white/.style={circle, draw, fill=white, inner sep=0pt, minimum size=6pt},
    black/.style={circle, draw, fill=black, inner sep=0pt, minimum size=6pt},
    red/.style={circle, draw, fill=red, inner sep=0pt, minimum size=6pt},
    label/.style={font=\small, midway, sloped, above},
    coord/.style={font=\scriptsize, anchor=south east},
    coordne/.style={font=\scriptsize, anchor=north east},
    coordnw/.style={font=\scriptsize, anchor=north west},
    coordsw/.style={font=\scriptsize, anchor=south west}
]
\coordinate (A) at (0,0);  
\coordinate (B) at (1,0); 
\coordinate (C) at (1,1);
\coordinate (D) at (0,1); 
\coordinate (Aa) at (0.04,0.04); 
\coordinate (Ba) at (0.96,0.04);  
\coordinate (Ca) at (0.96,0.96);  
\coordinate (Da) at (0.04,0.96);  
\node[coordne] at (A) {$x$};
\node[coordnw] at (B) {$$};
\node[coordsw] at (C) {$$};
\node[coord] at (D) {$$};
\draw (A) -- (B) node[midway, below] {$i$};
\draw (B) -- (C) node[midway, right] {$j$};
\draw (C) -- (D) node[midway, above] {$k$};
\draw (D) -- (A) node[midway, left] {$\l$};
\node[red] at (A) {};
\node[black] at (B) {};
\node[white] at (C) {};
\node[black] at (D) {};
\end{tikzpicture}}}
\rightarrow 
\vcenter{\hbox{\begin{tikzpicture}[scale=2,
    white/.style={circle, draw, fill=white, inner sep=0pt, minimum size=6pt},
    black/.style={circle, draw, fill=black, inner sep=0pt, minimum size=6pt},
    label/.style={font=\small, midway, sloped, above},
    coord/.style={font=\scriptsize, anchor=south east},
    coordne/.style={font=\scriptsize, anchor=north east},
    coordnw/.style={font=\scriptsize, anchor=north west},
    coordsw/.style={font=\scriptsize, anchor=south west}
]
\coordinate (A) at (0,0);  
\coordinate (B) at (1,0); 
\coordinate (C) at (1,1);
\coordinate (D) at (0,1); 
\coordinate (Aa) at (0.04,0.04); 
\coordinate (Ba) at (0.96,0.04);  
\coordinate (Ca) at (0.96,0.96);  
\coordinate (Da) at (0.04,0.96);  
\node[coordne] at (A) {$x$};
\node[coordnw] at (B) {$$};
\node[coordsw] at (C) {$$};
\node[coord] at (D) {$$};
\draw[red, midarrow, line width=1.5pt] (A) -- (B) node[midway, below] {$i$};
\draw[red, midarrow, line width=1.5pt] (Ba) -- (Aa) node[midway, above] {$i'$};
\draw (B) -- (C) node[midway, right] {$j$};
\draw (C) -- (D) node[midway, above] {$k$};
\draw (D) -- (A) node[midway, left] {$\l$};
\node[white] at (A) {};
\node[black] at (B) {};
\node[white] at (C) {};
\node[black] at (D) {};
\end{tikzpicture}}}
\rightarrow 
\vcenter{\hbox{\begin{tikzpicture}[scale=2,
    white/.style={circle, draw, fill=white, inner sep=0pt, minimum size=6pt},
    black/.style={circle, draw, fill=black, inner sep=0pt, minimum size=6pt},
    label/.style={font=\small, midway, sloped, above},
    coord/.style={font=\scriptsize, anchor=south east},
    coordne/.style={font=\scriptsize, anchor=north east},
    coordnw/.style={font=\scriptsize, anchor=north west},
    coordsw/.style={font=\scriptsize, anchor=south west}
]
\coordinate (A) at (0,0);  
\coordinate (B) at (1,0); 
\coordinate (C) at (1,1);
\coordinate (D) at (0,1); 
\coordinate (Aa) at (0.04,0.04); 
\coordinate (Ba) at (0.96,0.04);  
\coordinate (Ca) at (0.96,0.96);  
\coordinate (Da) at (0.04,0.96);  
\node[coordne] at (A) {$x$};
\node[coordnw] at (B) {$$};
\node[coordsw] at (C) {$$};
\node[coord] at (D) {$$};
\draw[red, midarrow, line width=1.5pt] (A) -- (B) node[midway, below] {$i$};
\draw[red, midarrow, line width=1.5pt] (Ba) -- (Aa) node[midway, above] {$i'$};
\draw[red, midarrow, line width=1.5pt] (B) -- (C) node[midway, right] {$j$};
\draw[red, midarrow, line width=1.5pt] (Ca) -- (Ba) node[midway, left] {$j'$};
\draw (C) -- (D) node[midway, above] {$k$};
\draw (D) -- (A) node[midway, left] {$\l$};
\node[white] at (A) {};
\node[black] at (B) {};
\node[white] at (C) {};
\node[black] at (D) {};
\end{tikzpicture}}}
\rightarrow 
\vcenter{\hbox{\begin{tikzpicture}[scale=2,
    white/.style={circle, draw, fill=white, inner sep=0pt, minimum size=6pt},
    black/.style={circle, draw, fill=black, inner sep=0pt, minimum size=6pt},
    label/.style={font=\small, midway, sloped, above},
    coord/.style={font=\scriptsize, anchor=south east},
    coordne/.style={font=\scriptsize, anchor=north east},
    coordnw/.style={font=\scriptsize, anchor=north west},
    coordsw/.style={font=\scriptsize, anchor=south west}
]
\coordinate (A) at (0,0);  
\coordinate (B) at (1,0); 
\coordinate (C) at (1,1);
\coordinate (D) at (0,1); 
\coordinate (Aa) at (0.04,0.04); 
\coordinate (Ba) at (0.96,0.04);  
\coordinate (Ca) at (0.96,0.96);  
\coordinate (Da) at (0.04,0.96);  
\node[coordne] at (A) {$x$};
\node[coordnw] at (B) {$$};
\node[coordsw] at (C) {$$};
\node[coord] at (D) {$$};
\draw[red, midarrow, line width=1.5pt] (A) -- (B) node[midway, below] {$i$};
\draw[red, midarrow, line width=1.5pt] (Ba) -- (Aa) node[midway, above] {$i'$};
\draw[red, midarrow, line width=1.5pt] (B) -- (C) node[midway, right] {$j$};
\draw[red, midarrow, line width=1.5pt] (Ca) -- (Ba) node[midway, left] {$j'$};
\draw[red, midarrow, line width=1.5pt] (C) -- (D) node[midway, above] {$k$};
\draw[red, midarrow, line width=1.5pt] (Da) -- (Ca) node[midway, below] {$k'$};
\draw (D) -- (A) node[midway, left] {$\l$};
\node[white] at (A) {};
\node[black] at (B) {};
\node[white] at (C) {};
\node[black] at (D) {};
\end{tikzpicture}}}
\eea
We may get some intuition what might be going on here, physically, when one compares the following two moves,
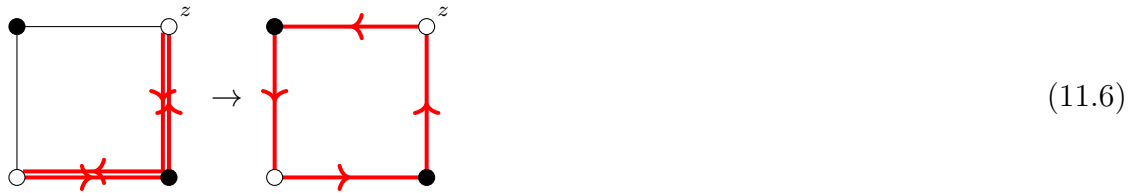
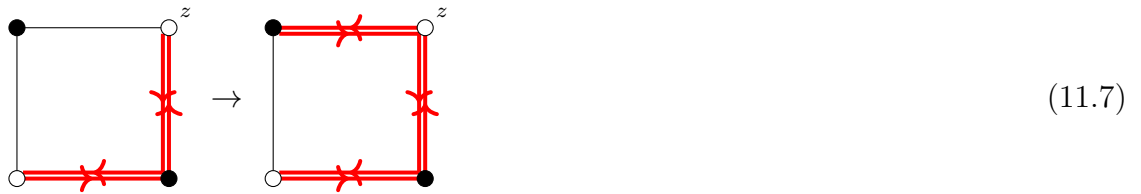
\bea
\vcenter{\hbox{\begin{tikzpicture}[scale=2,
    white/.style={circle, draw, fill=white, inner sep=0pt, minimum size=6pt},
    black/.style={circle, draw, fill=black, inner sep=0pt, minimum size=6pt},
    label/.style={font=\small, midway, sloped, above},
    coord/.style={font=\scriptsize, anchor=south east},
    coordne/.style={font=\scriptsize, anchor=north east},
    coordnw/.style={font=\scriptsize, anchor=north west},
    coordsw/.style={font=\scriptsize, anchor=south west}
]
\coordinate (A) at (0,0);  
\coordinate (B) at (1,0); 
\coordinate (C) at (1,1);
\coordinate (D) at (0,1); 
\coordinate (Aa) at (0.04,0.04); 
\coordinate (Ba) at (0.96,0.04);  
\coordinate (Ca) at (0.96,0.96);  
\coordinate (Da) at (0.04,0.96);  
\node[coordne] at (A) {$$};
\node[coordnw] at (B) {$$};
\node[coordsw] at (C) {$z$};
\node[coord] at (D) {$$};
\draw[red, midarrow, line width=1.5pt] (A) -- (B) node[midway, below] {$$};
\draw[red, midarrow, line width=1.5pt] (Ba) -- (Aa) node[midway, above] {$$};
\draw[red, midarrow, line width=1.5pt] (B) -- (C) node[midway, right] {$$};
\draw[red, midarrow, line width=1.5pt] (Ca) -- (Ba) node[midway, left] {$$};
\draw (C) -- (D) node[midway, above] {$$};
\draw (D) -- (A) node[midway, left] {$$};
\node[white] at (A) {};
\node[black] at (B) {};
\node[white] at (C) {};
\node[black] at (D) {};
\end{tikzpicture}}}
\rightarrow 
\vcenter{\hbox{\begin{tikzpicture}[scale=2,
    white/.style={circle, draw, fill=white, inner sep=0pt, minimum size=6pt},
    black/.style={circle, draw, fill=black, inner sep=0pt, minimum size=6pt},
    label/.style={font=\small, midway, sloped, above},
    coord/.style={font=\scriptsize, anchor=south east},
    coordne/.style={font=\scriptsize, anchor=north east},
    coordnw/.style={font=\scriptsize, anchor=north west},
    coordsw/.style={font=\scriptsize, anchor=south west}
]
\coordinate (A) at (0,0);  
\coordinate (B) at (1,0); 
\coordinate (C) at (1,1);
\coordinate (D) at (0,1); 
\coordinate (Aa) at (0.04,0.04); 
\coordinate (Ba) at (0.96,0.04);  
\coordinate (Ca) at (0.96,0.96);  
\coordinate (Da) at (0.04,0.96);  
\node[coordne] at (A) {$$};
\node[coordnw] at (B) {$$};
\node[coordsw] at (C) {$z$};
\node[coord] at (D) {$$};
\draw[red, midarrow, line width=1.5pt] (A) -- (B) node[midway, below] {$$};
\draw[red, midarrow, line width=1.5pt] (B) -- (C) node[midway, right] {$$};
\draw[red, midarrow, line width=1.5pt] (C) -- (D) node[midway, above] {$$};
\draw[red, midarrow, line width=1.5pt] (D) -- (A) node[midway, left] {$$};
\node[white] at (A) {};
\node[black] at (B) {};
\node[white] at (C) {};
\node[black] at (D) {};
\end{tikzpicture}}}
\eea
and
\bea
\vcenter{\hbox{\begin{tikzpicture}[scale=2,
    white/.style={circle, draw, fill=white, inner sep=0pt, minimum size=6pt},
    black/.style={circle, draw, fill=black, inner sep=0pt, minimum size=6pt},
    label/.style={font=\small, midway, sloped, above},
    coord/.style={font=\scriptsize, anchor=south east},
    coordne/.style={font=\scriptsize, anchor=north east},
    coordnw/.style={font=\scriptsize, anchor=north west},
    coordsw/.style={font=\scriptsize, anchor=south west}
]
\coordinate (A) at (0,0);  
\coordinate (B) at (1,0); 
\coordinate (C) at (1,1);
\coordinate (D) at (0,1); 
\coordinate (Aa) at (0.04,0.04); 
\coordinate (Ba) at (0.96,0.04);  
\coordinate (Ca) at (0.96,0.96);  
\coordinate (Da) at (0.04,0.96);  
\node[coordne] at (A) {$$};
\node[coordnw] at (B) {$$};
\node[coordsw] at (C) {$z$};
\node[coord] at (D) {$$};
\draw[red, midarrow, line width=1.5pt] (A) -- (B) node[midway, below] {$$};
\draw[red, midarrow, line width=1.5pt] (Ba) -- (Aa) node[midway, above] {$$};
\draw[red, midarrow, line width=1.5pt] (B) -- (C) node[midway, right] {$$};
\draw[red, midarrow, line width=1.5pt] (Ca) -- (Ba) node[midway, left] {$$};
\draw (C) -- (D) node[midway, above] {$$};
\draw (D) -- (A) node[midway, left] {$$};
\node[white] at (A) {};
\node[black] at (B) {};
\node[white] at (C) {};
\node[black] at (D) {};
\end{tikzpicture}}}
\rightarrow 
\vcenter{\hbox{\begin{tikzpicture}[scale=2,
    white/.style={circle, draw, fill=white, inner sep=0pt, minimum size=6pt},
    black/.style={circle, draw, fill=black, inner sep=0pt, minimum size=6pt},
    label/.style={font=\small, midway, sloped, above},
    coord/.style={font=\scriptsize, anchor=south east},
    coordne/.style={font=\scriptsize, anchor=north east},
    coordnw/.style={font=\scriptsize, anchor=north west},
    coordsw/.style={font=\scriptsize, anchor=south west}
]
\coordinate (A) at (0,0);  
\coordinate (B) at (1,0); 
\coordinate (C) at (1,1);
\coordinate (D) at (0,1); 
\coordinate (Aa) at (0.04,0.04); 
\coordinate (Ba) at (0.96,0.04);  
\coordinate (Ca) at (0.96,0.96);  
\coordinate (Da) at (0.04,0.96);  
\node[coordne] at (A) {$$};
\node[coordnw] at (B) {$$};
\node[coordsw] at (C) {$z$};
\node[coord] at (D) {$$};
\draw[red, midarrow, line width=1.5pt] (A) -- (B) node[midway, below] {$$};
\draw[red, midarrow, line width=1.5pt] (Ba) -- (Aa) node[midway, above] {$$};
\draw[red, midarrow, line width=1.5pt] (B) -- (C) node[midway, right] {$$};
\draw[red, midarrow, line width=1.5pt] (Ca) -- (Ba) node[midway, left] {$$};
\draw[red, midarrow, line width=1.5pt] (C) -- (D) node[midway, above] {$$};
\draw[red, midarrow, line width=1.5pt] (Da) -- (Ca) node[midway, below] {$$};
\draw (D) -- (A) node[midway, left] {$$};
\node[white] at (A) {};
\node[black] at (B) {};
\node[white] at (C) {};
\node[black] at (D) {};
\end{tikzpicture}}}
\eea
What seems to happen as we try pull the string out from the vertex $z$, is that the string wants to minimize its total length. As we try to pull out a spike from the vertex $z$, the string will find the other allowed configuration that costs less energy as that allowed configuration only requires four string segments instead of six for the disallowed configuration.

\section{The Wilson surface of a cube}
We will now obtain an expression for the Wilson surface of a cube by 'lassoing' the cube with our string. We may lasso the cube by a sequence of string moves as depicted in (\ref{lassoing}). The way we depict the lassoing process in (\ref{lassoing}) is by starting from a vertex point that some closed string goes through, and from this vertex point we extend the string by three subsequent K-spike moves in three different dimensions. We then apply $2\to 2$ moves to move three adjacent segments of this 3-spike string so that it wraps all across the cube and returns to the original configuration. The three segments that are moved across the cube thus sweep out a surface holonomy where the surface is the cube with a cut along the three edges from which we started and ended the round trip around the cube. The surface holonomy has the index structure 
\bea
U^{g''}{}_{j''}{}^{i''}{}_i{}^j{}_g
\eea
where $i,j,g$ are the indices that we put on the cut. We put some primes on the indices to tell that these index values on these three edges as we return back after the roundtrip may have been changed from their original values. While this is the structure of the surface holonomy, we are also interested in a gauge invariant object, namely the Wilson surface, of the cube. There is one obvious candidate, which is the trace. Physically the trace should be formed when the three returning segments get annihilated by the three segments that remained at the cut all along. This  physical process is similar to how we may define a Wilson loop in Yang-Mills-matter theory by letting a quark and anti-quark pair be created and then propagate for a certain period of time along two separate worldlines, and then rejoin where they then annihilate each other again. 

Below is a picture of the cube where on each face we have put a roman letter. We use the letters $A,B,C,a,b,c$ to denote the universal plaquette operator on each face. 
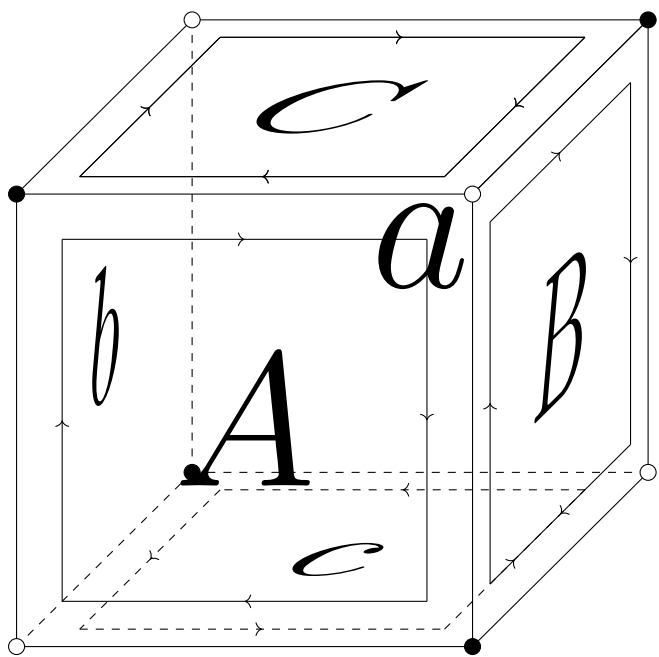
\bea
\begin{tikzpicture}[scale=4,
    white/.style={circle, draw, fill=white, inner sep=0pt, minimum size=6pt},
    black/.style={circle, draw, fill=black, inner sep=0pt, minimum size=6pt},
    label/.style={font=\small, midway, sloped, above},
    coord/.style={font=\scriptsize, anchor=south east},
    coordy/.style={font=\scriptsize, anchor=south west}
]
\coordinate (A) at (0,0,0);   
\coordinate (B) at (1,0,0);   
\coordinate (C) at (1,0,1);   
\coordinate (D) at (0,0,1);   
\coordinate (Aa) at (0.1,0,0.1);   
\coordinate (Ba) at (0.9,0,0.1);   
\coordinate (Ca) at (0.9,0,0.9);   
\coordinate (Da) at (0.1,0,0.9);   
\coordinate (E) at (0,1,0);   
\coordinate (F) at (1,1,0);   
\coordinate (G) at (1,1,1);   
\coordinate (H) at (0,1,1);   
\coordinate (Ea) at (0.1,1,0.1);   
\coordinate (Fa) at (0.9,1,0.1);   
\coordinate (Ga) at (0.9,1,0.9);   
\coordinate (Ha) at (0.1,1,0.9);   
\coordinate (DA) at (0.1,0.1,1);   
\coordinate (CA) at (0.9,0.1,1);   
\coordinate (HA) at (0.1,0.9,1);   
\coordinate (GA) at (0.9,0.9,1);   
\coordinate (FB) at (1,0.9,0.9);   
\coordinate (CB) at (1,0.1,0.9);   
\coordinate (BB) at (1,0.1,0.1);   
\coordinate (GB) at (1,0.9,0.1);   
\coordinate (I) at (0,2,0);   
\coordinate (J) at (1,2,0);   
\coordinate (K) at (1,2,1);   
\coordinate (L) at (0,2,1);   
\node[coord,align=left] at (D) {};
\node[coordy,align=left] at (F) {};
\draw (D) -- (C) node[midway, above] {$$};
\draw (C) -- (B) node[midway, right] {$$};
\draw[dashed] (B) -- (A) node[midway, above] {$$};
\draw[dashed] (A) -- (D) node[midway, right] {$$};
\draw[midarrow,dashed] (Da) -- (Ca) node[midway, above] {$$};
\draw[midarrow,dashed] (Ca) -- (Ba) node[midway, right] {$$};
\draw[midarrow,dashed] (Ba) -- (Aa) node[midway, above] {$$};
\draw[midarrow,dashed] (Aa) -- (Da) node[midway, right] {$$};
\draw[midarrow] (Ga) -- (Ha) node[midway, above] {$$};
\draw[midarrow] (Fa) -- (Ga) node[midway, right] {$$};
\draw[midarrow] (Ea) -- (Fa) node[midway, above] {$$};
\draw[midarrow] (Ha) -- (Ea) node[midway, right] {$$};
\draw[midarrow] (Ga) -- (Ha) node[midway, above] {$$};
\draw[midarrow] (Fa) -- (Ga) node[midway, right] {$$};
\draw[midarrow] (Ea) -- (Fa) node[midway, above] {$$};
\draw[midarrow] (Ha) -- (Ea) node[midway, right] {$$};
\draw[midarrow] (CA) -- (DA) node[midway, above] {$$};
\draw[midarrow] (GA) -- (CA) node[midway, right] {$$};
\draw[midarrow] (HA) -- (GA) node[midway, above] {$$};
\draw[midarrow] (DA) -- (HA) node[midway, right] {$$};
\draw[midarrow] (BB) -- (CB) node[midway, above] {$$};
\draw[midarrow] (GB) -- (BB) node[midway, right] {$$};
\draw[midarrow] (FB) -- (GB) node[midway, above] {$$};
\draw[midarrow] (CB) -- (FB) node[midway, right] {$$};
\draw[dashed] (A) -- (E) node[midway, above] {$$};
\draw (F) -- (B) node[midway, above] {$$};
\draw (C) -- (G) node[midway, above] {$$};
\draw (H) -- (D) node[midway, above] {$$};
\draw (F) -- (E) node[midway, right] {$$};
\draw (E) -- (H) node[midway, right] {$$};
\draw (H) -- (G) node[midway, right] {$$};
\draw (G) -- (F) node[midway, right] {$$};
\draw (F) -- (G) node[midway, above] {$$};
\node[black] at (A) {};
\node[white] at (B) {};
\node[black] at (C) {};
\node[white] at (D) {};
\node[white] at (E) {};
\node[black] at (F) {};
\node[white] at (G) {};
\node[black] at (H) {};

\node[canvas is xy plane at z=0, transform shape] at (0.5,0.5,0) {$a$};
\node[canvas is xy plane at z=1, transform shape] at (0.5,0.5,0) {$A$};
\node[canvas is yz plane at x=0, transform shape,rotate=-90] at (0,0.5,0.5) {$b$};
\node[canvas is yz plane at x=1, transform shape,rotate=-90] at (0,0.5,0.5) {$B$};
\node[canvas is zx plane at y=0, transform shape,rotate=90] at (0.5,0,0.5) {$c$};
\node[canvas is zx plane at y=1, transform shape,rotate=90] at (0.5,0,0.5) {$C$};
\end{tikzpicture}
\eea
As a two-dimensional surface, it has an intrinsic orientation. The orientation is represented by an arrow of closed loops drawn on the surface. In the picture we have drawn a few such loops. There are two choices for the direction of the arrow and hence of the orientation of the surface. Once we pick an orientation of one loop, then the orientation of all the other loops on the surface is uniquely fixed by requiring that the arrows of any two loops that meet at a point have arrows that are anti-parallel.  
 
Now we also like to put the color indices on the edges. Therefore we will draw the same cube again, this time with color indices on the edges,
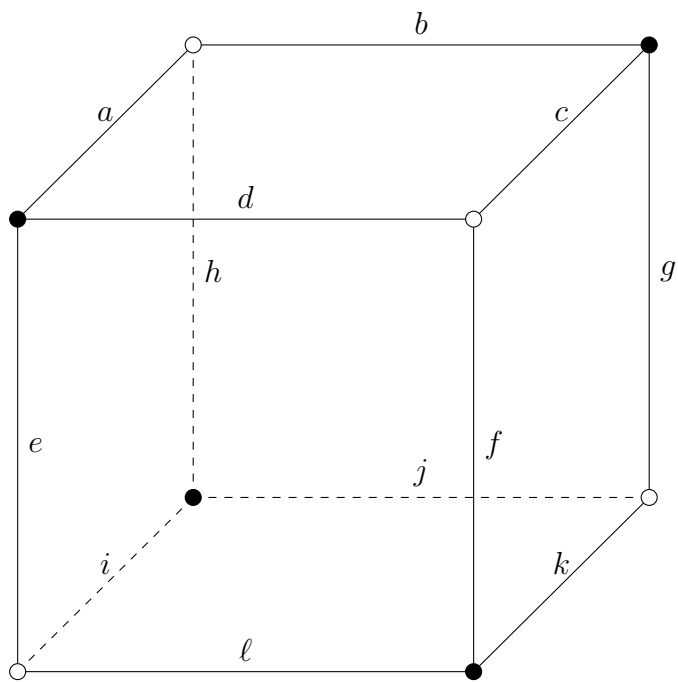
\bea
\begin{tikzpicture}[scale=4,
    white/.style={circle, draw, fill=white, inner sep=0pt, minimum size=6pt},
    black/.style={circle, draw, fill=black, inner sep=0pt, minimum size=6pt},
    label/.style={font=\small, midway, sloped, above},
    coord/.style={font=\scriptsize, anchor=south east}
 ]
\coordinate (A) at (0,0,0);   
\coordinate (B) at (1,0,0);   
\coordinate (C) at (1,0,1);   
\coordinate (D) at (0,0,1);   
\coordinate (E) at (0,1,0);   
\coordinate (F) at (1,1,0);   
\coordinate (G) at (1,1,1);   
\coordinate (H) at (0,1,1);   
\draw (D) -- (C) node[midway, above] {$\l$};
\draw (C) -- (B) node[midway, above] {$k$};
\draw[dashed] (B) -- (A) node[midway, above] {$j$};
\draw[dashed] (A) -- (D) node[midway, above] {$i$};
\draw[dashed] (A) -- (E) node[midway, right] {$h$};
\draw (F) -- (B) node[midway, right] {$g$};
\draw (C) -- (G) node[midway, right] {$f$};
\draw (H) -- (G) node[midway, above] {$d$};
\draw (G) -- (F) node[midway, above] {$c$};
\draw (F) -- (E) node[midway, above] {$b$};
\draw (E) -- (H) node[midway, above] {$a$};
\draw (H) -- (D) node[midway, right] {$e$};
\node[black] at (A) {};
\node[white] at (B) {};
\node[black] at (C) {};
\node[white] at (D) {};
\node[white] at (E) {};
\node[black] at (F) {};
\node[white] at (G) {};
\node[black] at (H) {};
\end{tikzpicture}
\eea
Now we shall pull a closed string over the cube. We may imagine that we shall throw out a lasso from some base point and then wrap the lasso around the entire cube and then pull back the lasso to the base point again. There are various ways that one can do this. One way is as pictured in the sequence below, where the base point is the vertex at which the string starts out as a pointlike string,
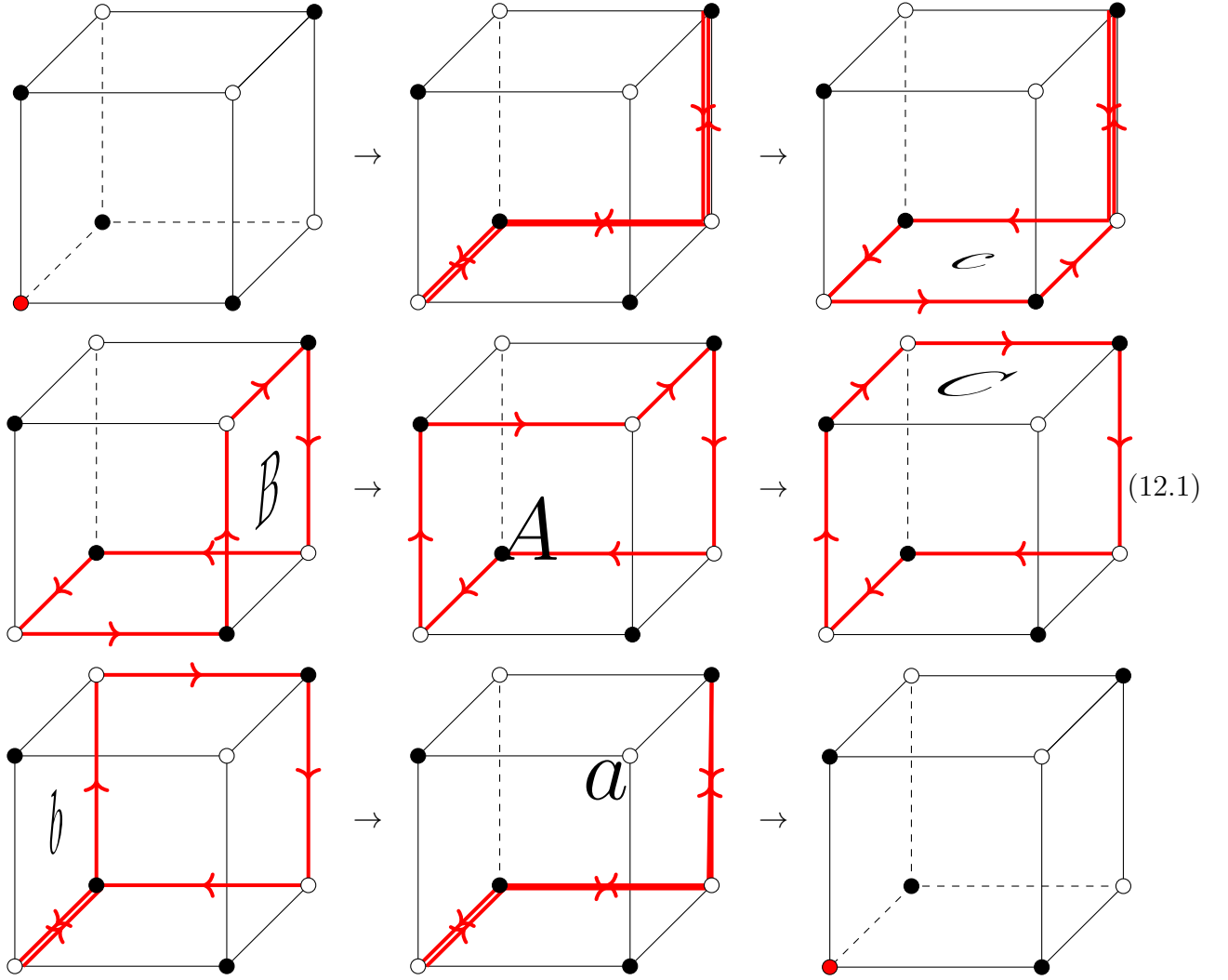
\bea
\begin{array}{ccccc}
\vcenter{\hbox{\begin{tikzpicture}[scale=3,
    white/.style={circle, draw, fill=white, inner sep=0pt, minimum size=6pt},
    black/.style={circle, draw, fill=black, inner sep=0pt, minimum size=6pt},
    red/.style={circle, draw, fill=red, inner sep=0pt, minimum size=6pt},
    label/.style={font=\small, midway, sloped, above},
    coord/.style={font=\scriptsize, anchor=south east}
]
\coordinate (A) at (0,0,0);   
\coordinate (B) at (1,0,0);   
\coordinate (C) at (1,0,1);   
\coordinate (D) at (0,0,1);   

\coordinate (Aa) at (0.1,0,0.1);   
\coordinate (Ba) at (0.9,0,0.1);   
\coordinate (Ca) at (0.9,0,0.9);   
\coordinate (Da) at (0.1,0,0.9);   

\coordinate (E) at (0,1,0);   
\coordinate (F) at (1,1,0);   
\coordinate (G) at (1,1,1);   
\coordinate (H) at (0,1,1);   

\coordinate (Ea) at (0.1,1,0.1);   
\coordinate (Fa) at (0.9,1,0.1);   
\coordinate (Ga) at (0.9,1,0.9);   
\coordinate (Ha) at (0.1,1,0.9);   

\coordinate (DA) at (0.1,0.1,1);   
\coordinate (CA) at (0.9,0.1,1);   
\coordinate (HA) at (0.1,0.9,1);   
\coordinate (GA) at (0.9,0.9,1);   

\coordinate (FB) at (1,0.9,0.9);   
\coordinate (CB) at (1,0.1,0.9);   
\coordinate (BB) at (1,0.1,0.1);   
\coordinate (GB) at (1,0.9,0.1);   

\coordinate (I) at (0,2,0);   
\coordinate (J) at (1,2,0);   
\coordinate (K) at (1,2,1);   
\coordinate (L) at (0,2,1);   
\draw (D) -- (C) node[midway, above] {$$};
\draw (C) -- (B) node[midway, right] {$$};
\draw[dashed] (B) -- (A) node[midway, above] {$$};
\draw[dashed] (A) -- (D) node[midway, right] {$$};

\draw[dashed] (A) -- (E) node[midway, above] {$$};
\draw (F) -- (B) node[midway, above] {$$};
\draw (C) -- (G) node[midway, above] {$$};
\draw (H) -- (D) node[midway, above] {$$};

\draw (F) -- (E) node[midway, right] {$$};

\draw (E) -- (H) node[midway, right] {$$};
\draw (H) -- (G) node[midway, right] {$$};
\draw (G) -- (F) node[midway, right] {$$};
\draw (F) -- (G) node[midway, above] {$$};

\node[black] at (A) {};
\node[white] at (B) {};
\node[black] at (C) {};
\node[red] at (D) {};
\node[white] at (E) {};
\node[black] at (F) {};
\node[white] at (G) {};
\node[black] at (H) {};
\end{tikzpicture}}}
&\rightarrow &
\vcenter{\hbox{\begin{tikzpicture}[scale=3,
    white/.style={circle, draw, fill=white, inner sep=0pt, minimum size=6pt},
    black/.style={circle, draw, fill=black, inner sep=0pt, minimum size=6pt},
    label/.style={font=\small, midway, sloped, above},
    coord/.style={font=\scriptsize, anchor=south east}
 ]
\coordinate (A) at (0,0,0);   
\coordinate (Aa) at (0.04,0,0.04); 
\coordinate (B) at (1,0,0);   
\coordinate (Ba) at (1,0,0.04);  
\coordinate (Bb) at (0.96,0,0);  
\coordinate (C) at (1,0,1);   
\coordinate (D) at (0,0,1);   
\coordinate (Da) at (0.04,0,1);   
\coordinate (E) at (0,1,0);   
\coordinate (Fa) at (1,1,0.04);   
\coordinate (Fb) at (0.96,1,0);
\coordinate (G) at (1,1,1);   
\coordinate (Gb) at (1,1,0.96);
\coordinate (Hb) at (0.04,1,0.96);
 \coordinate (Eb) at (0.04,1,0);
\coordinate (H) at (0,1,1);   
\coordinate (Ha) at (0.04,1,1);   
\draw (D) -- (C) node[midway, below] {$$};
\draw (C) -- (B) node[midway, right] {$$};
\draw[dashed] (B) -- (A) node[midway, above] {$$};
\draw[dashed] (A) -- (D) node[midway, right] {$$};
\draw[dashed] (A) -- (E) node[midway, right] {$$};
\draw (F) -- (B) node[midway, right] {$$};
\draw (C) -- (G) node[midway, right] {$$};
\draw (G) -- (H);
\draw (H) -- (D);
\draw (F) -- (E) node[midway, above] {$$};
\draw (E) -- (H) node[midway, right] {$$};
\draw (G) -- (F) node[midway, right] {$$};

\draw[midarrow,red, line width=1.5pt] (Da) -- (Aa) node[midway, below] {$$};
\draw[midarrow,red, line width=1.5pt] (Aa) -- (Ba) node[midway, below] {$$};
\draw[midarrow,red, line width=1.5pt] (Ba) -- (Fa) node[midway, right] {$$};
\draw[midarrow,red, line width=1.5pt] (Fb) -- (Bb) node[midway, left] {$$};
\draw[midarrow,red, line width=1.5pt] (Bb) -- (A) node[midway, above] {$$};
\draw[midarrow,red, line width=1.5pt] (A) -- (D) node[midway, left] {$$};
\node[black] at (A) {};
\node[white] at (B) {};
\node[black] at (C) {};
\node[white] at (D) {};
\node[white] at (E) {};
\node[black] at (F) {};
\node[white] at (G) {};
\node[black] at (H) {};
\end{tikzpicture}}}
&\rightarrow &
\vcenter{\hbox{\begin{tikzpicture}[scale=3,
    white/.style={circle, draw, fill=white, inner sep=0pt, minimum size=6pt},
    black/.style={circle, draw, fill=black, inner sep=0pt, minimum size=6pt},
    label/.style={font=\small, midway, sloped, above},
    coord/.style={font=\scriptsize, anchor=south east}
 ]
\coordinate (A) at (0,0,0);   
\coordinate (B) at (1,0,0);   
\coordinate (C) at (1,0,1);   
\coordinate (D) at (0,0,1);   
\coordinate (Da) at (0.04,0,1);   
\coordinate (E) at (0,1,0);   
\coordinate (F) at (1,1,0);   
\coordinate (G) at (1,1,1);   
\coordinate (Gb) at (1,1,0.96);
\coordinate (Hb) at (0.04,1,0.96);
 \coordinate (Eb) at (0.04,1,0);
\coordinate (H) at (0,1,1);   
\coordinate (Ha) at (0.04,1,1);   
\draw (D) -- (C) node[midway, below] {$$};
\draw (C) -- (B) node[midway, right] {$$};
\draw[dashed] (B) -- (A) node[midway, above] {$$};
\draw[dashed] (A) -- (D) node[midway, right] {$$};
\draw[dashed] (A) -- (E) node[midway, right] {$$};
\draw (F) -- (B) node[midway, right] {$$};
\draw (C) -- (G) node[midway, right] {$$};
\draw (G) -- (H);
\draw (H) -- (D);
\draw (F) -- (E) node[midway, above] {$$};
\draw (E) -- (H) node[midway, right] {$$};
\draw (G) -- (F) node[midway, right] {$$};

\draw[midarrow,red, line width=1.5pt] (Ba) -- (Fa) node[midway, right] {$$};
\draw[midarrow,red, line width=1.5pt] (Fb) -- (Bb) node[midway, left] {$$};
\draw[midarrow,red, line width=1.5pt] (Bb) -- (A) node[midway, above] {$$};
\draw[midarrow,red, line width=1.5pt] (A) -- (D) node[midway, left] {$$};

\draw[midarrow,red, line width=1.5pt] (D) -- (C) node[midway, below] {$$};
\draw[midarrow,red, line width=1.5pt] (C) -- (B) node[midway, below] {$$};
\draw[midarrow,red, line width=1.5pt] (A) -- (D) node[midway, left] {$$};
\node[black] at (A) {};
\node[white] at (B) {};
\node[black] at (C) {};
\node[white] at (D) {};
\node[white] at (E) {};
\node[black] at (F) {};
\node[white] at (G) {};
\node[black] at (H) {};


\node[canvas is zx plane at y=0, transform shape,rotate=90] at (0.5,0,0.5) {$c$};
\end{tikzpicture}}}\cr
\vcenter{\hbox{\begin{tikzpicture}[scale=3,
    white/.style={circle, draw, fill=white, inner sep=0pt, minimum size=6pt},
    black/.style={circle, draw, fill=black, inner sep=0pt, minimum size=6pt},
    label/.style={font=\small, midway, sloped, above},
    coord/.style={font=\scriptsize, anchor=south east}
 ]
\coordinate (A) at (0,0,0);   
\coordinate (B) at (1,0,0);   
\coordinate (C) at (1,0,1);   
\coordinate (D) at (0,0,1);   
\coordinate (Da) at (0.04,0,1);   
\coordinate (E) at (0,1,0);   
\coordinate (F) at (1,1,0);   
\coordinate (G) at (1,1,1);   
\coordinate (Gb) at (1,1,0.96);
\coordinate (Hb) at (0.04,1,0.96);
 \coordinate (Eb) at (0.04,1,0);
\coordinate (H) at (0,1,1);   
\coordinate (Ha) at (0.04,1,1);   
\draw (D) -- (C) node[midway, below] {$$};
\draw (C) -- (B) node[midway, right] {$$};
\draw[dashed] (B) -- (A) node[midway, above] {$$};
\draw[dashed] (A) -- (D) node[midway, right] {$$};
\draw[dashed] (A) -- (E) node[midway, right] {$$};
\draw (F) -- (B) node[midway, right] {$$};
\draw (C) -- (G) node[midway, right] {$$};
\draw (G) -- (H);
\draw (H) -- (D);
\draw (F) -- (E) node[midway, above] {$$};
\draw (E) -- (H) node[midway, right] {$$};
\draw (G) -- (F) node[midway, right] {$$};

\draw[midarrow,red, line width=1.5pt] (F) -- (B) node[midway, left] {$$};
\draw[midarrow,red, line width=1.5pt] (B) -- (A) node[midway, above] {$$};
\draw[midarrow,red, line width=1.5pt] (A) -- (D) node[midway, left] {$$};
\draw[midarrow,red, line width=1.5pt] (D) -- (C) node[midway, below] {$$};
\draw[midarrow,red, line width=1.5pt] (C) -- (G) node[midway, right] {$$};
\draw[midarrow,red, line width=1.5pt] (G) -- (F) node[midway, right] {$$};
\node[black] at (A) {};
\node[white] at (B) {};
\node[black] at (C) {};
\node[white] at (D) {};
\node[white] at (E) {};
\node[black] at (F) {};
\node[white] at (G) {};
\node[black] at (H) {};

\node[canvas is yz plane at x=1, transform shape,rotate=-90] at (0,0.5,0.5) {$B$};
\end{tikzpicture}}}
&\rightarrow &
\vcenter{\hbox{\begin{tikzpicture}[scale=3,
    white/.style={circle, draw, fill=white, inner sep=0pt, minimum size=6pt},
    black/.style={circle, draw, fill=black, inner sep=0pt, minimum size=6pt},
    label/.style={font=\small, midway, sloped, above},
    coord/.style={font=\scriptsize, anchor=south east}
 ]
\coordinate (A) at (0,0,0);   
\coordinate (B) at (1,0,0);   
\coordinate (C) at (1,0,1);   
\coordinate (D) at (0,0,1);   
\coordinate (Da) at (0.04,0,1);   
\coordinate (E) at (0,1,0);   
\coordinate (F) at (1,1,0);   
\coordinate (Fa) at (1,0.96,0);  
\coordinate (G) at (1,1,1);   
\coordinate (Ga) at (1,0.96,1); 
\coordinate (Gb) at (1,1,0.96);
\coordinate (Hb) at (0.04,1,0.96);
 \coordinate (Eb) at (0.04,1,0);
\coordinate (H) at (0,1,1);   
\coordinate (Ha) at (0.04,0.96,1);   
\draw (D) -- (C) node[midway, below] {$$};
\draw (C) -- (B) node[midway, right] {$$};
\draw[dashed] (B) -- (A) node[midway, above] {$$};
\draw[dashed] (A) -- (D) node[midway, right] {$$};
\draw[dashed] (A) -- (E) node[midway, right] {$$};
\draw (F) -- (B) node[midway, right] {$$};
\draw (C) -- (G) node[midway, right] {$$};
\draw (H) -- (D);
\draw (F) -- (E) node[midway, above] {$$};
\draw (E) -- (H) node[midway, right] {$$};
\draw (G) -- (F) node[midway, right] {$$};

\draw[midarrow,red, line width=1.5pt] (F) -- (B) node[midway, left] {$$};
\draw[midarrow,red, line width=1.5pt] (B) -- (A) node[midway, above] {$$};
\draw[midarrow,red, line width=1.5pt] (A) -- (D) node[midway, left] {$$};

\draw[midarrow,red, line width=1.5pt] (D) -- (H) node[midway, right] {$$};
\draw[midarrow,red, line width=1.5pt] (H) -- (G) node[midway, below] {$$};
\draw[midarrow,red, line width=1.5pt] (G) -- (F) node[midway, right] {$$};

\node[black] at (A) {};
\node[white] at (B) {};
\node[black] at (C) {};
\node[white] at (D) {};
\node[white] at (E) {};
\node[black] at (F) {};
\node[white] at (G) {};
\node[black] at (H) {};

\node[canvas is xy plane at z=1, transform shape] at (0.5,0.5,0) {$A$};

\end{tikzpicture}}}
&\rightarrow &
\vcenter{\hbox{\begin{tikzpicture}[scale=3,
    white/.style={circle, draw, fill=white, inner sep=0pt, minimum size=6pt},
    black/.style={circle, draw, fill=black, inner sep=0pt, minimum size=6pt},
    label/.style={font=\small, midway, sloped, above},
    coord/.style={font=\scriptsize, anchor=south east}
 ]
\coordinate (A) at (0,0,0);   
\coordinate (B) at (1,0,0);   
\coordinate (C) at (1,0,1);   
\coordinate (D) at (0,0,1);   
\coordinate (Da) at (0.04,0,1);   
\coordinate (E) at (0,1,0);   
\coordinate (F) at (1,1,0);   
\coordinate (Fa) at (1,0.96,0);  
\coordinate (G) at (1,1,1);   
\coordinate (Ga) at (1,0.96,1); 
\coordinate (Gb) at (1,1,0.96);
\coordinate (Hb) at (0.04,1,0.96);
 \coordinate (Eb) at (0.04,1,0);
\coordinate (H) at (0,1,1);   
\coordinate (Ha) at (0.04,0.96,1);   
\draw (D) -- (C) node[midway, below] {$$};
\draw (C) -- (B) node[midway, right] {$$};
\draw[dashed] (B) -- (A) node[midway, above] {$$};
\draw[dashed] (A) -- (D) node[midway, right] {$$};
\draw[dashed] (A) -- (E) node[midway, right] {$$};
\draw (F) -- (B) node[midway, right] {$$};
\draw (C) -- (G) node[midway, right] {$$};
\draw (H) -- (D);
\draw (F) -- (E) node[midway, above] {$$};
\draw (E) -- (H) node[midway, right] {$$};
\draw (G) -- (F) node[midway, right] {$$};
\draw (G) -- (H) node[midway, right] {$$};

\draw[midarrow,red, line width=1.5pt] (F) -- (B) node[midway, left] {$$};
\draw[midarrow,red, line width=1.5pt] (B) -- (A) node[midway, above] {$$};
\draw[midarrow,red, line width=1.5pt] (A) -- (D) node[midway, left] {$$};

\draw[midarrow,red, line width=1.5pt] (D) -- (H) node[midway, right] {$$};
\draw[midarrow,red, line width=1.5pt] (H) -- (E) node[midway, below] {$$};
\draw[midarrow,red, line width=1.5pt] (E) -- (F) node[midway, right] {$$};
\node[black] at (A) {};
\node[white] at (B) {};
\node[black] at (C) {};
\node[white] at (D) {};
\node[white] at (E) {};
\node[black] at (F) {};
\node[white] at (G) {};
\node[black] at (H) {};
\node[canvas is zx plane at y=1, transform shape,rotate=90] at (0.5,0,0.5) {$C$};
\end{tikzpicture}}}\cr
\vcenter{\hbox{\begin{tikzpicture}[scale=3,
    white/.style={circle, draw, fill=white, inner sep=0pt, minimum size=6pt},
    black/.style={circle, draw, fill=black, inner sep=0pt, minimum size=6pt},
    label/.style={font=\small, midway, sloped, above},
    coord/.style={font=\scriptsize, anchor=south east}
 ]
\coordinate (A) at (0,0,0);   
\coordinate (B) at (1,0,0);   
\coordinate (C) at (1,0,1);   
\coordinate (D) at (0,0,1);   
\coordinate (Da) at (0.04,0,1);   
\coordinate (E) at (0,1,0);   
\coordinate (F) at (1,1,0);   
\coordinate (Fa) at (1,0.96,0);  
\coordinate (G) at (1,1,1);   
\coordinate (Ga) at (1,0.96,1); 
\coordinate (Gb) at (1,1,0.96);
\coordinate (Hb) at (0.04,1,0.96);
 \coordinate (Eb) at (0.04,1,0);
\coordinate (H) at (0,1,1);   
\coordinate (Ha) at (0.04,0.96,1);   
\draw (D) -- (C) node[midway, below] {$$};
\draw (C) -- (B) node[midway, right] {$$};
\draw[dashed] (B) -- (A) node[midway, above] {$$};
\draw[dashed] (A) -- (D) node[midway, right] {$$};
\draw[dashed] (A) -- (E) node[midway, right] {$$};
\draw (F) -- (B) node[midway, right] {$$};
\draw (C) -- (G) node[midway, right] {$$};
\draw (H) -- (D);
\draw (F) -- (E) node[midway, above] {$$};
\draw (E) -- (H) node[midway, right] {$$};
\draw (G) -- (F) node[midway, right] {$$};
\draw (G) -- (H) node[midway, right] {$$};

\draw[midarrow,red, line width=1.5pt] (F) -- (B) node[midway, left] {$$};
\draw[midarrow,red, line width=1.5pt] (B) -- (A) node[midway, above] {$$};
\draw[midarrow,red, line width=1.5pt] (A) -- (D) node[midway, left] {$$};

\draw[midarrow,red, line width=1.5pt] (Da) -- (Aa) node[midway, right] {$$};
\draw[midarrow,red, line width=1.5pt] (A) -- (E) node[midway, below] {$$};
\draw[midarrow,red, line width=1.5pt] (E) -- (F) node[midway, right] {$$};
\node[black] at (A) {};
\node[white] at (B) {};
\node[black] at (C) {};
\node[white] at (D) {};
\node[white] at (E) {};
\node[black] at (F) {};
\node[white] at (G) {};
\node[black] at (H) {};
\node[canvas is yz plane at x=0, transform shape,rotate=-90] at (0,0.5,0.5) {$b$};
\end{tikzpicture}}}
&\rightarrow &
\vcenter{\hbox{\begin{tikzpicture}[scale=3,
    white/.style={circle, draw, fill=white, inner sep=0pt, minimum size=6pt},
    black/.style={circle, draw, fill=black, inner sep=0pt, minimum size=6pt},
    label/.style={font=\small, midway, sloped, above},
    coord/.style={font=\scriptsize, anchor=south east}
 ]
\coordinate (A) at (0,0,0);   
\coordinate (B) at (1,0,0);   
\coordinate (C) at (1,0,1);   
\coordinate (D) at (0,0,1);   
\coordinate (Da) at (0.04,0,1);   
\coordinate (E) at (0,1,0);   
\coordinate (F) at (1,1,0);   
\coordinate (Fa) at (1,0.96,0);  
\coordinate (G) at (1,1,1);   
\coordinate (Ga) at (1,0.96,1); 
\coordinate (Gb) at (1,1,0.96);
\coordinate (Hb) at (0.04,1,0.96);
 \coordinate (Eb) at (0.04,1,0);
\coordinate (H) at (0,1,1);   
\coordinate (Ha) at (0.04,0.96,1);   
\draw (D) -- (C) node[midway, below] {$$};
\draw (C) -- (B) node[midway, right] {$$};
\draw[dashed] (B) -- (A) node[midway, above] {$$};
\draw[dashed] (A) -- (D) node[midway, right] {$$};
\draw[dashed] (A) -- (E) node[midway, right] {$$};
\draw (F) -- (B) node[midway, right] {$$};
\draw (C) -- (G) node[midway, right] {$$};
\draw (H) -- (D);
\draw (F) -- (E) node[midway, above] {$$};
\draw (E) -- (H) node[midway, right] {$$};
\draw (G) -- (F) node[midway, right] {$$};
\draw (G) -- (H) node[midway, right] {$$};

\draw[midarrow,red, line width=1.5pt] (F) -- (B) node[midway, left] {$$};
\draw[midarrow,red, line width=1.5pt] (B) -- (A) node[midway, above] {$$};
\draw[midarrow,red, line width=1.5pt] (A) -- (D) node[midway, left] {$$};

\draw[midarrow,red, line width=1.5pt] (Da) -- (Aa) node[midway, right] {$$};
\draw[midarrow,red, line width=1.5pt] (Aa) -- (Ba) node[midway, below] {$$};
\draw[midarrow,red, line width=1.5pt] (Ba) -- (Fa) node[midway, right] {$$};

\node[black] at (A) {};
\node[white] at (B) {};
\node[black] at (C) {};
\node[white] at (D) {};
\node[white] at (E) {};
\node[black] at (F) {};
\node[white] at (G) {};
\node[black] at (H) {};

\node[canvas is xy plane at z=0, transform shape] at (0.5,0.5,0) {$a$};
\end{tikzpicture}}}
&\rightarrow &
\vcenter{\hbox{\begin{tikzpicture}[scale=3,
    white/.style={circle, draw, fill=white, inner sep=0pt, minimum size=6pt},
    black/.style={circle, draw, fill=black, inner sep=0pt, minimum size=6pt},
    red/.style={circle, draw, fill=red, inner sep=0pt, minimum size=6pt},
    label/.style={font=\small, midway, sloped, above},
    coord/.style={font=\scriptsize, anchor=south east}
]
\coordinate (A) at (0,0,0);   
\coordinate (B) at (1,0,0);   
\coordinate (C) at (1,0,1);   
\coordinate (D) at (0,0,1);   

\coordinate (Aa) at (0.1,0,0.1);   
\coordinate (Ba) at (0.9,0,0.1);   
\coordinate (Ca) at (0.9,0,0.9);   
\coordinate (Da) at (0.1,0,0.9);   

\coordinate (E) at (0,1,0);   
\coordinate (F) at (1,1,0);   
\coordinate (G) at (1,1,1);   
\coordinate (H) at (0,1,1);   

\coordinate (Ea) at (0.1,1,0.1);   
\coordinate (Fa) at (0.9,1,0.1);   
\coordinate (Ga) at (0.9,1,0.9);   
\coordinate (Ha) at (0.1,1,0.9);   

\coordinate (DA) at (0.1,0.1,1);   
\coordinate (CA) at (0.9,0.1,1);   
\coordinate (HA) at (0.1,0.9,1);   
\coordinate (GA) at (0.9,0.9,1);   

\coordinate (FB) at (1,0.9,0.9);   
\coordinate (CB) at (1,0.1,0.9);   
\coordinate (BB) at (1,0.1,0.1);   
\coordinate (GB) at (1,0.9,0.1);   

\coordinate (I) at (0,2,0);   
\coordinate (J) at (1,2,0);   
\coordinate (K) at (1,2,1);   
\coordinate (L) at (0,2,1);   
\draw (D) -- (C) node[midway, above] {$$};
\draw (C) -- (B) node[midway, right] {$$};
\draw[dashed] (B) -- (A) node[midway, above] {$$};
\draw[dashed] (A) -- (D) node[midway, right] {$$};

\draw[dashed] (A) -- (E) node[midway, above] {$$};
\draw (F) -- (B) node[midway, above] {$$};
\draw (C) -- (G) node[midway, above] {$$};
\draw (H) -- (D) node[midway, above] {$$};

\draw (F) -- (E) node[midway, right] {$$};

\draw (E) -- (H) node[midway, right] {$$};
\draw (H) -- (G) node[midway, right] {$$};
\draw (G) -- (F) node[midway, right] {$$};
\draw (F) -- (G) node[midway, above] {$$};

\node[black] at (A) {};
\node[white] at (B) {};
\node[black] at (C) {};
\node[red] at (D) {};
\node[white] at (E) {};
\node[black] at (F) {};
\node[white] at (G) {};
\node[black] at (H) {};
\end{tikzpicture}}}
\end{array}\label{lassoing}
\eea
The wave function transforms as
\bea
\psi(x) &\rightarrow & \psi{}_{i'}{}^{j'}{}_{g'}{}^g{}_j{}^i \cr
&\rightarrow & \psi{}_{i'}{}^{j'}{}_{g'}{}^{g''}{}_{j''}{}^{i''} = U{}^{g''}{}_{j''}{}^{i''}{}_i{}^j{}_g \psi{}_{i'}{}^{j'}{}_{g'}{}^g{}_j{}^i
\eea
where the surface holonomy is given by
\bea
U{}^{g''}{}_{j''}{}^{i''}{}_i{}^j{}_g &=& 
a^{g''}{}_{j''}{}^h{}_b 
b_h{}^{i''}{}_e{}^a 
C_a{}^d{}_c{}^b
A^e{}_{\l}{}^f{}_d 
B_f{}^k{}_g{}^c  
c_i{}^j{}_k{}^{\l} 
\eea
We shall define the gauge invariant Wilson surface of the cube as the trace, with a normalization constant of $1/N^3$, thus
\bea
W &=& \frac{1}{N^3} U{}^{g}{}_{j}{}^{i}{}_i{}^j{}_g
\eea
or more explicitly,
\bea
W &=& \frac{1}{N^3} a^{g}{}_{j}{}^h{}_b 
b_h{}^{i}{}_e{}^a 
C_a{}^d{}_c{}^b
A^e{}_{\l}{}^f{}_d 
B_f{}^k{}_g{}^c  
c_i{}^j{}_k{}^{\l} 
\eea
One way to motivate the normalization is by taking the plaquette operators as unit operators, in which case this normalization gives the result
\bea
W &=& 1
\eea

The normalization can also be obtained in a more straightforward way if we note that the initial wave function must take the particular form of a six-segmented K-spike
\bea
\psi{}_{i'}{}^{j'}{}_{g'}{}^g{}_j{}^i = \(\frac{1}{\sqrt{N}}\)^3 \delta^g_{g'} \delta_j^{j'} \delta^i_{i'} \psi(x) 
\eea
as dictated by $U(N)$ gauge symmetry. Then if we also assume that the final wave function takes a similar form from calibration
\bea
\psi{}_{i'}{}^{j'}{}_{g'}{}^{g''}{}_{j''}{}^{i''} = \(\frac{1}{\sqrt{N}}\)^3 \delta^{g''}_{g'} \delta_{j''}^{j'} \delta^{i''}_{i'} \psi''(x) 
\eea
then we find that the Wilson surface with this normalization transforms the gauge invariant initial wave function to the final gauge invariant wave function as
\bea
\psi''(x) &=& W \psi(x)
\eea
Let us comment that the calibrated wave function at the final six-segmented spike string configuration is a necessary assumption if we shall be able to pull back the string to the vertex $x$ again by three subsequent R-moves. So, in the regime where we can compute anything, this is the result that we get.

We have an analogous result in Yang-Mills gauge theory on a 4d hypercubic lattice. The Wilson loop on a plaquette in such a lattice gauge theory is defined as
\bea
W &=& \frac{1}{N} A^i{}_j B^j{}_k a^k{}_{\l} b^{\l}{}_i 
\eea
with the normalization constant $1/N$. Then again for the unit plaquette operator (this is up to gauge, but at the end of the day the Wilson loop is gauge invariant) $U^i{}_j = \delta^i_j$ this reduces to $W = 1$. Below is a picture of this Wilson loop,
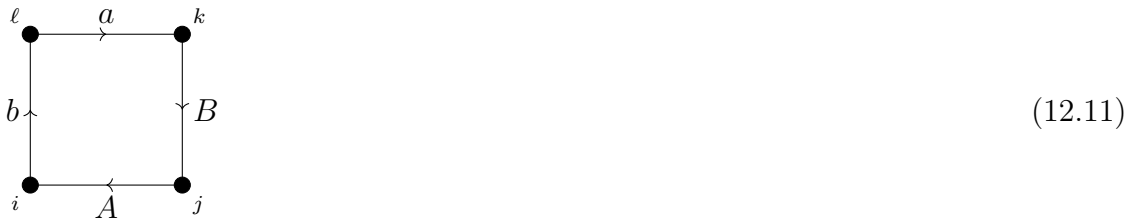
\bea
\vcenter{\hbox{\begin{tikzpicture}[scale=2,
    white/.style={circle, draw, fill=white, inner sep=0pt, minimum size=6pt},
    black/.style={circle, draw, fill=black, inner sep=0pt, minimum size=6pt},
    label/.style={font=\small, midway, sloped, above},
    coord/.style={font=\scriptsize, anchor=south east},
    coordne/.style={font=\scriptsize, anchor=north east},
    coordnw/.style={font=\scriptsize, anchor=north west},
    coordsw/.style={font=\scriptsize, anchor=south west}
]
\coordinate (A) at (0,0);  
\coordinate (B) at (1,0); 
\coordinate (C) at (1,1);
\coordinate (D) at (0,1); 
\node[coordne] at (A) {$i$};
\node[coordnw] at (B) {$j$};
\node[coordsw] at (C) {$k$};
\node[coord] at (D) {$\l$};
\draw[midarrow] (B) -- (A) node[midway, below] {$A$};
\draw[midarrow] (C) -- (B) node[midway, right] {$B$};
\draw[midarrow] (D) -- (C) node[midway, above] {$a$};
\draw[midarrow] (A) -- (D) node[midway, left] {$b$};
\node[black] at (A) {};
\node[black] at (B) {};
\node[black] at (C) {};
\node[black] at (D) {};
\end{tikzpicture}}}
\eea

\section{The generic Wilson surface} 
The generalization to a generic Wilson surface is now immediate. We no longer need to think of how the string moves across the cube, but can immediately write down the Wilson surface once we know the holonomies on each of its faces (plaquettes) and have specified its orientation. Then all we need to do is to put together the holonomies, sum over adjacent color indices on each edge and the result will be independent of how the string is being pulled over the surface. 

But the string picture is essential since it describes how an actual closed string is moved by the surface holonomies. Thus although the string is not necessary for the computation of the Wilson surface \cite{Rey:2010uz}, just as little as quarks are necessary for the computation of a Wilson loop in Yang-Mills theory, they are an essential ingredient in the theory. 

The basic idea of placing four color indices on the four edges of a plaquette in a hypercubic lattice to form some sort of plaquette operator to be used to build up specifically a Wilson surface of a cube by gluing together plaquette operators on the six faces of the cube, is not new. Such a construction first emerged in \cite{Nepomechie:1982rb} and quite some time later again in \cite{Rey:2010uz} and \cite{Lipstein:2014vca}. However, upon closer inspection, our construction seems o be different from these references. Our universal plaquette operator has cyclic symmetry. None of these references made such a cyclic symmetry explicit. Instead, in these references, a base point was picked in one vertices of the plaquette, and this base point selected the first index of the plaquette operator. Then different base points could lead to seemingly different and unrelated plaquette operators. Our plaquette operator on the other hand selects no base point in anyone of the four vertices and it is cyclically symmetric. Another difference that is a consequence of this over-looked cyclic symmetry is that in these references the choice of plaquette operator on each face of the cube depends on the choice of adjacent plaquette operators. This extra nonlocal structure, which was explicitly drawn as dashed lines in an unwrapped cube in \cite{Nepomechie:1982rb}, is necessary in order to consistently glue together the plaquette operators across the six faces of the cube. In our construction, we do not need such an additional nonlocal data from adjacent faces. Our plaquette operators are determined by the plaquette itself and gluing is automatically taken care of by the way that each plaquette carries its own orientation and bipartite structure. Those two features determines the index structure of each plaquette operator and takes care of gluing of adjacent plaquette operators automatically so that a gauge invariant Wilson surface is formed upon index contraction. In this way, our constructiion of the nonabelian WIlson surface seems to be a new construction, compared to these older proposals. 

\section{Gauge symmetry}
We have introduced color indices in fundamental and anti-fundamental representations of $U(N)$ gauge symmetry. But so far we have not spelled out how the gauge symmetry acts. So let us present this here. Since we have associated one color index $i$ with a string segment, it is natural to ascribe a $U(N)$ gauge group element to this string segment. There are two cases that we need to distinguish depending on the orientation of the string in relation to the bipartite lattice.
\bea
\vcenter{\hbox{\begin{tikzpicture}[scale=3,
    white/.style={circle, draw, fill=white, inner sep=0pt, minimum size=6pt},
    black/.style={circle, draw, fill=black, inner sep=0pt, minimum size=6pt},
    label/.style={font=\small, midway, sloped, above},
    coord/.style={font=\scriptsize, anchor=south east},
    coordne/.style={font=\scriptsize, anchor=north east},
    coordnw/.style={font=\scriptsize, anchor=north west},
    coordsw/.style={font=\scriptsize, anchor=south west}
]
\coordinate (A) at (0,0);   
\coordinate (B) at (1,0);  
\node[coordne] at (A) {$$};
\node[coordnw] at (B) {$$};
\draw[red,midarrow, line width=1.5pt] (A) -- (B) node[midway, below] {$i$};
\node[white] at (A) {};
\node[black] at (B) {};
\end{tikzpicture}}}
&=& \psi^i\cr
\vcenter{\hbox{\begin{tikzpicture}[scale=3,
    white/.style={circle, draw, fill=white, inner sep=0pt, minimum size=6pt},
    black/.style={circle, draw, fill=black, inner sep=0pt, minimum size=6pt},
    label/.style={font=\small, midway, sloped, above},
    coord/.style={font=\scriptsize, anchor=south east},
    coordne/.style={font=\scriptsize, anchor=north east},
    coordnw/.style={font=\scriptsize, anchor=north west},
    coordsw/.style={font=\scriptsize, anchor=south west}
]
\coordinate (A) at (0,0);   
\coordinate (B) at (1,0);  
\node[coordne] at (A) {$$};
\node[coordnw] at (B) {$$};
\draw[red,midarrow, line width=1.5pt] (B) -- (A) node[midway, below] {$i$};
\node[white] at (A) {};
\node[black] at (B) {};
\end{tikzpicture}}}
&=& \psi_i
\eea
The $U(N)$ gauge symmetry acts on these corresponding wave functions as
\bea
\psi^i &\to & \t\psi^{i'} = g^{i'}{}_i \psi^i\cr
\psi_i &\to & \t\psi_{i'} = \psi_i \bar{g}^i{}_{i'}
\eea
where we define
\bea
\bar{g}^{i}{}_{j} &=& \(g^j{}_i\)^*
\eea
and we have the unitary compactness condition
\bea
\bar{g}^i{}_{i'} g^{i'}{}_j &=& \delta^i_j
\eea
Gauge symmetry is thus implemented of the string wave function by assigning a gauge group element $g^{i'}{}_i$ or its complex conjugate $\bar{g}^i{}_{i'}$ to each segment of the string in the hypercubic lattice. The gauge transformation of the surface holonomy is then derived from gauge covariance under parallel transport of the wave function.

\section{Dimensional reduction}
Dimensional reduction of a lattice model for tensor gauge fields was considered, apparently independently, in both \cite{Rey:2010uz} and \cite{Lipstein:2014vca}. In both these references the compactification radius is set equal to the lattice spacing parameter. The next step, is to perform the dimensional reduction. It is suggested in these references that upon dimensional reduction we shall 'squash' the plaquette in the direction of dimensional reduction such that it becomes a line holonomy. This intuition is a bit misleading, and the reason why \cite{Lipstein:2014vca} ended up concluding that dimensional reduction does not commute with taking the continuum limit. Let us begin with clarifying what what we shall really do in order to perform dimensional reduction of our lattice gauge theory in a consistent manner. 

We have two parameters, the lattice spacing $a$ and the compactification radius $R$. We start by fixing the radius $R$ at some finite value and then we adjust the lattice spacing in the direction of the compact circle so that $a_{S^1} = R$. This assumption means that we reduce the plaquette surface holonomy to a line holonomy that transforms in the fundamental representation and it needs to be adjusted accordingly if we want to get a higher-dimensional representation. Where we depart from these references is in how we carry out the subsequent dimensional reduction. Let us start with a rectangular lattice with lattice spacing $a$ in all six direction. If we want to compactify one direction on a circle of radius $R>>a$ in such a way that upon dimensional reduction we end up with a line holonomy in the fundamental representation, then we need to stretch the lattice along the circle so that the lattice spacing $a_{S^1}$ in the direction of the circle becomes very large, $a_{S^1} >> a$ and equal to the compactification radius $R$, while the lattice spacing in the other five directions is held fixed and at the value $a$ much smaller than $R$. Now that we have compactified our lattice theory we can perform dimensional reduction, which amounts to studying the theory at energy levels much smaller than the first Kaluza-Klein excitation at energy $1/R$ while we restrict ourselves to an energy level $E$ that are well above $1/a$ such that our processes are not influenced by the discreteness of the lattice. From the aforementioned two requirements $1/a << E << 1/R$, it follows that $a_{S^1} = R>>a$ and therefore we need to stretch (as opposed to shrink) the plaquette, in the direction of the dimensional reduction.

We may introduce a periodic identification in one direction on the lattice. Since the lattice is bipartite, there is a problem to overcome if we want to obtain the smallest possible peridic identification that extends over one edge because that edge should, at least initially, due to the bipartite nature of the lattice, end on one white and one black vertex, respectively. It will be much easier to understand how to identify, say, two white vertices. But this requires an even number of edges along the compact circle direction. 

Under dimensional reduction the string that wraps the compact dimension is supposed to be straight. We expect that any wiggle of the string will result in a massive mode in the lower dimensional theory and under dimensional reduction these modes will be suppressed by $1/R$ where $R$ is the radius of the compact circle and we can discard them. The general idea with distributing color indices along the string is that we place one color index on each segment of the string that we intend to deform locally. If we do not deform the string locally, then we may just place one color index on the entire string. That corresponds to assigning one edge to the entire compact dimension. We then have one white vertex on the compact direction and one color index on the edge, the compact circle. The basic plaquette is replaced with a square that has  three different color indices. There are three and not four since the top and the bottom edges are identical upon identification. This plaquette is drawn below,
\bea
\vcenter{\hbox{\begin{tikzpicture}[scale=3,
    white/.style={circle, draw, fill=white, inner sep=0pt, minimum size=6pt},
    black/.style={circle, draw, fill=black, inner sep=0pt, minimum size=6pt},
    label/.style={font=\small, midway, sloped, above},
    coord/.style={font=\scriptsize, anchor=south east},
    coordne/.style={font=\scriptsize, anchor=north east},
    coordnw/.style={font=\scriptsize, anchor=north west},
    coordsw/.style={font=\scriptsize, anchor=south west}
]
\coordinate (A) at (0,0);   
\coordinate (B) at (1,0);   
\coordinate (C) at (1,1);   
\coordinate (D) at (0,1);
\draw[midarrow, red, line width=1.5pt] (A) -- (D) node[midway, left] {$i$};
\draw (A) -- (B) node[midway,below] {$k$};
\draw (C) -- (D) node[midway,below] {$k$};
\draw (B) -- (C) node[midway,left] {$j$};
\node[white] at (A) {};
\node[black] at (B) {};
\node[white] at (D) {};
\node[black] at (C) {};
\end{tikzpicture}}}
& \rightarrow &
\vcenter{\hbox{\begin{tikzpicture}[scale=3,
    white/.style={circle, draw, fill=white, inner sep=0pt, minimum size=6pt},
    black/.style={circle, draw, fill=black, inner sep=0pt, minimum size=6pt},
    label/.style={font=\small, midway, sloped, above},
    coord/.style={font=\scriptsize, anchor=south east},
    coordne/.style={font=\scriptsize, anchor=north east},
    coordnw/.style={font=\scriptsize, anchor=north west},
    coordsw/.style={font=\scriptsize, anchor=south west}
]
\coordinate (A) at (0,0);   
\coordinate (B) at (1,0);   
\coordinate (C) at (1,1);   
\coordinate (D) at (0,1);
\draw (A) -- (D) node[midway,left] {$i$};
\draw (A) -- (B) node[midway,below] {$k$};
\draw (C) -- (D) node[midway,below] {$k$};
\draw[midarrow, red, line width=1.5pt] (B) -- (C) node[midway, left] {$j$};
\node[white] at (A) {};
\node[black] at (B) {};
\node[white] at (D) {};
\node[black] at (C) {};
\end{tikzpicture}}}
\eea
The problem we need to now address is how we can understand in a systematic way what the surface holonomy operator shall be on this plaquette. This is somewhat confusing, since in the vertical direction we have only one white vertex. It appears to be two, but those two are identified by the periodic identification. We can of course guess what the plaquette operator shall be. Since $k$ seems to be contracted, it seems like the plaquette operator shall be either $U^j{}_i$ or $U_j{}^i$. But in fact, since we have lost the bipartite structure, we have no way to tell how the string is oriented with respect to the lattice. That means, we have no longer the ability to decide if the color index shall be placed upstairs or downstairs since regardless of how we orient the initial string, it will always start from a white vertex (and it will also end on the same white vertex). In this situation, we shall draw the string without an orientation, since the orientation has lost its significance. Thus we shall replace the above graphs with the following unoriented graphs,
\bea
\vcenter{\hbox{\begin{tikzpicture}[scale=3,
    white/.style={circle, draw, fill=white, inner sep=0pt, minimum size=6pt},
    black/.style={circle, draw, fill=black, inner sep=0pt, minimum size=6pt},
    label/.style={font=\small, midway, sloped, above},
    coord/.style={font=\scriptsize, anchor=south east},
    coordne/.style={font=\scriptsize, anchor=north east},
    coordnw/.style={font=\scriptsize, anchor=north west},
    coordsw/.style={font=\scriptsize, anchor=south west}
]
\coordinate (A) at (0,0);   
\coordinate (B) at (1,0);   
\coordinate (C) at (1,1);   
\coordinate (D) at (0,1);
\draw[red, line width=1.5pt] (A) -- (D) node[midway, left] {$i$};
\draw (A) -- (B) node[midway,below] {$k$};
\draw (C) -- (D) node[midway,below] {$k$};
\draw (B) -- (C) node[midway,left] {$j$};
\node[white] at (A) {};
\node[black] at (B) {};
\node[white] at (D) {};
\node[black] at (C) {};
\end{tikzpicture}}}
& \rightarrow &
\vcenter{\hbox{\begin{tikzpicture}[scale=3,
    white/.style={circle, draw, fill=white, inner sep=0pt, minimum size=6pt},
    black/.style={circle, draw, fill=black, inner sep=0pt, minimum size=6pt},
    label/.style={font=\small, midway, sloped, above},
    coord/.style={font=\scriptsize, anchor=south east},
    coordne/.style={font=\scriptsize, anchor=north east},
    coordnw/.style={font=\scriptsize, anchor=north west},
    coordsw/.style={font=\scriptsize, anchor=south west}
]
\coordinate (A) at (0,0);   
\coordinate (B) at (1,0);   
\coordinate (C) at (1,1);   
\coordinate (D) at (0,1);
\draw (A) -- (D) node[midway,left] {$i$};
\draw (A) -- (B) node[midway,below] {$k$};
\draw (C) -- (D) node[midway,below] {$k$};
\draw[red, line width=1.5pt] (B) -- (C) node[midway, left] {$j$};
\node[white] at (A) {};
\node[black] at (B) {};
\node[white] at (D) {};
\node[black] at (C) {};
\end{tikzpicture}}}
\eea
We now notice that $i$ is associated with white vertex and $j$ with the black vertex. So these could now instead determine whether these color indices shall be placed upstair or downstairs. But how can we make up our mind? What shall be the precise rule? 

In fact, this is not a very nice strategy, since it would mean that as we move the string, that should now be viewed as a particle in the dimensionally reduced theory, its representation will flip for each lattice step that it moves (to the right),
\bea
\psi^i \rightarrow \psi_j = U_{ji} \psi^i \rightarrow \psi^k = U^{kj} \psi_j \rightarrow ...
\eea
This is not what we expect to see in a five-dimensional lattice discretized theory with a Wilson line connecting each vertex. Thus we shall redraw the graphs one more time, as
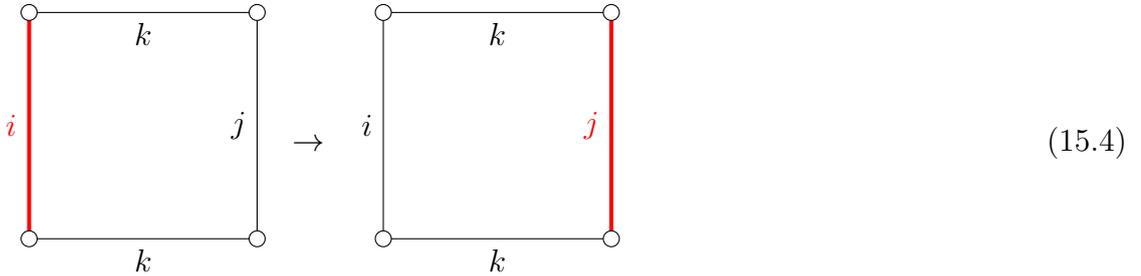
\bea
\vcenter{\hbox{\begin{tikzpicture}[scale=3,
    white/.style={circle, draw, fill=white, inner sep=0pt, minimum size=6pt},
    black/.style={circle, draw, fill=black, inner sep=0pt, minimum size=6pt},
    label/.style={font=\small, midway, sloped, above},
    coord/.style={font=\scriptsize, anchor=south east},
    coordne/.style={font=\scriptsize, anchor=north east},
    coordnw/.style={font=\scriptsize, anchor=north west},
    coordsw/.style={font=\scriptsize, anchor=south west}
]
\coordinate (A) at (0,0);   
\coordinate (B) at (1,0);   
\coordinate (C) at (1,1);   
\coordinate (D) at (0,1);
\draw[red, line width=1.5pt] (A) -- (D) node[midway, left] {$i$};
\draw (A) -- (B) node[midway,below] {$k$};
\draw (C) -- (D) node[midway,below] {$k$};
\draw (B) -- (C) node[midway,left] {$j$};
\node[white] at (A) {};
\node[white] at (B) {};
\node[white] at (D) {};
\node[white] at (C) {};
\end{tikzpicture}}}
& \rightarrow &
\vcenter{\hbox{\begin{tikzpicture}[scale=3,
    white/.style={circle, draw, fill=white, inner sep=0pt, minimum size=6pt},
    black/.style={circle, draw, fill=black, inner sep=0pt, minimum size=6pt},
    label/.style={font=\small, midway, sloped, above},
    coord/.style={font=\scriptsize, anchor=south east},
    coordne/.style={font=\scriptsize, anchor=north east},
    coordnw/.style={font=\scriptsize, anchor=north west},
    coordsw/.style={font=\scriptsize, anchor=south west}
]
\coordinate (A) at (0,0);   
\coordinate (B) at (1,0);   
\coordinate (C) at (1,1);   
\coordinate (D) at (0,1);
\draw (A) -- (D) node[midway,left] {$i$};
\draw (A) -- (B) node[midway,below] {$k$};
\draw (C) -- (D) node[midway,below] {$k$};
\draw[red, line width=1.5pt] (B) -- (C) node[midway, left] {$j$};
\node[white] at (A) {};
\node[white] at (B) {};
\node[white] at (D) {};
\node[white] at (C) {};
\end{tikzpicture}}}
\eea
This could correspond to a wave function transforming all the time in the fundamental representation,
\bea
\psi^i \rightarrow \psi^j = U^j{}_i \psi^i \rightarrow \psi^k = U^k{}_j \psi^j \rightarrow ...
\eea
and then we could assign the rule that if all these four vertices are instead black, then the wave function transforms all the time in the antifundamental representation. But there is still an uncertainty since we could also declare black to correspond to fundamental representation and white to antifundamental. How can we make up our mind? 

We will now show that we can make up our mind. Namely we will derive these rules from principles that we have already introduced and this will then show that there is no new rule that we need to implement ad hoc. 

To this end, we begin with two plaquettes that we stack on top of each other since then we can naturally identify the bottom and top edges while preserving the bipartite structure of the lattice. This is drawn below before we identify the top and bottom, and therefore we put two different color indices $k$ and $k'$ there,
\bea
\vcenter{\hbox{\begin{tikzpicture}[scale=3,
    white/.style={circle, draw, fill=white, inner sep=0pt, minimum size=6pt},
    black/.style={circle, draw, fill=black, inner sep=0pt, minimum size=6pt},
    label/.style={font=\small, midway, sloped, above},
    coord/.style={font=\scriptsize, anchor=south east},
    coordne/.style={font=\scriptsize, anchor=north east},
    coordnw/.style={font=\scriptsize, anchor=north west},
    coordsw/.style={font=\scriptsize, anchor=south west}
]
\coordinate (A) at (0,0);   
\coordinate (B) at (1,0);   
\coordinate (C) at (1,1);   
\coordinate (D) at (0,1);
\coordinate (E) at (0,2);
\coordinate (F) at (1,2);

\draw[midarrow, red, line width=1.5pt] (A) -- (D) node[midway, left] {$i$};
\draw (A) -- (B) node[midway,below] {$k$};
\draw (C) -- (D) node[midway, below] {$\l$};
\draw (B) -- (C) node[midway, left] {$j$};
\draw (E) -- (F) node[midway,below] {$k'$};
\draw[midarrow, red, line width=1.5pt] (D) -- (E) node[midway, left] {$i'$};
\draw (C) -- (F) node[midway, left] {$j'$};
\node[white] at (A) {};
\node[black] at (B) {};
\node[black] at (D) {};
\node[white] at (C) {};
\node[white] at (E) {};
\node[black] at (F) {};
\end{tikzpicture}}}
&\rightarrow &
\vcenter{\hbox{\begin{tikzpicture}[scale=3,
    white/.style={circle, draw, fill=white, inner sep=0pt, minimum size=6pt},
    black/.style={circle, draw, fill=black, inner sep=0pt, minimum size=6pt},
    label/.style={font=\small, midway, sloped, above},
    coord/.style={font=\scriptsize, anchor=south east},
    coordne/.style={font=\scriptsize, anchor=north east},
    coordnw/.style={font=\scriptsize, anchor=north west},
    coordsw/.style={font=\scriptsize, anchor=south west}
]
\coordinate (A) at (0,0);   
\coordinate (B) at (1,0);   
\coordinate (C) at (1,1);   
\coordinate (D) at (0,1);
\coordinate (E) at (0,2);
\coordinate (F) at (1,2);

\draw (A) -- (D) node[midway, left] {$i$};
\draw[midarrow, red, line width=1.5pt] (A) -- (B) node[midway,below] {$k$};
\draw (C) -- (D) node[midway, below] {$\l$} ;
\draw[midarrow, red, line width=1.5pt] (B) -- (C) node[midway, left] {$j$};
\draw[midarrow, red, line width=1.5pt] (F) -- (E) node[midway,below] {$k'$};
\draw (D) -- (E) node[midway, left] {$i'$};
\draw[midarrow, red, line width=1.5pt] (C) -- (F) node[midway, left] {$j'$};
\node[white] at (A) {};
\node[black] at (B) {};
\node[black] at (D) {};
\node[white] at (C) {};
\node[white] at (E) {};
\node[black] at (F) {};
\end{tikzpicture}}}
\eea
We now have no problems writing down the combined surface holonomy as
\bea
U_{k'}{}^{j'}{}_j{}^{k}{}_{i}{}^{i'} &=& U_{k'}{}^{j'}{}_{\l}{}^{i'} U_i{}^{\l}{}_{j}{}^k 
\eea
where on the right-hand side the index $\l$ is contracted. The initial wave function of a string along $i$ and $i'$ transforms into
\bea
\psi_{i'}{}^i \rightarrow \psi_{k'}{}^{j'}{}_j{}^{k} &=& U_{k'}{}^{j'}{}_j{}^{k}{}_{i}{}^{i'} \psi_{i'}{}^i
\eea
Now if we move the top and bottom edges together upon circle compactification, then the portions of the string that go along the edges $k$ and $k'$ will meet on the same edge with opposite orientations. Let us assume that we form a K-spike when they meet by calibration such that the wave function transforms as 
\bea
\psi_{k'}{}^{j'}{}_j{}^k \rightarrow  \frac{1}{\sqrt{N}} \delta_{k'}^{k} \psi^{j'}{}_j \rightarrow  \psi^{j'}{}_j
\eea
In the second transformation the spike is self-annihilating and that corresponds to take the Kronecker delta off the wave function.  By looking at how the wave function has transformed, we can deduce the surface holonomy,
\bea
\psi^{j'}{}_j &=& U^{j'}{}_j{}_{i}{}^{i'} \psi_{i'}{}^i
\eea
The transformed wave function now has a different type of index structure compared to the initial wave function. This is a consequence of the bipartite nature of the lattice. We may resolve that issue either by repeating the same kind of transformation one more step. However, there is also another way. We may cyclically rotate the color indices which we can do here because of the circle identification, so we have $\psi^{j'}{}_j = \psi_j{}^{j'}$. That means that we can equivalently express the same transformation as
\bea
\psi_{j}{}^{j'} &=& U_j{}^{j'}{}_i{}^{i'}  \psi_{i'}{}^i
\eea
This implies that this particular surface holonomy must have the property
\bea
U_j{}^{j'}{}_i{}^{i'} &=& U^{j'}{}_j{}_{i}{}^{i'}
\eea
as it follows from cyclic symmetry of a closed string. This is a special property that arises solely due to the fact that we have compactified the lattice on one circle direction, and we do not have such a property otherwise of the surface holonomy. For the noncompact case the plaquette operator only has the overall cyclicity that acts on all four indices cyclically. Perhaps we should have used a different name on the above surface holonomy to not confuse it with our earlier introduce plaquette operator, but we expect it shall be clear from the context what operator we refer to. The important point is that the operator depends on the placement of color indices in a graph. Thus the operator always needs to be accompanied with the corresponding graph. That is why we draw many graphs. The operator without also drawing the graph that shows where the color indices are placed, would thus not have any meaning to us. 

If we now think of the lattice and the surface holonomies as topological entities, then there is no difficulties in pushing identification of edges further such that we equate the edges $i'$ and $j$, together with our earlier identification of $k'$ with $k$. With this identification, we see that the string segment along the edge $i'$ does not move. The string gets stuck on the edge $i'$ while it only can move from the edge $i$ to the edge $j'$. Since the string is stuck on the edge $i'$ it can not undergo any transformation on that portion, which means that the surface holonomy must now act on this string in the form
\bea
U_j{}^{j'}{}_i{}^{i'} &=& \delta_j^{i'} U^{j'}{}_i
\eea
and the wave function transforms as
\bea
\psi_{j}{}^{j'} &=& U^{j'}{}_i \psi_{j}{}^i
\eea
We may now want to just drop the index $j$ since it does not participate in the transformation, but only produces $N$ copies of essentially the same  wave function and then we get precisely the situation of four white vertices and the transformation rule as we stated in an ad hoc manner above. If instead of identifying $i'$ and $j$, we identify $i$ and $j'$ we get the antifundamental situation with four black vertices instead and the transformation rule
\bea
\psi_{j}{}^{i} &=& U_j{}^{i'}  \psi_{i'}{}^i
\eea
and we may want drop the index $i$. 

It is important to note that these structures are topological. The lattice can be deformed arbitrarily. What we have after any of the identifications above is still a cylinder topologically. The fact that we now only consider one edge in the compact direction does not restrict the radius of compactification in any way. The edge along the compact circle direction can be stretched out or shrunken to any radius we want.

\section{Open strings}
We have assumed that our strings are closed. This seems to be motivated by M5 brane physics, where an M2 brane can stretch between a probe M5 brane and a stack of coincident and parallel M5 branes. The intersection between these branes is a tensile closed string that requires a surface holonomy as it moves adiabically inside the stack of M5 branes. 

But it seems like the surface holonomies can also act on open strings if the endpoints of an open string carries no separate charge. Locally, away from its endpoints, the open string is indistinguishabel from a closed string and everything we have done so far goes through equally well with an open string. 

An open selfdual string can exist inside the worldvolume of coincident M5 branes if there are in addition two orthogonally intersecting M5 branes and an M2 brane stretching between them. The intersecting M5 branes are perceived as two parallel and separated three-branes inside the worldvolume theory of the coincident M5 branes with a selfdual string stretching between them. 

However, if we like to describe a process where one open string splits into two open strings, then we seem to be in trouble. Why this seems problematic is illustrated below:
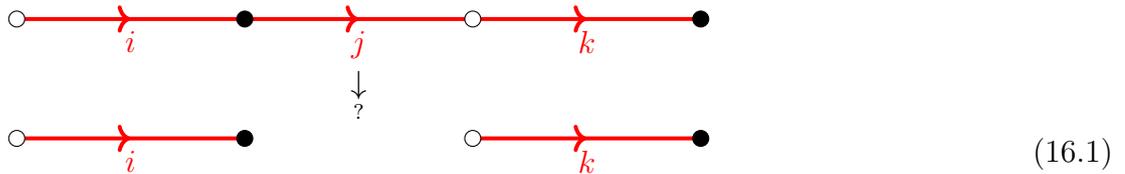
\bea
&\vcenter{\hbox{\begin{tikzpicture}[scale=3,
        white/.style={circle, draw, fill=white, inner sep=0pt, minimum size=6pt},
        black/.style={circle, draw, fill=black, inner sep=0pt, minimum size=6pt},
        coordne/.style={font=\scriptsize, anchor=north east}
    ]
      \coordinate (A) at (0,0);
      \coordinate (B) at (1,0);
      \coordinate (C) at (2,0);
      \coordinate (D) at (3,0);
      \draw[red, line width=1.5pt, midarrow] (A) -- (B) node[midway, below] {$i$};
      \draw[red, line width=1.5pt, midarrow] (B) -- (C) node[midway, below] {$j$};
      \draw[red, line width=1.5pt, midarrow] (C) -- (D) node[midway, below] {$k$};
      \node[white] at (A) {};
      \node[black] at (B) {};
      \node[white] at (C) {};
      \node[black] at (D) {};
    \end{tikzpicture}}}&\cr
&\underset{?}{\downarrow} &\cr
&\vcenter{\hbox{\begin{tikzpicture}[scale=3,
        white/.style={circle, draw, fill=white, inner sep=0pt, minimum size=6pt},
        black/.style={circle, draw, fill=black, inner sep=0pt, minimum size=6pt},
        coordne/.style={font=\scriptsize, anchor=north east}
    ]
      \coordinate (A) at (0,0);
      \coordinate (B) at (1,0);
      \coordinate (C) at (2,0);
      \coordinate (D) at (3,0);
      \draw[red, line width=1.5pt, midarrow] (A) -- (B) node[midway, below] {$i$};
   
      \draw[red, line width=1.5pt, midarrow] (C) -- (D) node[midway, below] {$k$};
      \node[white] at (A) {};
      \node[black] at (B) {};
      \node[white] at (C) {};
      \node[black] at (D) {};
    \end{tikzpicture}}}&
\eea
The problem is now obvious. Removing the segment $j$ from the string would violate charge conservation. While a single open string can not split into two, we can achieve a splitting if two open strings come together and meet along the segment $j$ with opposite orientations and form a K-spike there, which will annihilate segment $j$. 

Let us now discuss the brane picture in which we have $N=2$ coincident M5 branes and one M2 brane that stretches from these two M5 branes to a very distant M5 branes. We have argued that the nonabelian selfdual string in a lattice becomes a discrete set of color indices where each color index takes two values $i = 1,2$ and these values correspond to each of these M5 branes. Now let us suppose that we have a selfdual string with the wave function 
\bea
\psi_j{}^i &=& \delta^i_1 \delta_j^2
\eea
This corresponds to a string that is confined to $M5_1$ and then makes a jump to $M5_2$ at the vertex point in a lattice. Now what happens if we slightly would separate these two M5 branes? The string is the boundary of the M2 brane. But if we separate the M5 branes, the string seems to make  a jump from one M5 brane to another. But of course that is not the correct picture since the string can not leave one M5 brane and exist in the space between two M5 branes. One possibility can be that the two M5 branes are joined together at the cross over point. Then the M5 branes will act as a bridge for the selfdual string enabling it to cross over from one M5 brane to another.

In the continuum limit, this suggests that a generic tensile nonabelian selfdual string will glue together the M5 branes along the entire selfdual string while these M5 branes can still be separated from each other as we are away from the location of the selfdual string. What we are imagining is not so different from how two sheets of textile are sewn together by a thread (the selfdual string). However, the selfdual string may also not hold together all these M5 branes very strongly. This depends on how the selfdual string is sewing together the M5 branes. For instance, if the wave function of the selfdual is string is such that it is nonvanishing only when all color indices are equal, then that corresponds to a string that is confined inside a single M5 brane and then we may be able to separate the M5 branes at the location of the selfdual string.

\section{The continuum limit} 
The surface holonomies that we have discussed do not depend on a spacetime metric. They only depend on the topology of the lattice. In this paper we have made use of a bipartite structure, which is closely tied to a  particular kind of lattice, the hypercubic lattice. A lattice with triangular plaquettes \cite{Akhmedov:2005tn} may instead be assigned a tripartite structure, in which case our construction does not work. 

We are not tied to a regular lattice that is embedded in flat background euclidean spacetime. Instead, we do not have to assume that there is a continuum spacetime at all as long as we only concern about the surface holonomies. However, we still need to require that the lattice is six-dimensional in a topological sense. That is, we shall have at least 12 edges connected to each vertex, those being the lattice remnants of the six dimensions and they need to go in both directions. For the minimal choice that means that there need to be exactly 12 edges connected to each vertex. It is plausible that the only lattice with this property will have the topology of a hypercubic lattice. If each edge would cost money to buy, then it feels plausible to think that any construction company would prefer to build a hypercubic building to any other building since we would naturally think that this would be the cheapest way to construct a building. Bees on the other hand prefer hexagonal rooms because bees have a metric and they pay for building materials. In 2d that leads to the honeycomb structure \cite{Hales1999Honeycomb} as the cheapest construction. Without a metric, the cost can only be measured by counting the number of items, that is, the number of edges for a given number of vertices. 

Once we have a lattice that extends indefinitely in all 'directions' (in a topological sense), we may compactify the lattice in various ways by taking  quotients, by which we identify edges and vertices of the lattice in certain ways. If we do not think on the lattice as embedded in a spacetime, then we only have the topological relations how vertices are connected by edges and so on. There is no distance and no lattice spacing parameter. But if we insist to think on the lattice as embedded in a spacetime geometry, then we may still deform the lattice as long as we do not break it apart. For example, we may stretch edges of the lattice indefinitely in one direction, while we let edges in all the other five direction shrink towards zero. Any such deformation will not affect how the surface holonomies act on strings as they move across the lattice. 

Let us now discuss the problem of the continuum limit. Let us begin with an abelian gauge group and let us assume that the lattice is embedded in $\mb{R}^{1,5}$ with the euclidean metric $ds^2 = \delta_{MN}dx^M dx^N$ as a uniform hypercubic lattice with lattice spacing $a$ in all six dimensions. On a cube that extends in the directions $x^1$, $x^2$ and $x^3$, we put the following universal plaquette operators
\bea
A &=& e^{i a^2 B_{23}(a,0,0,0,0,0)}\cr
B &=& e^{i a^2 B_{31}(0,a,0,0,0,0)}\cr
C &=& e^{i a^2 B_{12}(0,0,a,0,0,0)}\cr
a &=& e^{- i a^2 B_{23}(0,0,0,0,0,0)}\cr
b &=& e^{- i a^2 B_{31}(0,0,0,0,0,0)}\cr
c &=& e^{- i a^2 B_{12}(0,0,0,0,0,0)}
\eea
where we put the electric charge $e =1$ for a more convenient notation. We put these holonomies on the six faces or plaquettes of the cube. Here $a^2$ is the area of one plaquette. By Taylor expanding to first order in the lattice spacing parameter $a$, we find that 
\bea
A a &=& e^{i a^3 \partial_1 B_{23} + \O(a^4)}\cr
B b &=& e^{i a^3 \partial_2 B_{31} + \O(a^4)}\cr
C c &=& e^{i a^3 \partial_3 B_{12} + \O(a^4)}
\eea
all of which are evaluated at the origin. The Wilson surface for the cube is given by 
\bea
W_{123} &=& ABC abc
\eea
where for abelian gauge group the ordering of these six factors does not matter. We evaluate this Wilson surface to be 
\bea
W_{123} &=& e^{i a^3 H_{123} + \O(a^4)}\cr
&=& 1 + i a^3 H_{123} - \frac{a^6}{2} H_{123}^2 + \O(a^8) 
\eea
where 
\bea
H_{123} &=& \partial_1 B_{23} + \partial_2 B_{31} + \partial_3 B_{12}
\eea
is the abelian field strength evaluated at the origin. The euclidean Wilson action is given by 
\bea
S_{Wilson} = - \frac{1}{12} \sum_{x\in \mb{Z}^6} \sum_{M,N,P = 1,...,6} \(W_{MNP} + W_{MNP}^{\dag} - 2\) = \frac{a^6}{12} \sum_{x\in \mb{Z}^6} H_{MNP}^2
\eea
where the sum over $x$ indicates the sum over lattice cubes. This Wilson action is a discretization of the euclidean Maxwell action $\frac{1}{12} \int d^6 x H_{MNP}^2$. 

For the nonabelian generalization, we may try a similar approach. That is, we may expand the nonabelian surface operators to the lowest nontrivial leading order in the lattice spacing parameter as
\bea
A_a{}^e{}_i{}^h &=& \I_a{}^e{}_i{}^h + i a^2 \(B_{23}(a,0,0,0,0,0)\){}_a{}^e{}_i{}^h + \O(a^3)\cr
B_e{}^b{}_f{}^j &=& \I_e{}^b{}_f{}^j + i a^2 \(B_{31}(0,a,0,0,0,0)\){}_e{}^b{}_f{}^j + \O(a^3)\cr
C_i{}^{\l}{}_k{}^j &=& \I_i{}^{\l}{}_k{}^j + i a^2 \(B_{12}(0,0,a,0,0,0)\){}_i{}^{\l}{}_k{}^j + \O(a^3)\cr
a_k{}^f{}_c{}^g &=& \I_k{}^f{}_c{}^g + i a^2 \(B_{23}(0,0,0,0,0,0)\){}_k{}^f{}_c{}^g  + \O(a^3)\cr
b_g{}^d{}_h{}^{\l} &=& \I_g{}^d{}_h{}^{\l} + i a^2 \(B_{31}(0,0,0,0,0,0)\){}_g{}^d{}_h{}^{\l}  + \O(a^3)\cr
c_d{}^c{}_b{}^a &=& \I_d{}^c{}_b{}^a + i a^2 \(B_{12}(0,0,0,0,0,0)\){}_d{}^c{}_b{}^a + \O(a^3)
\eea
Then we may expand the Wilson surface on the cube, denoted as $W_{123}$, that will involve several index contractions, one such term would be
\bea
i a^2 \(B_{23}(a,0,0,0,0,0)\){}_i{}^h{}_a{}^e    \I_e{}^b{}_f{}^j (\I{}_b{}^a{}_d{}^c \I{}^f{}_c{}^g{}_k \I_j{}^k{}_{\l}{}^i)  \I_g{}^d{}_h{}^{\l} 
\eea
We will then get the usual type of derivative terms of same structure as we got for an abelian gauge group, $H = dB$. These are multiplied by the factor of $a^3 = a^2 a$ where $a^2$ is the area of the plaquette and $a$ is from the Taylor expansion, the difference between two opposite sides of the cube being one lattice spacing. Next we may also want to look at higher nonabelian commutator-like terms. Dimensional analysis shows that these should take the form
\bea
H &=& dB + e_2 a B^2 + e_3 a^3 B^3 + ... + e_n a^{2n-3} B^n + ...
\eea
where $e_n$ for $n=2,3,...$ are some dimensionless parameters. Since there is no commutator-like term at order $a^0$, we see that this leads to the field strength $H = dB$, indicating that we may get just a free abelian theory in the continuum limit $a\rightarrow 0$. 

There conclusion that we would get a free theory described by free abelian tensor multiplets from a lattice formulation of a nonabelian tensor gauge field as we take the continuum limit is not new. The same conclusion was also reached in \cite{Nepomechie:1982rb},\cite{Lipstein:2014vca}, \cite{Akhmedov:2005tn}. 

It is however not possible that our lattice discretized theory in the continuum limit would become a free theory described by a set of abelian tensor multiplets, since that would contradict the fact that we have been able to reduce a surface holonomy to a line holonomy of nonabelian Yang-Mills type under dimensional reduction. 

Our intuition for why we get a locally abelian description in the continuum limit is as follows. If we consider a closed tensile string that shrinks to zero size, then it becomes neutral under the gauge group because of charge annihilation and after that, it can move freely as a neutral point-particle without aquiring a surface holonomy as it moves. But the theory also has Wilson surface observables and these suggest that there shall also be extended closed strings and these strings are interacting. They can not move freely since they are charged under the gauge group and aquire a surface holonomy as they move. They would thus be the nonabelian degrees of freedom. 

Incorporating the metric in the discussion of dimensional reduction, what we shall do in order to obtain a W boson in the fundamental representation is that we shall fix the size of the edges that wrap the circle dimension to some radius $R$, while we take the lattice spacing in the five other dimensions to be $a$ that we take to zero while we keep $R$ fixed. Only after we have taken $a$ to zero, can we take $R$ to zero. The two limits may not necessarily commute.

More generally, we may consider some number $n$ of edges wrapped around the compact dimension, in which case we get a W boson transforming in some higher-dimensional representation of dimension $d\leq N^n$. If we do not perform dimensional reduction, but want to stay in six noncompact dimensions, we still have a possibility of having noncommuting objects, these are extended objects, such as surface operators and strings. For such extended objects, the theory appears to be nonabelian, while for pointlike strings, the dynamics appears to be abelian. 

Let us assume that we do not keep any single edge large, as in the case of dimensional reduction. Instead we may consider say a lattice with lattice spacing $a$ in all six euclidean dimensions. As we then take $a$ to zero, the theory can remain a nontrivial interacting theory involving strings. Let us consider a string that goes over $n$ edges so that it has a total lenght $L = n a$ that we keep fixed as we take $a$ to zero. That means that in the limit, $n$ goes to infinity. For such strings, we expect them to remain interacting in the continuum limit. On the other hand, for a string with a number $n$ that is kept fixed, its length $L = n a$ will shrink to zero, and we expect that it will become a noninteracting point-particle in the continuum limit. 

One type of computation that one might try to do is to take the Wilson action of the same form as for abelian gauge group, but now generalized to a nonabelian gauge group, by using the Wilson surface of a cube that we have presented. We may then use this action to compute a correlation function of two such cubes $C_1$ and $C_2$ 
\bea
\frac{\<W(C_1) W(C_2)\>}{\<W(C_1)\>\<W(C_2)\>} 
\eea
Each cube has a side length $a$ and we may assume that these are separated by a large distance $r>>a$ in some direction. What we expect to then find is a correlation function that at a very large distance $r$ behaves essentially like the abelian correlation function. For an abelian gauge group, we can compute this correlation function rather explicitly. We have
\bea
\frac{\<W(C_1) W(C_2)\>}{\<W(C_1)\>\<W(C_2)\>} \sim e^{- \int_{x_1 \in C_1}  \int_{x_2 \in C_2} \<B(x_1) B(x_2)\>}
\eea
and by using the propagator for the abelian gauge potential, $\<B(x_1) B(x_2)\> \sim 1/|x_1-x_2|^4 \sim 1/r^4$ in Feynman gauge, we get
\bea
\frac{\<W(C_1) W(C_2)\>}{\<W(C_1)\>\<W(C_2)\>} = 1 - f \frac{a^4}{r^4} + \O\(\frac{a^5}{r^5}\)
\eea
where $f$ is some numerical constant. 

We may also perform a computation on the gravity dual side. Instead of cubes we may consider two spherical Wilson surfaces $S_1$ and $S_2$, each with radius $R$, and separated by a distance $r$ in some direction. We expect to see a phase transition \cite{Zarembo:1999bu} at a particular numerical value $r/R = r_*/R$. For $r>r_*$ we expect that the correlation function will  behave like an abelian Wilson surface correlation function. Thus the expansion that we expect to see on the gravity side for $r>r_*$ would take the form
\bea
\frac{\<W(S_1) W(S_2)\>}{\<W(S_1)\>\<W(S_2)\>} &=& 1 - f(N) \frac{R^4}{r^4} + \O\(\frac{R^8}{r^8}\)\label{miss}
\eea
The first term would come from the disconnected geometric saddle point that describes two isolated minimal volumes each of which ends on each of the two spheres. The second term is what we would expect to be generated by supergraviton exchanges. In either case this would be a tiny correction in the large $N$ limit. Nevertheless that tiny correction could further confirm whether pointlike strings will behave like abelian particles. On the other hand, if there is a phase transition at a finite value of $R/r$ \cite{Liu:2013uoa}, then that could be an indication that these strings, as they expand, could start to behave like nonabelian interacting strings. 

The second term in (\ref{miss}) was not included in \cite{Corrado:1999pi}. However, there is a scalar mode on the gravity side for spherical harmonic mode number $k = 1$ in their notation that would give this term, but it  was excluded from the gravity side by hand by the argument that there would be no chiral primary operator in the 6d (2,0) theory with conformal dimension $\Delta =2$ that it could match. However, for an abelian 6d (2,0) theory we have five scalar fields $\phi^A$ each of dimension $\Delta =2$ where $A$ transforms in the representation $5$ of the $SO(5)$ R symmetry and these are gauge invariant chiral primary operators. For the nonabelian generalization, one might argue that an operator with $\Delta =2$ would not be gauge invariant. However, we have argued that for a tensile string, when it shrinks to a pointlike string, it loses all its color indices and becomes a gauge singlet abelian point-particle. Whether there can be a chiral primary operator with dimension $\Delta =2$ in the nonabelian theory might then be open for debate.

\section{Taming the rapid charge oscillation}
We may not understand the selfdual string perfectly well yet. But for a selfdual string in a single M5 brane, we can understand its electric and magnetic charges that are uniquely fixed by selfduality and a Dirac-Zwanziger like charge quantization condition. For a straight selfdual string we shall also, at least in the continuum theory, have translation symmetry along the direction of the string. 

In the formution that we have presented here on a lattice, the wave function of the string would take the form $\psi^i{}_j{}^k{}_{\;}...$ in the theory of $N$ coincident M5 branes. 

Let us now take $N = 1$. In that case these indices take just one value. This effective theory of a selfdual string came from $N+1 = 2$ parallel M5 branes with a single M2 brane stretching between them. We break $U(2)$ spontaneously down to $U(1)_1 \times U(1)_2$ by separating the two M5 branes. We have the Lie algebra relations of $U(2)$ 
\bea
[H_1,E_{12}] &=& E_{12}\cr
[H_2,E_{12}] &=& - E_{12}
\eea
that we may decompose into one center of mass generator $H_1+H_2$ satisfying  
\bea
[H_1 + H_2,E_{12}] &=& 0\cr
[H_1 + H_2,E_{21}] &=& 0
\eea
and one $SU(2)$ generator $H := (H_1 - H_2)/2$ satisfying 
\bea
[H,E_{12}] &=& E_{12}\cr
[H,E_{21}] &=& - E_{21}
\eea
Here we are not going to study this decomposition. Instead we are only interested in $M5_1$ and the string inside this single M5 brane. What happens on the distant $M5_2$ is not of our concern. For $M5_1$, the relevant commutation relations are 
\bea
[H_1,E_{12}] &=& E_{12}\cr
[H_1,E_{21}] &=& - E_{21}
\eea
Despite we do not consider an $SU(2)$ Lie algebra here -- in fact these relations do not form any nontrivial Lie algebra at all since clearly $[E_{12},E_{21}] \sim H_1 - H_2$ do not close back onto $H_1$ alone -- we only have the unbroken $U(1)_1$ symmetry generated by its single generator $H_1$, we see that two distinct electric charges are inherited from the underlying $U(2)$ symmetry nevertheless. Regardless we use the $SU(2)$ generator $H := (H_1 - H_2)/2$ or the $U(1)_1$ generator $H_1$ to measure the charges, we get the same charges in both cases, simply because the charge that is measured by $H_2$ is always the opposite to the charge measured by $H_1$. 

When $N=1$ we have that the wave function that we have introduced on the lattice will necessarily take the form 
\bea
\psi^1{}_1{}^1{}_1...\label{problem}
\eea
because all color indices in this case can take only one possible value $i,j,k,... = 1$. These color indices represent the $U(1)_1$ charges. An index upstairs means $H_1 \psi^1 = \psi^1$ and an index downstairs means $H_1 \psi_1 = - \psi_1$. The language is not that of $SU(2)$ representation theory, but the outcome, the charges are the same as those we would measure by the $SU(2)$ charge operator $H = (H_1 - H_2)/2$. 

This is now a problem because clearly the selfdual string in a single M5 brane obtained from an M2 brane stretching between two M5 branes, should be translationally symmetric along its direction. The electric charge should not oscillate between positive and negative charge as in the wave function (\ref{problem}). Thus we should rather have that the wave function takes the form
\bea
\psi^{1111...}
\eea
or 
\bea
\psi_{1111...}
\eea
depending on how the string is oriented. There should be no possibility of having an electric charge that is rapidly oscillating as we move along the string. 

The oscillating structure is tied to the string. We do not manipulate the lattice, but we can manipulate the way that we present the wave function of the string. We can define a physical wave function as follows,
\bea
\psi_{{\mbox{phys}}}^{ijkl...} &=& S^{jj'} S^{\l\l'} \psi^i{}_{j'}{}^k{}_{\l'}...
\eea
The definition can be made unique as follows. We start by reading the color indices of the string from the right to the left. These indices oscillate between up and down for each segment of the string that we traverse. We bring the indices that are downstairs up by applying $S^{ij}$ that raises this index and it will transform an index from the anti-fundamental to the fundamental representation. No gauge invariant such matrix $S^{ij}$ exists. We do not need this matrix to be gauge invariant, we only need to pick one such matrix and then follow it along as it transforms through the computations. We then declare that the charges, or the color indices, on the physical wave function are the physical charges that we can measure in an experiment, while the color indices on the original wave function are just tools for putting together strings and surface holonomies in a consistent manner on the lattice. For $N = 1$ there is accidentally a gauge invariant choice, namely $S^{ij} = \eps^{ij}$, but for $N>1$ this is not the case. 

We now see that not only do we need to introduce $S^{ij}$ to obtain a unit surface holonomy. We also need it to obtain physical charges. 

It is only by transforming to the physical wave function that we can have any hope of finding a continuum limit. With a very rapid charge oscillation as we traverse the string from one segment to the next it would seem hopeless to take the continuum limit.

\subsection*{Acknowledgments}
I would like to thank Dongsu Bak for collaboration at an early stage of this work. This work was supported in part by Basic Science Research Program through NRF funded by the Ministry of Education (2018R1A6A1A06024977).

\appendix
\section{The hexeract}
We take the lattice to be bipartite and densily packed with six-dimensional cubes, called 'hexeracts'. As far as surface holonomies concern, we do not need a metric and only the topological structure of the lattice will be important. What is quite interesting is that this topological structure of the hexaract can actually be completely visualized in a graph that we can draw on a paper, although we can not visualize geometric objects in higher dimensions than three dimensions. 

Since the hexeract as a geometric object lives in six dimensions, let us start with something that we can visualize, namely the three-dimensional cube. Below we will draw the cube in such a way that its graph can be generalized to higher dimensions,
\bea
\begin{tikzpicture}[scale=2.5, 
    black_node/.style={circle, draw=black, fill=black, minimum size=0.18cm, inner sep=0pt},
    white_node/.style={circle, draw=black, fill=white, minimum size=0.18cm, inner sep=0pt, thick},
    every label/.style={font=\tiny, inner sep=3pt, color=black}]


  \node[black_node, label=above:{(000)}] (n000) at (0, 1.5) {};

  \node[white_node, label=left:{(001)}] (n001) at (-1, 0.5) {};
  \node[white_node, label=above:{(010)}] (n010) at (0, 0.5) {};
  \node[white_node, label=right:{(100)}] (n100) at (1, 0.5) {};

  \node[black_node, label=left:{(011)}] (n011) at (-1, -0.5) {};
  \node[black_node, label=below:{(101)}] (n101) at (0, -0.5) {};
  \node[black_node, label=right:{(110)}] (n110) at (1, -0.5) {};

  \node[white_node, label=below:{(111)}] (n111) at (0, -1.5) {};

  \begin{scope}[thick, gray!40]
    \draw (n000) -- (n001);
    \draw (n000) -- (n010);
    \draw (n000) -- (n100);

    \draw (n001) -- (n011);
    \draw (n001) -- (n101);
    \draw (n010) -- (n011);
    \draw (n010) -- (n110);
    \draw (n100) -- (n101);
    \draw (n100) -- (n110);

    \draw (n011) -- (n111);
    \draw (n101) -- (n111);
    \draw (n110) -- (n111);
  \end{scope}
\end{tikzpicture}
\eea
With just a small amount of imagination we can easily see the cube as an embedding in three dimensions from the above two-dimensional graph. But the two-dimensional graph itself can be generalized to any higher dimensions. In higher dimensions our ability to visualize gets lost since we can not visualize objects in higher dimensions. But we can still draw the graphs. For example, the graph below captures the topology of a four-dimensional cube (4-cube or tesseract). We have highlighted two of its eight 3-cubes in red and blue colors respectively. These are located at vertices $(abc0)$ and $(abc1)$, with $a,b,c = 0,1$, respectively,
\bea
\begin{tikzpicture}[scale=2.5, 
    black_node/.style={circle, draw=black, fill=black, minimum size=0.18cm, inner sep=0pt},
    white_node/.style={circle, draw=black, fill=white, minimum size=0.18cm, inner sep=0pt, thick},
    every label/.style={font=\tiny, inner sep=3pt, color=black}]


  \node[black_node, label=above:{(0000)}] (n0000) at (0, 2) {};

  \node[white_node, label=above:{(0001)}] (n0001) at (-1.5, 1) {};
  \node[white_node, label=above:{(0010)}] (n0010) at (-0.5, 1) {};
  \node[white_node, label=above:{(0100)}] (n0100) at (0.5, 1) {};
  \node[white_node, label=above:{(1000)}] (n1000) at (1.5, 1) {};

  \node[black_node, label=left:{(0011)}] (n0011) at (-2.5, 0) {};
  \node[black_node, label=below:{(0101)}] (n0101) at (-1.5, 0) {};
  \node[black_node, label=below:{(1001)}] (n1001) at (-0.5, 0) {};
  \node[black_node, label=below:{(0110)}] (n0110) at (0.5, 0) {};
  \node[black_node, label=below:{(1010)}] (n1010) at (1.5, 0) {};
  \node[black_node, label=right:{(1100)}] (n1100) at (2.5, 0) {};

  \node[white_node, label=below:{(0111)}] (n0111) at (-1.5, -1) {};
  \node[white_node, label=below:{(1011)}] (n1011) at (-0.5, -1) {};
  \node[white_node, label=below:{(1101)}] (n1101) at (0.5, -1) {};
  \node[white_node, label=below:{(1110)}] (n1110) at (1.5, -1) {};

  \node[black_node, label=below:{(1111)}] (n1111) at (0, -2) {};

  \begin{scope}[thick, black!40]
    \foreach \d in {0001, 0010, 0100, 1000} {\draw (n0000) -- (n\d);}
    
    \draw (n0001) -- (n0011); \draw (n0001) -- (n0101); \draw (n0001) -- (n1001);
    \draw (n0010) -- (n0011); \draw (n0010) -- (n0110); \draw (n0010) -- (n1010);
    \draw (n0100) -- (n0101); \draw (n0100) -- (n0110); \draw (n0100) -- (n1100);
    \draw (n1000) -- (n1001); \draw (n1000) -- (n1010); \draw (n1000) -- (n1100);

    \draw (n0011) -- (n0111); \draw (n0011) -- (n1011);
    \draw (n0101) -- (n0111); \draw (n0101) -- (n1101);
    \draw (n1001) -- (n1011); \draw (n1001) -- (n1101);
    \draw (n0110) -- (n0111); \draw (n0110) -- (n1110);
    \draw (n1010) -- (n1011); \draw (n1010) -- (n1110);
    \draw (n1100) -- (n1101); \draw (n1100) -- (n1110);

    \foreach \s in {0111, 1011, 1101, 1110} {\draw (n\s) -- (n1111);}
  \end{scope}

  \begin{scope}[ultra thick, red, opacity=0.8]
    \draw (n0001) -- (n0011); \draw (n0001) -- (n0101); \draw (n0001) -- (n1001);
    \draw (n1111) -- (n0111); \draw[dashed] (n1111) -- (n1011); \draw (n1111) -- (n1101);
    \draw (n0011) -- (n0111); \draw[dashed] (n0011) -- (n1011);
    \draw (n0101) -- (n0111); \draw (n0101) -- (n1101);
    \draw[dashed] (n1001) -- (n1011); \draw (n1001) -- (n1101);
  \end{scope}
  \begin{scope}[ultra thick, blue, opacity=0.8]
    \draw (n0000) -- (n0010); \draw (n0000) -- (n0100); \draw (n0000) -- (n1000);
    \draw (n1110) -- (n0110); \draw[dashed] (n1110) -- (n1010); \draw (n1110) -- (n1100);
    \draw (n0010) -- (n0110); \draw[dashed] (n0010) -- (n1010);
    \draw (n0100) -- (n0110); \draw (n0100) -- (n1100);
    \draw[dashed] (n1000) -- (n1010); \draw (n1000) -- (n1100);
  \end{scope}
\end{tikzpicture}
\eea
There are eight $3$-cubes in the $4$-cube in total. They correspond to the number of ways we pick the vertices that form a cube. To obtain these vertices, we need to first select one slot among four entries and in that slot insert either 0 or 1. This leads to
\bea
8 &=& \binom{4}{1} 2
\eea
possibilies. This result generalizes to an $n$-cube for which we have the following number of $k$-cubes,
\bea
\binom{n}{k} 2^{n-k}
\eea
This formula is valid for $k = 0,1,...,n$. One may check that the number of $0$-cubes are the $2^n$ number of vertices, and the number of $n$-cubes is equal to one. 

We are now specifically interested in the hypercubic lattice whose building blocks are hexeracts. The topological structure of the hexeract is drawn below, 

\begin{tikzpicture}[scale=0.8, 
    b/.style={circle, fill=black, minimum size=5pt, inner sep=0pt}, 
    w/.style={circle, draw=black, fill=white, minimum size=5pt, inner sep=0pt, thick},
    e/.style={black!40, line width=0.3pt}]

\node[b] (v0) at (0,6) {}; 
\foreach \i [count=\x] in {1,2,4,8,16,32} \node[w] (v\i) at (-3.5+\x, 4) {}; 
\foreach \i [count=\x] in {3,5,6,9,10,12,17,18,20,24,33,34,36,40,48} \node[b] (v\i) at (-8+\x, 2) {}; 
\foreach \i [count=\x] in {7,11,13,14,19,21,22,25,26,28,35,37,38,41,42,44,49,50,52,56} \node[w] (v\i) at (-10.5+\x, 0) {}; 
\foreach \i [count=\x] in {15,23,27,29,30,39,43,45,46,51,53,54,57,58,60} \node[b] (v\i) at (-8+\x, -2) {}; 
\foreach \i [count=\x] in {31,47,55,59,61,62} \node[w] (v\i) at (-3.5+\x, -4) {}; 
\node[b] (v63) at (0,-6) {}; 


\begin{scope}[e]
\draw (v0)--(v1); \draw (v0)--(v2); \draw (v0)--(v4); \draw (v0)--(v8); \draw (v0)--(v16); \draw (v0)--(v32);

\draw (v1)--(v3); \draw (v1)--(v5); \draw (v1)--(v9); \draw (v1)--(v17); \draw (v1)--(v33);
\draw (v2)--(v3); \draw (v2)--(v6); \draw (v2)--(v10); \draw (v2)--(v18); \draw (v2)--(v34);
\draw (v4)--(v5); \draw (v4)--(v6); \draw (v4)--(v12); \draw (v4)--(v20); \draw (v4)--(v36);
\draw (v8)--(v9); \draw (v8)--(v10); \draw (v8)--(v12); \draw (v8)--(v24); \draw (v8)--(v40);
\draw (v16)--(v17); \draw (v16)--(v18); \draw (v16)--(v20); \draw (v16)--(v24); \draw (v16)--(v48);
\draw (v32)--(v33); \draw (v32)--(v34); \draw (v32)--(v36); \draw (v32)--(v40); \draw (v32)--(v48);

\draw (v3)--(v7); \draw (v3)--(v11); \draw (v3)--(v19); \draw (v3)--(v35);
\draw (v5)--(v7); \draw (v5)--(v13); \draw (v5)--(v21); \draw (v5)--(v37);
\draw (v6)--(v7); \draw (v6)--(v14); \draw (v6)--(v22); \draw (v6)--(v38);
\draw (v9)--(v11); \draw (v9)--(v13); \draw (v9)--(v25); \draw (v9)--(v41);
\draw (v10)--(v11); \draw (v10)--(v14); \draw (v10)--(v26); \draw (v10)--(v42);
\draw (v12)--(v13); \draw (v12)--(v14); \draw (v12)--(v28); \draw (v12)--(v44);
\draw (v17)--(v19); \draw (v17)--(v21); \draw (v17)--(v25); \draw (v17)--(v49);
\draw (v18)--(v19); \draw (v18)--(v22); \draw (v18)--(v26); \draw (v18)--(v50);
\draw (v20)--(v21); \draw (v20)--(v22); \draw (v20)--(v28); \draw (v20)--(v52);
\draw (v24)--(v25); \draw (v24)--(v26); \draw (v24)--(v28); \draw (v24)--(v56);
\draw (v33)--(v35); \draw (v33)--(v37); \draw (v33)--(v41); \draw (v33)--(v49);
\draw (v34)--(v35); \draw (v34)--(v38); \draw (v34)--(v42); \draw (v34)--(v50);
\draw (v36)--(v37); \draw (v36)--(v38); \draw (v36)--(v44); \draw (v36)--(v52);
\draw (v40)--(v41); \draw (v40)--(v42); \draw (v40)--(v44); \draw (v40)--(v56);
\draw (v48)--(v49); \draw (v48)--(v50); \draw (v48)--(v52); \draw (v48)--(v56);

\draw (v7)--(v15); \draw (v7)--(v23); \draw (v7)--(v31); 
\draw (v7)--(v15); \draw (v7)--(v23); \draw (v7)--(v31); 
\draw (v7)--(v15); \draw (v7)--(v23); \draw (v7)--(v39);
\draw (v11)--(v15); \draw (v11)--(v27); \draw (v11)--(v43);
\draw (v13)--(v15); \draw (v13)--(v29); \draw (v13)--(v45);
\draw (v14)--(v15); \draw (v14)--(v30); \draw (v14)--(v46);
\draw (v19)--(v23); \draw (v19)--(v27); \draw (v19)--(v51);
\draw (v21)--(v23); \draw (v21)--(v29); \draw (v21)--(v53);
\draw (v22)--(v23); \draw (v22)--(v30); \draw (v22)--(v54);
\draw (v25)--(v27); \draw (v25)--(v29); \draw (v25)--(v57);
\draw (v26)--(v27); \draw (v26)--(v30); \draw (v26)--(v58);
\draw (v28)--(v29); \draw (v28)--(v30); \draw (v28)--(v60);
\draw (v35)--(v39); \draw (v35)--(v43); \draw (v35)--(v51);
\draw (v37)--(v39); \draw (v37)--(v45); \draw (v37)--(v53);
\draw (v38)--(v39); \draw (v38)--(v46); \draw (v38)--(v54);
\draw (v41)--(v43); \draw (v41)--(v45); \draw (v41)--(v57);
\draw (v42)--(v43); \draw (v42)--(v46); \draw (v42)--(v58);
\draw (v44)--(v45); \draw (v44)--(v46); \draw (v44)--(v60);
\draw (v49)--(v51); \draw (v49)--(v53); \draw (v49)--(v57);
\draw (v50)--(v51); \draw (v50)--(v54); \draw (v50)--(v58);
\draw (v52)--(v53); \draw (v52)--(v54); \draw (v52)--(v60);
\draw (v56)--(v57); \draw (v56)--(v58); \draw (v56)--(v60);

\draw (v15)--(v31); \draw (v15)--(v47); \draw (v23)--(v31); \draw (v23)--(v55);
\draw (v27)--(v31); \draw (v27)--(v59); \draw (v29)--(v31); \draw (v29)--(v61);
\draw (v30)--(v31); \draw (v30)--(v62); \draw (v39)--(v47); \draw (v39)--(v55);
\draw (v43)--(v47); \draw (v43)--(v59); \draw (v45)--(v47); \draw (v45)--(v61);
\draw (v46)--(v47); \draw (v46)--(v62); \draw (v51)--(v55); \draw (v51)--(v59);
\draw (v53)--(v55); \draw (v53)--(v61); \draw (v54)--(v55); \draw (v54)--(v62);
\draw (v57)--(v59); \draw (v57)--(v61); \draw (v58)--(v59); \draw (v58)--(v62);
\draw (v60)--(v61); \draw (v60)--(v62);

\draw (v31)--(v63); \draw (v47)--(v63); \draw (v55)--(v63); \draw (v59)--(v63); \draw (v61)--(v63); \draw (v62)--(v63);
\end{scope}

\end{tikzpicture}

The hexeract has 64 vertices that are connected in certain ways by 192 edges so that each vertex connects to six other vertices of the opposite type (white or black). These edges form 240 plaquettes (2-cubes), 160 cubes (3-cubes), 60 tesseracts (4-cubes), and 12 penteracts (5-cubes). 

If we stretch out the above net of lines in all six Euclidean dimensions then we get a regular hexeract. This is a very simple object if one looks at it from afar along anyone of the six Euclidean coordinate directions. One will then see only one plaquette of the hexeract. There are in total $\binom{6}{2} = 15$ coordinate planes in which this plaquette can be extended. We also need to specify if that is seen from the top or from the bottom in the given transverse direction. Since there are four transverse direction to anyone of the planes, there are in total $16$ different ways to assign tops and bottoms in each of those four directions. In total there are therefore $240$ different plaquettes of the hexeract that we may see by looking at this object from afar. If then we were to rotate the hexeract a little bit in our view, then we will see a tremendous complexity arises from this single plaquette, and it may appear to us in a shape similar to what is drawn above.

\end{document}